\documentclass[a4paper,fleqn,usenatbib]{mnras}
\usepackage{aasmacros}
\usepackage{times}
\usepackage{amsmath}
\usepackage{graphicx}
\usepackage{amssymb}
\usepackage[usenames,dvipsnames]{color}
\usepackage{url}
\usepackage{fixltx2e}
\usepackage{mathtools}
\usepackage{amsmath}
\usepackage{float}
\usepackage{cite}
\usepackage{fixltx2e}
\bibpunct{(}{)}{;}{a}{,}{,}
\usepackage{longtable}
\usepackage{hyperref}
\usepackage{morefloats}
\usepackage{gensymb}
\usepackage{enumitem}
\usepackage{tabu}
\usepackage{totpages}
\usepackage{amssymb}
\usepackage{xcolor}
\usepackage[T1]{fontenc}
\usepackage{ae,aecompl}

\let\cite=\citen

\def\apj{ApJ}
\def\aj{AJ}
\def\mnras{MNRAS}
\def\pasa{PASA}

\def\na{New Astronomy}
\def\pasp{PASP}

\bibpunct{(}{)}{;}{a}{}{,}

\def\full{{\it FULL}}
\def\galr{{\it SUB}}
\def\minur{{\it SUB}-{\it FT}}

\makeatletter
\newcommand*{\rom}[1]{\expandafter\@slowromancap\romannumeral #1@}
\makeatother

\title[radio cosmic web]{Low Frequency Radio Constraints on the Synchrotron Cosmic Web}

\author[Vernstrom et. al]{T. Vernstrom\thanks{E-mail:vernstrom@dunlap.utoronto.ca}$^1$, B.M. Gaensler$^1$, S. Brown$^2$, E. Lenc $^{3,4}$, R.P. Norris$^{5,6}$ \\
$^1$Dunlap Institute for Astronomy and Astrophysics University of Toronto, Toronto, ON M5S 3H4, Canada\\
$^2$Department of Physics and Astronomy, University of Iowa, 203 Van Allen Hall, Iowa City, IA 52242, U.S.A.\\
$^3$Sydney Institute for Astronomy, School of Physics, The University of Sydney, NSW 2006, Australia\\
$^4$ARC Centre for Excellence for All-sky Astrophysics (CAASTRO)\\
$^5$CSIRO Astronomy $\&$ Space Science, PO Box 76, Epping, NSW 1710, Australia\\
$^6$Western Sydney University, Locked Bag 1797, Penrith South, NSW 1797, Australia\\
 }
\begin{document}
  
\pagerange{\pageref{firstpage}--\pageref{lastpage}} \pubyear{2016}

\maketitle

\label{firstpage}
\begin{abstract}
We present a search for the synchrotron emission from the synchrotron cosmic web by cross correlating 180MHz radio images from the Murchison Widefield Array with tracers of large scale structure (LSS). We use two versions of the radio image covering $21.76\degr \times 21.76\degr$ with point sources brighter than $0.05\,$Jy subtracted, with and without filtering of Galactic emission. As tracers of the LSS we use the Two-Micron-All-Sky-Survey (2MASS) and the Widefield InfraRed Explorer (WISE) redshift catalogues to produce galaxy number density maps. The cross correlation functions all show peak amplitudes at zero degrees, decreasing with varying slopes towards zero correlation over a range of one degree. The cross correlation signals include components from point source, Galactic, and extragalactic diffuse emission. We use models of the diffuse emission from smoothing the density maps with Gaussians of sizes 1--4 Mpc to find limits on the cosmic web components. From these models we find surface brightness 99.7$\,$per cent upper limits in the range of 0.09--$2.20\,$mJy beam$^{-1}$ (average beam size of $2.6\,$arcmin), corresponding to 0.01--$0.30\,$mJy arcmin$^{-2}$. Assuming equipartition between energy densities of cosmic rays and the magnetic field, the flux density limits translate to magnetic field strength limits of 0.03--$1.98\, \mu$G, depending heavily on the spectral index. We conclude that for a $3\sigma$ detection of $0.1\, \mu$G magnetic field strengths via cross correlations, image depths of sub-mJy to sub-$\mu$Jy are necessary. We include discussion on the treatment and effect of extragalactic point sources and Galactic emission, and next steps for building on this work. 
\end{abstract}


\section{Introduction}
\label{sec:introduction}

According to current cosmological theories, $5\,$per cent of the Universe mass/energy density is composed of normal baryons. In the local Universe (redshifts of $z\la2$), stars, neutral atomic and molecular gas, and the diffuse hot gas within galaxy clusters, account for only roughly one third of the baryons predicted by Big Bang nucleosynthesis \citep{Fukugita98}. Some fraction of the baryons not accounted for lie in the low redshift Ly$\alpha$ forest \citep[e.g.][]{Penton00b}. However, it is believed that roughly half of all the baryons in the Universe reside in the shocked warm-hot ($T\simeq 10^5$--$10^7\,$K) intergalactic medium \citep[WHIM, e.g.][]{Dave01,Gheller15}. However, this has yet to be observationally confirmed, with the exception of detections of O VI $\lambda \lambda$1032, 1038 absorbers \citep[the coolest WHIM components,][]{Tripp00a,Tripp00b}. Simulations predict that most of the WHIM resides in the filamentary network that characterizes the large-scale structure (LSS) of the Universe \citep[e.g.][]{Cen99}. 

Strong accretion shocks from matter falling into and along filaments between clusters accelerate the particles of the shock-heated cosmic gas to relativistic energies, tracing the WHIM distribution, known as the cosmic web \citep{Keshet04,Ryu08,Skillman08}. Cosmological accretion shocks, due to their large Mach number (${\cal M}\sim 10$--$10^2$), should be efficient accelerators of cosmic ray protons \citep{Ryu03,Pfrommer06} and electrons \citep{Hoeft07,Skillman11}. Merger, infall, and accretion shocks have been observed at the edges of dense clusters \citep[radio relics, e.g.][]{Bagchi02,Brown11a,Feretti12}. Similar, fainter, emission should be present further from cluster cores and out into the cosmic web filaments. Typical shock velocities of 500--$1000\,$km s$^{-1}$ are expected in filaments given reasonable cosmological properties. Such velocities are high enough to accelerate particles to relativistic energies, which would produce synchrotron emission in the presence of even a weak magnetic field \citep{Keshet03}, resulting in the {\it synchrotron} cosmic web.

This diffuse synchrotron emission should be observable at radio wavelengths \citep{Wilcots04}. The detection of this synchrotron emission would not only map the cosmic web, but also measure the energy distribution and electron density, allowing for an inferred measurement of the cosmic magnetic field strengths \citep{Rudnick09}. Direct imaging of the synchrotron cosmic web emission will likely not be possible until future radio telescopes such as the Square Kilometre Array \citep[SKA,][]{Wilcots04}. \citet{Vazza15} used simulations of the cosmic web to predict the sensitivity of different current and future telescopes and surveys to the synchrotron cosmic web signal. They found that before SKA1-LOW, the LOFAR and Murchison Widefield Array telescopes have the best chance of detection. The difficulty of this measurement is due to Galactic and extragalactic foregrounds, the latter being comprised mainly of unresolved radio point sources \citep{Dimatteo02}. The Galactic foreground is brighter than any predictions for the synchrotron cosmic web. Faint (sub-mJy) extragalactic point sources will become confused with low surface brightness diffuse emission at larger ($\sim$ arcmin) angular scales \citep{Vernstrom14}. 

A statistical detection (or constraints) could be possible using current radio data. A technique that may alleviate the foreground and confusion problems is cross correlation of radio images with optical or infrared (IR) tracers of large-scale structure \citep{Keshet04,Brown10}. The cross correlation method has been used previously to detect the Integrated Sachs-Wolfe effect by cross correlating CMB data with galaxy number density maps \citep[e.g.][]{Planck14xix}. Cross correlation is sensitive to signals that are spatially correlated and therefore can detect signals below the (uncorrelated) noise level. Assuming the galaxy number densities do not correlate with the noise or Galactic emission, all that should be left is any correlation between the LSS tracers and extragalactic emission. In this paper we use low-frequency radio data and IR galaxy catalogues to measure the radio-IR cross correlation function in an attempt to constrain the synchrotron cosmic web properties.

The outline for this paper is as follows. In Section~\ref{sec:data} we describe the radio data used, including the data calibration and imaging process, as well as a discussion about the image noise properties. Section~\ref{sec:odata} provides details on the tracers of large-scale structure used to cross correlate with the radio data. Section~\ref{sec:crx} details the method for computing the cross correlation function and the process for generating null results from randomly generated data to obtain limits on a zero-correlation result. In Section~\ref{sec:results} we provide the results of cross correlating the radio images with the galaxy number density maps. In Section~\ref{sec:discussion} we discuss the results including the effects from point sources and Galactic emission, as well as providing limits on the flux density, $S$, and magnetic field strength of the synchrotron cosmic web derived from the results. Also in Section~\ref{sec:discussion} we discuss the limitations of current diffuse emission models and detail other possible future tests for detecting or constraining the synchrotron cosmic web. The intrinsic parameters quoted in this paper are computed assuming a WMAP 7-year  $\Lambda$CDM cosmology with $H_0 = 70.2\,$km s$^{-1}$ Mpc$^{-1}$, $\Omega_{\rm m} = 0.272$, and $\Omega_{\Lambda} = 0.728$ \citep{Komatsu11}.

\section{Radio Data}
\label{sec:data}

We use data from the Murchison Widefield Array (MWA). The MWA is an interferometer made up of 128 16-crossed-pair dipole antenna `tiles'. For details of the technical design and specifications we refer readers to \citet{Lonsdale09} and \citet{Tingay13}, with \citet{Bowman13} detailing the primary science objectives. 

The MWA is well suited for cosmic web studies for several reasons. First, it has a large field of view ($15\degr$--$50\degr$) allowing us to study a large part of the sky. It covers a large frequency range at low frequencies (80--$300\,$MHz), with low frequencies being desirable because if the cosmic web has even a moderately steep spectral index $\alpha \la -0.8$ (with flux density $S(\nu)\propto \nu^{\alpha}$) the signal should be stronger at lower frequencies. Also, and importantly, the MWA has very good sensitivity to large angular scales due to its large number of short baselines. 

For this paper we use data from the MWA Epoch of Reionization field 0 (EoR0), centred on $\alpha = 00^{\rm{h}}00^{\rm{m}}00^{\rm{s}}$, $\delta = -27^{\circ}00\arcmin00\arcsec$ (J2000). This field was chosen for its low expected Galactic contamination (with a Galactic latitude of $-78^{\degr}$), overlap with other surveys, and large amount of data already taken with the MWA for the EoR experiment ($\sim 45\,$hours). Below we provide details on the calibration and imaging process. 

\subsection{Calibration and imaging}
\label{sec:rdatcal}

\begin{figure*}
\includegraphics[scale=0.395]{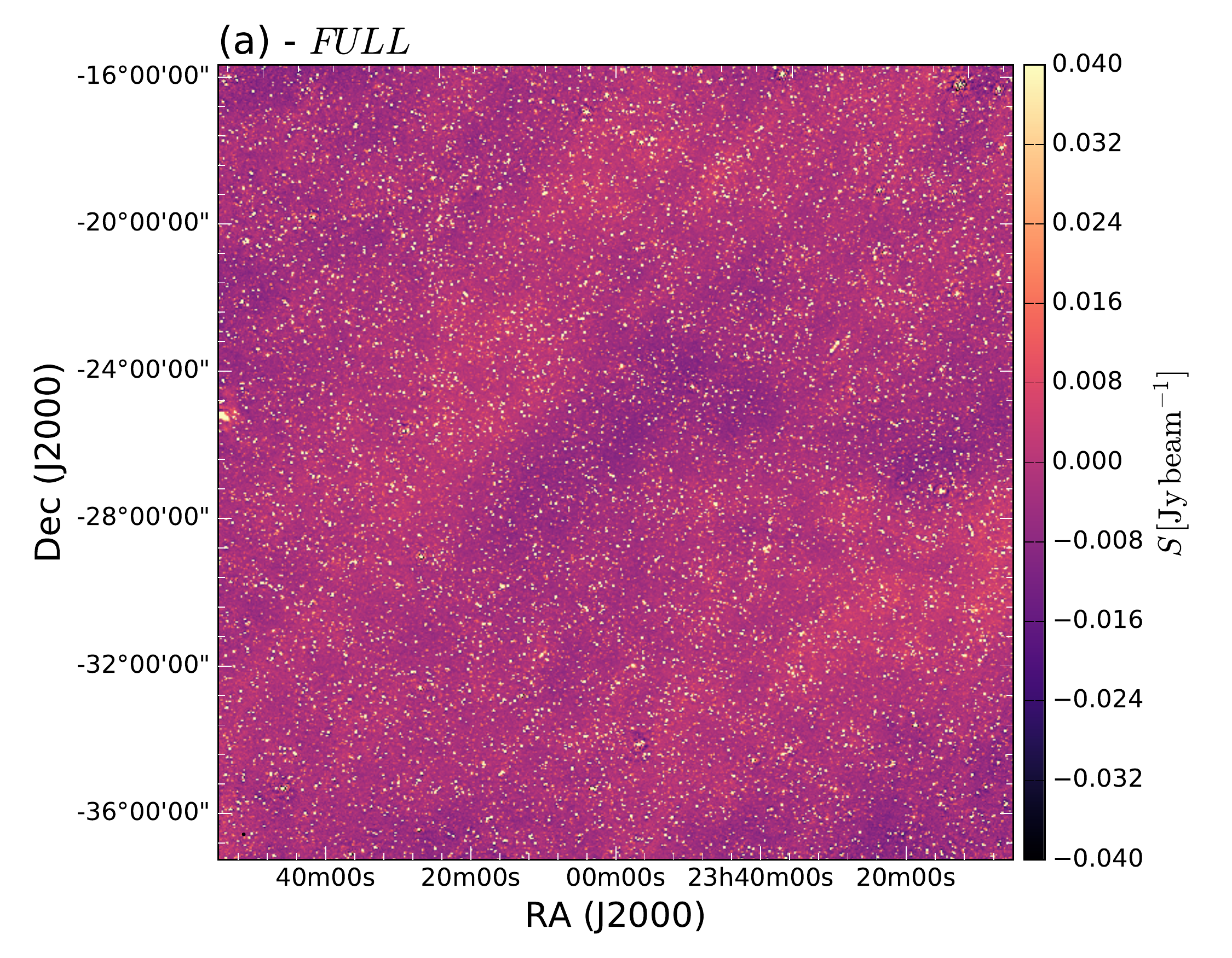}\includegraphics[scale=0.395]{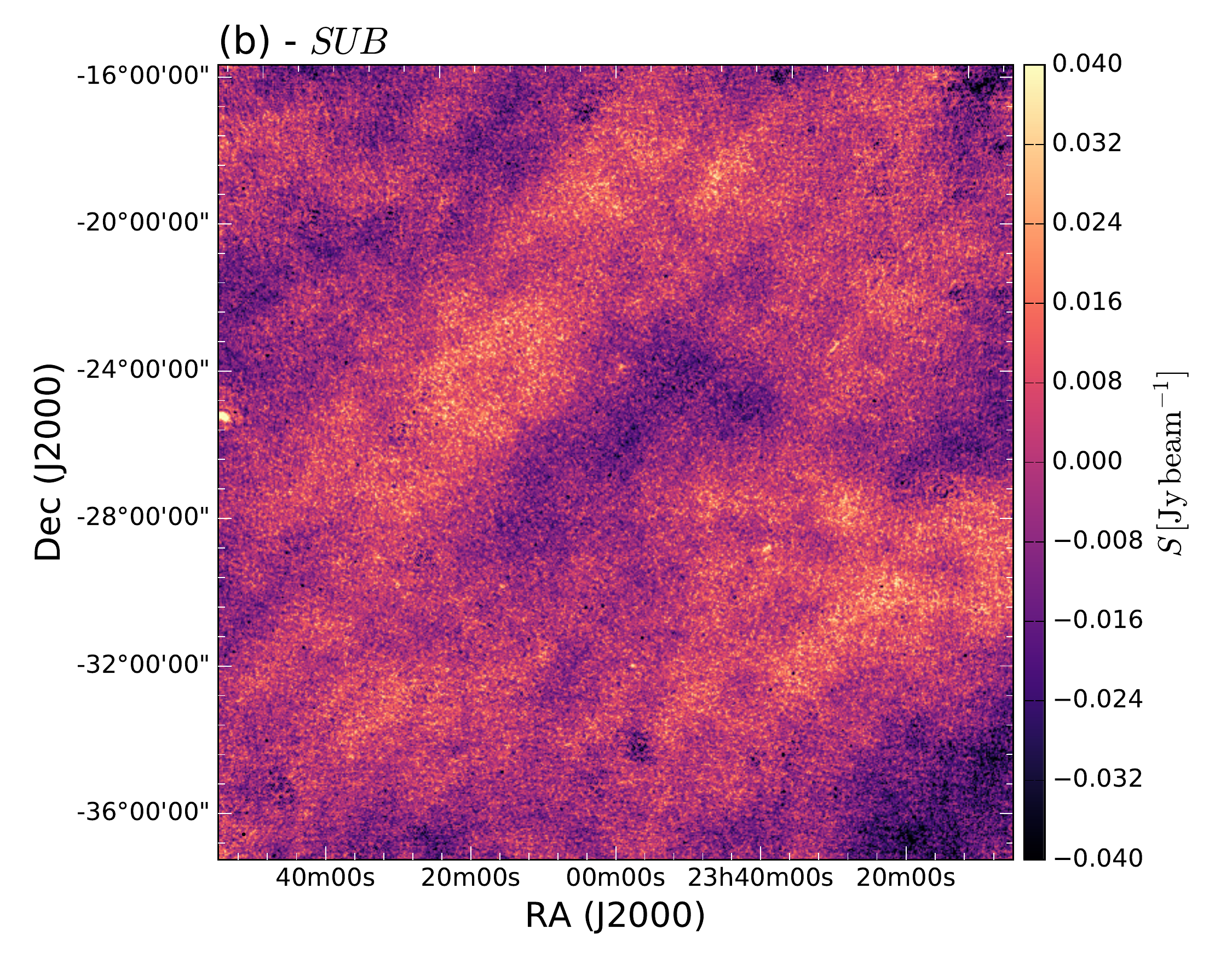}
\includegraphics[scale=0.395]{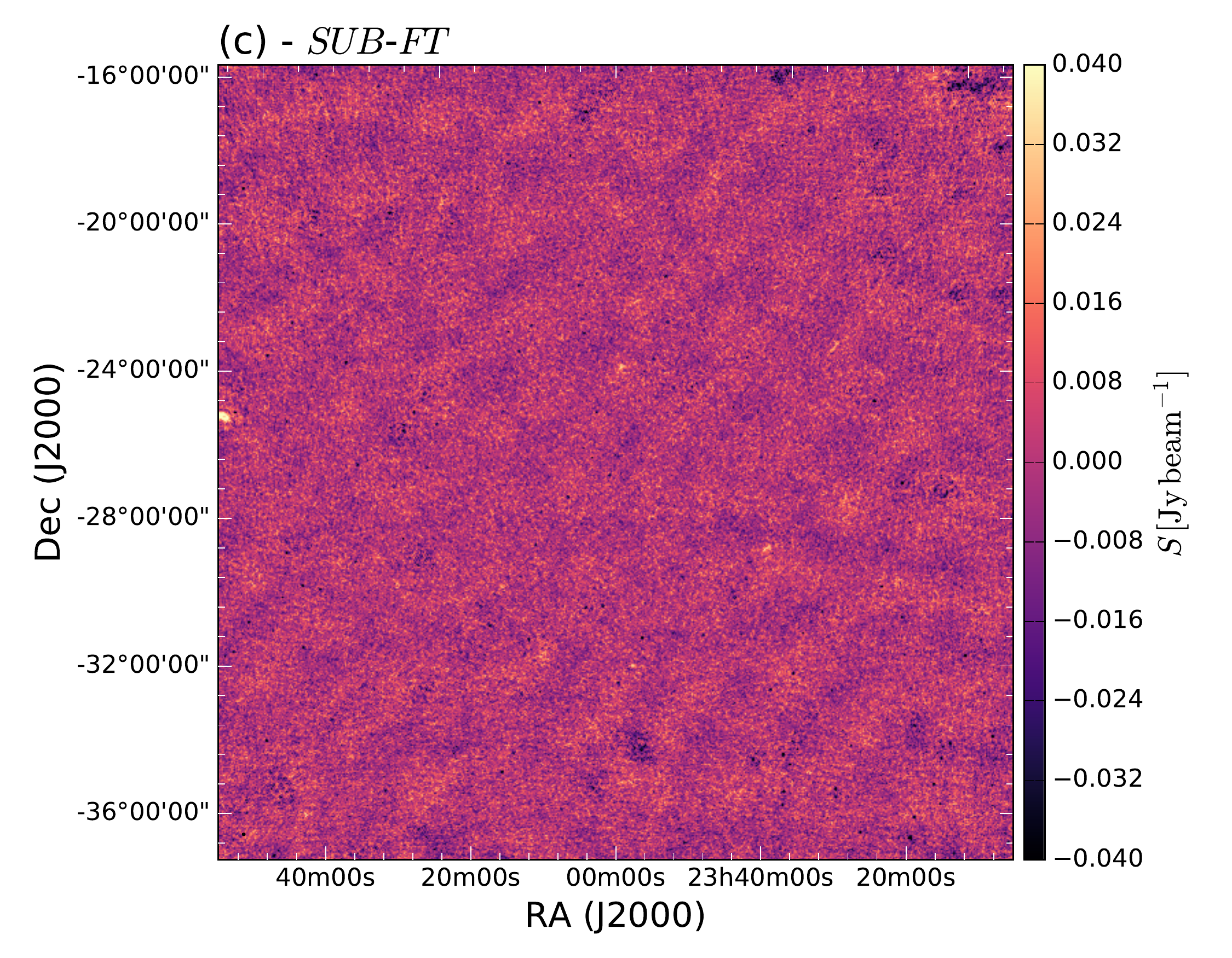}\includegraphics[scale=0.395]{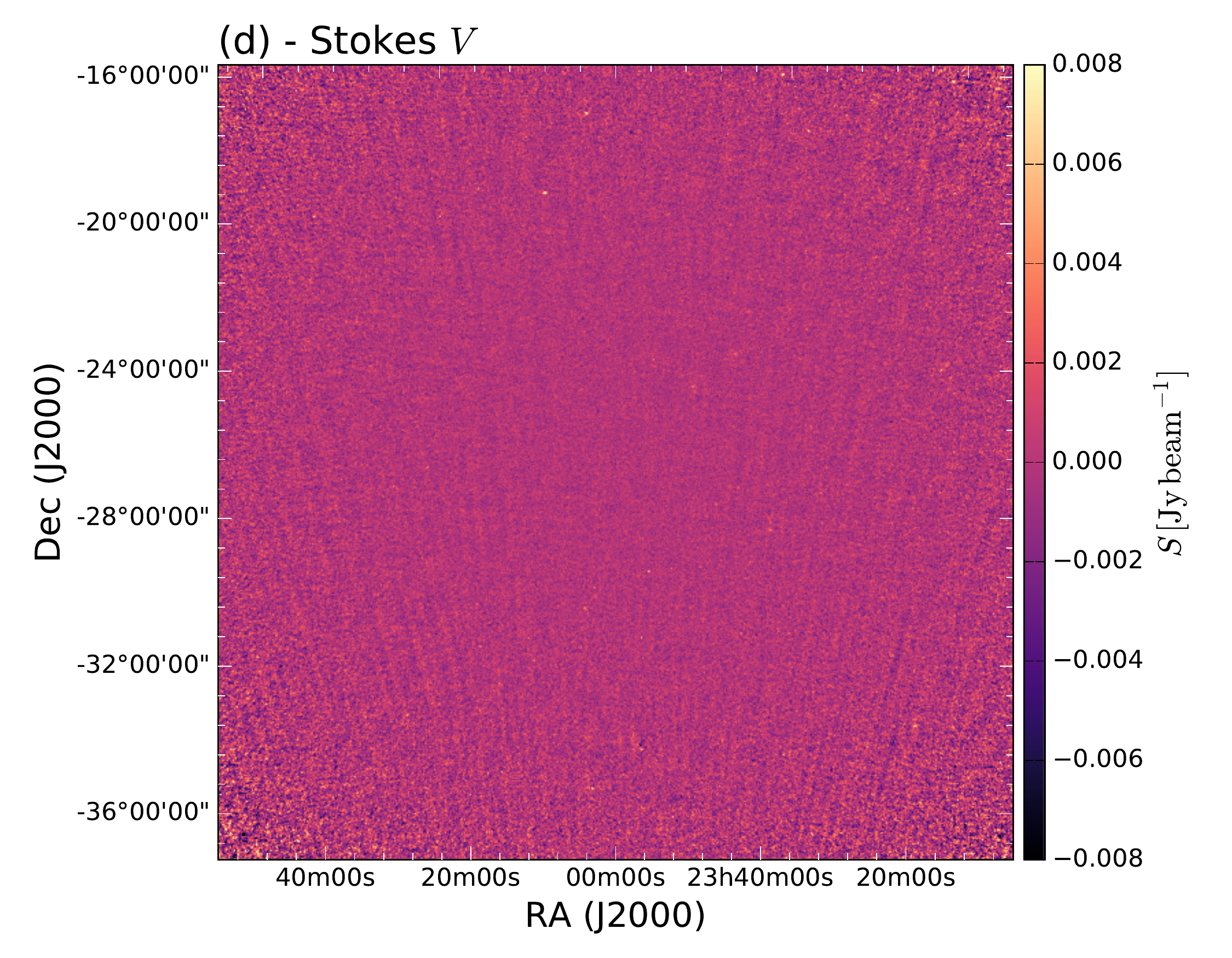}
\caption{MWA EoR0 field images. All images are $21.76\degr \times 21.76\degr$. Panel (a) is the {\full} field at $169\,$MHz with no point source subtraction and made with uniform weighting ($\theta_{\rm B}=2.3^\prime$). Panel (b) is the $180\,$MHz {\galr} image, made with robust 0.25 weighting ($\theta_{\rm B_{maj}}=2.9\,$arcmin) and with sources with $S \gtrapprox 0.1\,$Jy subtracted out. Panel (c) is the $180\,$MHz {\minur} image, and is also made with robust 0.25, point sources $S \gtrapprox 0.1\,$Jy subtracted, but with a minimum baseline of $34\,$m used ($\theta_{\rm B_{maj}}=2.8\,$arcmin). Panel (d) shows the {\minur} Stokes $V$ image.}
\label{fig:radims}
\end{figure*}

The data were processed and imaged in several different ways. The first set of data, known as the ``\full'' set was processed and imaged by \citet{Offringa16}. This full set contains approximately $45\,$hours of data, observed between August and December of 2013 \citep[see table 1 from][ for a full list of the dates]{Offringa16}. The data is a combination of ``low-band'' (138.9--$169.6\,$MHz) and ``high-band'' (167.0--$197.7\,$MHz) data.

For full details on the calibration process see \citet{Offringa16}. The first steps in the data processing were to flag RFI, average the data in time to $4\,$s and convert the raw data to measurement sets. These steps were performed by the \textsc{COTTER} preprocessing pipeline \citep{Offringa15}, which uses an \textsc{AOFLAGGER} strategy for RFI detection \citep{Offringa10,Offringa12}. Each measurement set is a snapshot of $112\,$s, and each snapshot was calibrated using a source model in which the spectral energy distribution of each source is assumed to follow a power law; the spectral index in the model is independent for each source. For brighter sources the model is a point source model which is bootstrapped from cross-matching the MWA commissioning survey \citep{Hurley-Walker14} at $180\,$MHz to the SUMMS catalogue at 843 MHz \citep{Mauch03}. Fainter sources are given a power law formed from their measured flux density combined with a measurement from other catalogues covering the source. For this, also the $408$-MHz Molonglo Reference Catalogue \citep[MRC][]{Large81} is used. The model contains approximately 16,000 sources.

The calibration uses sources from the source model within $18\degr$ of the field centre with $S\ge0.25 \,$Jy. The first calibration was performed as a direction-independent full-polarization self-calibration. After global calibration, 2500 sources were peeled using a clustered peeling procedure that mitigates the ionosphere by fitting positions and gains in 25 directions, which are the centres of the 25 clusters. Clusters were made by using an angular k-means clustering algorithm to group the modelled sources. The peeling was performed by a tool named IONPEEL, which was also specifically written for the MWA \citep{Offringa16}. IONPEEL performs a Levenberg-Marquardt (LM) least-squares optimisation between the model and data for the parameters  $\Delta l$, $\Delta m$ and $g$, with $l$ and $m$ the position offsets and the $g$ being the gain factor. 

\begin{figure*}
\includegraphics[scale=0.27]{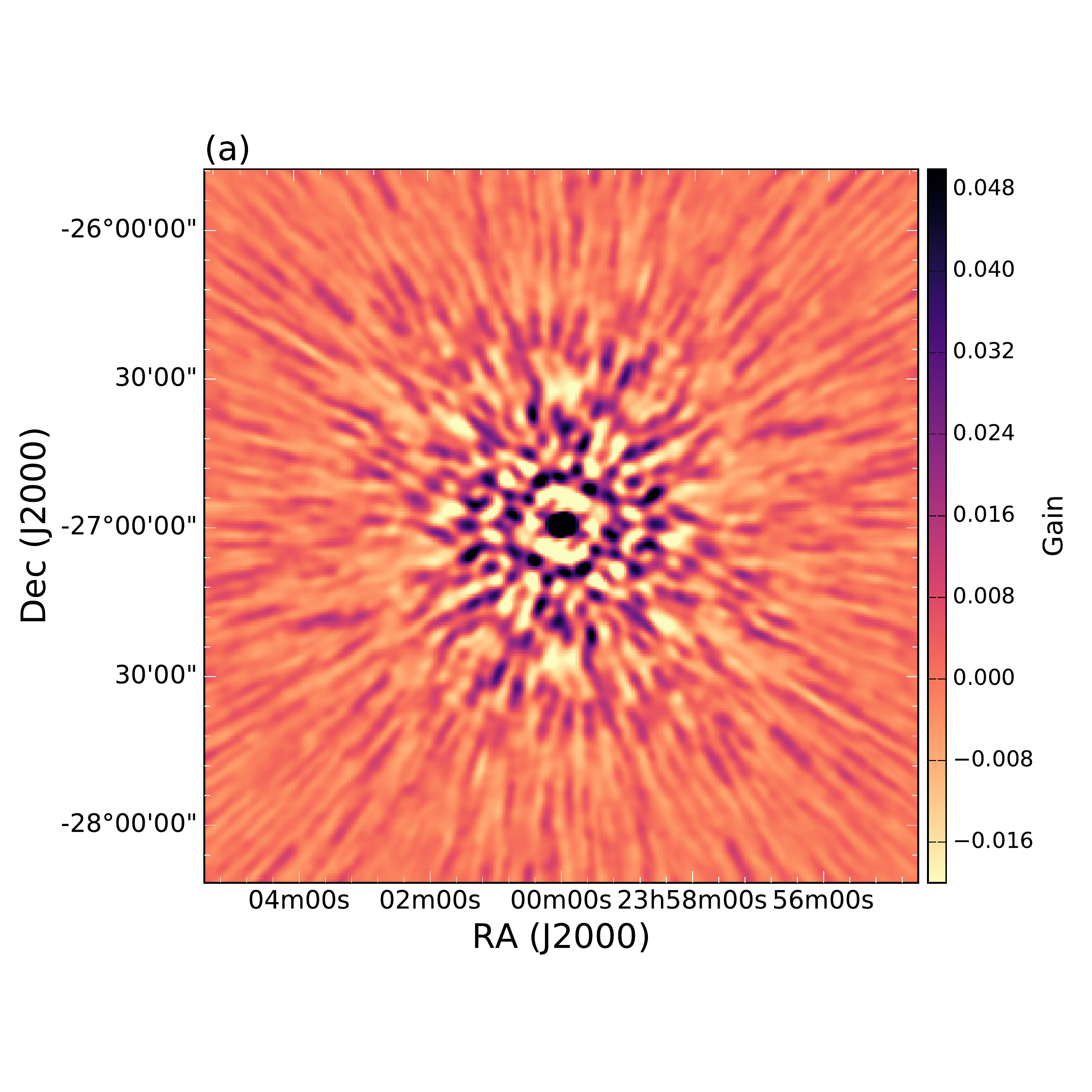}\includegraphics[scale=0.27]{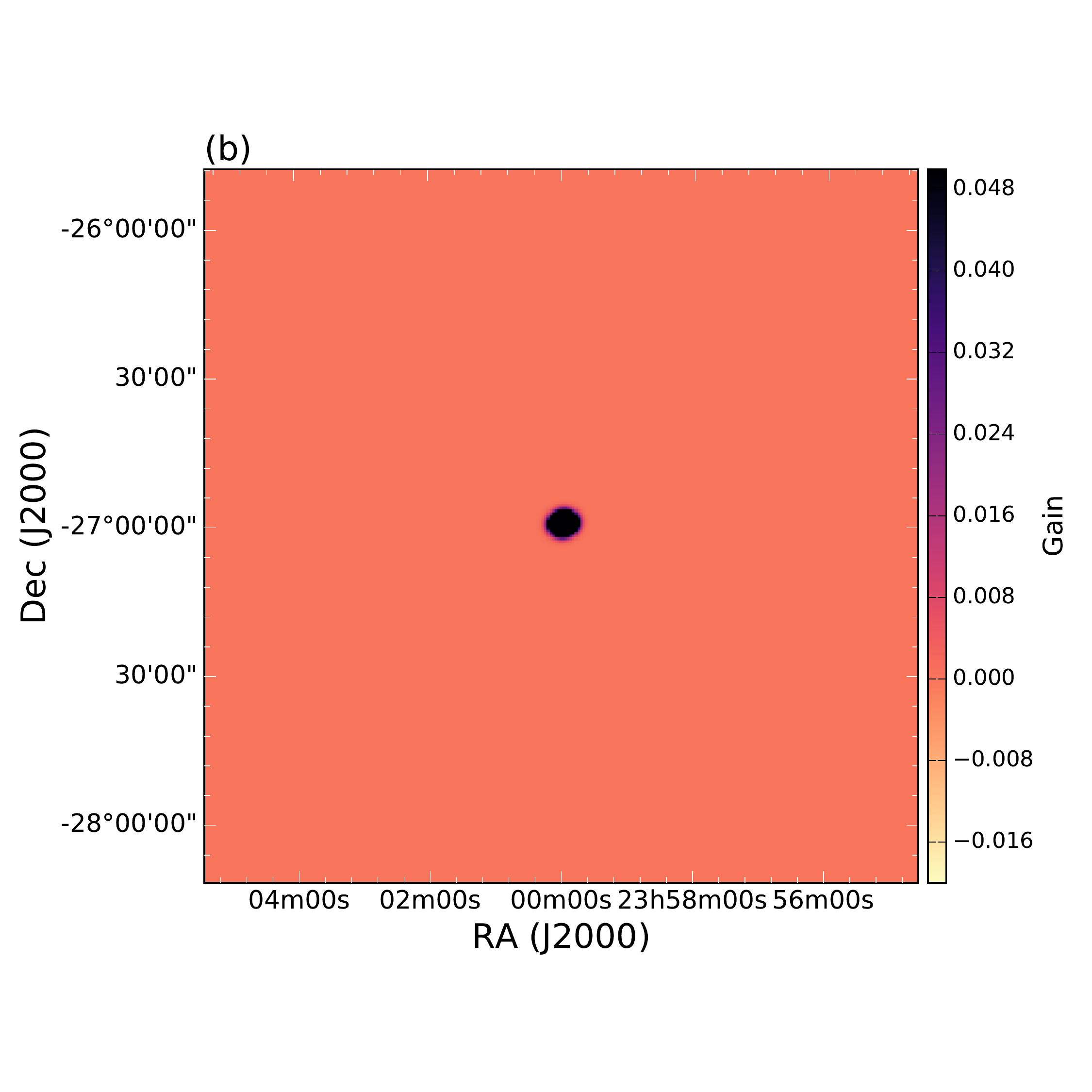}\includegraphics[scale=0.27]{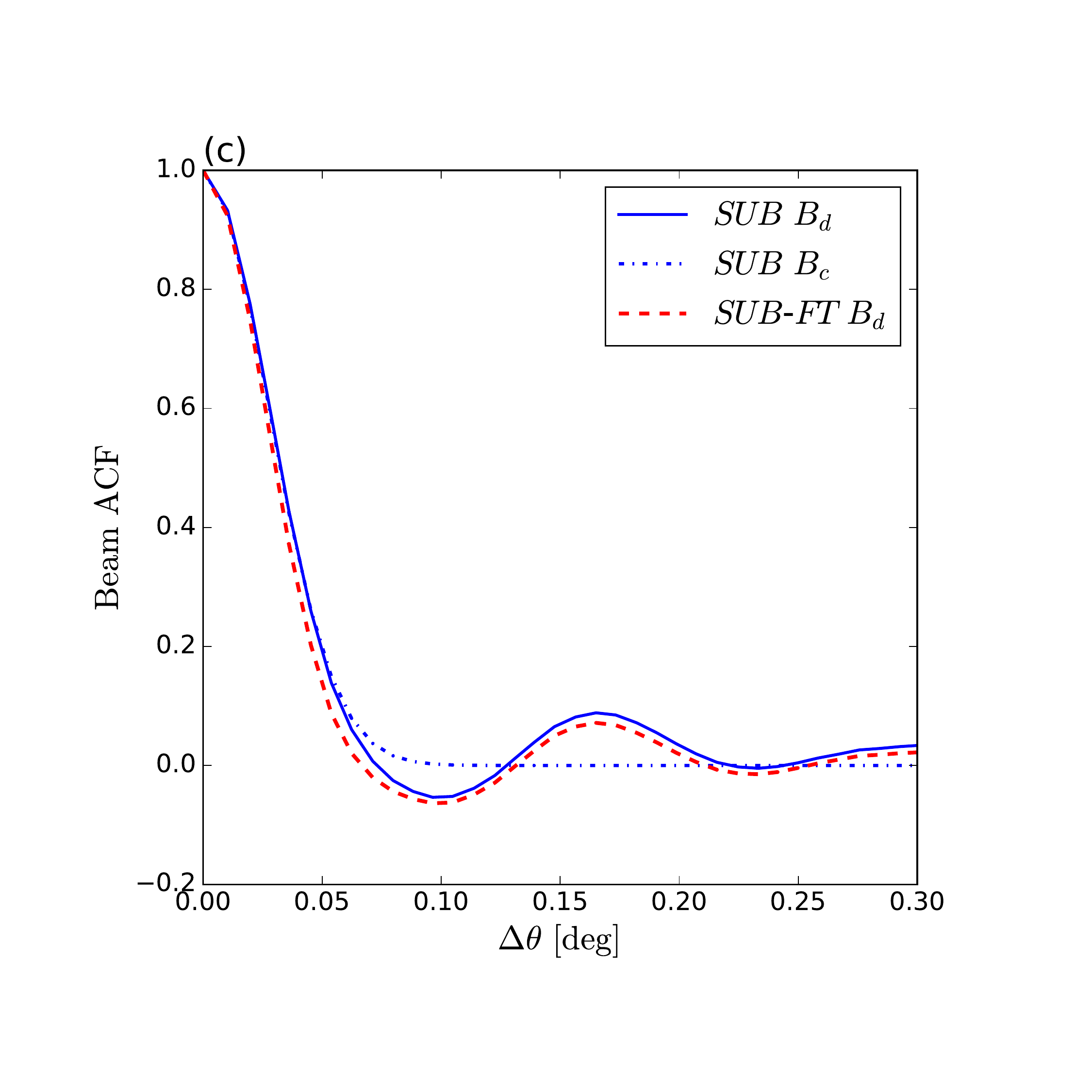}
\caption{MWA synthesized beams from 2-minute snapshots. Panel (a) shows the ``dirty'' beam, $B_d$, from the {\galr} image. Panel (b) shows the corresponding ``clean'' beam, $B_c$, with $\theta_{\rm B_{maj}}=2.9\,$arcmin and $\theta_{\rm B_{maj}}=2.6\,$arcmin. Panel (c) is the autocorrelation functions (ACF) of both the {\galr} and {\minur} dirty beams and the {\galr} clean beam. }
\label{fig:bms}
\end{figure*}

The snapshots were imaged using \textsc{WSCLEAN} \citep{Offringa14}.\footnote{\url{https://sourceforge.net/projects/wsclean/}} The snapshots were imaged with a size of $43.5\degr \times 43.5\degr$, pixel size of $0.51\,$arcmin, and cleaned down to $100\,$mJy using uniform weighting (the theoretical instrumental noise for a single snapshot using natural weighting is $\sim12\,$mJy beam$^{-1}$). For the final integrated {\full} image, the peeled sources were restored, the images corrected for the MWA primary beam and mosaicked together. The central $21.76\degr \times 21.76\degr$ of the image can be seen in Fig.~\ref{fig:radims}(a). The image shows a large number of point sources, as well as some more complex sources such as active galactic nuclei (AGN) jets. There is also visible large-scale diffuse emission such as a streak a few degrees wide starting in the lower left running northwest as well as another region in the lower right corner. The image contains negative pixels for two reasons. First, due to the incomplete sampling of the {\it uv} plane the dirty synthesized beam has negative sidelobes. Second, being from an interferometer, the data is missing the zeroth spacing, or ``dc'' component, which determines the mean of the signal. Without the zero spacing the data is insensitive to the absolute flux of the signal, resulting in a zero mean and negative pixel values. The {\full} image has a clean synthesized beam major and minor axis full width at half maximum (FWHM), $\theta_{\rm B_{\rm maj}}$ and $\theta_{\rm B_{\rm min}}$, of $2.31\,$arcmin. 

For the purposes of detecting the cosmic web, we want as little contamination from point sources (and beam side lobes from point sources) as possible. For this reason we chose to reprocess a subset of the data. We reprocess only a subset of the data as the decrease in instrumental noise from reprocessing all 45 hours of data is not large enough to outweigh the additional processing time needed (see the following subsection on the image noise properties for further details). The subset contains $5\,$hours of data with only high-band frequencies which were observed on August 23 and 27 and September 2 and 4 of 2013. 

For the subset, after the ionospheric peeling, a round of imaging and cleaning was performed down to approximately $100\,$mJy, using uniform weighting. The clean model for each snapshot was subtracted from the {\it uv} data. Then another round of cleaning was performed with a robust parameter of +0.25 (resulting in $\theta_{\rm B_{maj}}=2.9\,$arcmin and $\theta_{\rm B_{min}}=2.6\,$arcmin) down to a flux density limit of $S\simeq 50\,$mJy. The robust value was chosen to be larger than used for the {\full} image and was chosen as a balance between increasing sensitivity to extended emission (higher robust values) and having a cleaner beam shape (lower robust values). The peeled sources were not restored. We created a mosaic of the final residual images (without restoring the clean models). This image is herein referred to as {\galr} (for a {\it sub}set of the data and {\it sub}tracted point sources) and can be seen in Fig.~\ref{fig:radims}(b), which shows the absence of the bright point sources seen in the {\full} image, while the regions of diffuse emission are accentuated.

The {\galr} image shows a large amount of degree-scale diffuse emission; which could be Galactic synchrotron emission and/or diffuse cosmic web emission. If it is indeed Galactic in origin, this Galactic emission may interfere with the cosmic web detection. In case of this possibility we created an additional subset image by repeating the above procedure, except this time a minimum baseline constraint of $34\,$m was applied. This effectively filters out emission on scales larger than roughly $3\degr$ (we say roughly because the exact angular size depends on the frequency, $\simeq \lambda/b$, where $b$ is the baseline length, and is thus slightly different for each frequency channel of the data). We chose this value in an attempt to balance filtering as much Galactic emission as possible and leaving as much cosmic web signal as possible. This is discussed further in Sec.~\ref{sec:odata} and \ref{sec:crx}, but given the redshift range of the sample of infrared galaxies to be used in this analysis, scales of $3\degr$ correspond to a size of $\sim3.5\,$--$\,80\,$Mpc ($0.01 \le z \le 0.4$). From simulations the average width of a filament is estimated to be $3 \sim 6\,$Mpc, with lengths of $\sim 100\,$Mpc \citep{Cautun14,Gheller15,Gheller16}, which means for this analysis the filtering would only effect the cosmic web signal at the lowest redshifts or when looking along the length of a filament.  

The filtering altered the synthesized beam shape slightly, resulting in $\theta_{\rm B_{maj}}=2.8\,$arcmin and $\theta_{\rm B_{min}}=2.4\,$arcmin. The mosaic of residuals from this is referred to as {\minur} ($FT$ for filtered) and is shown in Fig.~\ref{fig:radims}(c). The {\minur} image is lacking the point sources and diffuse emission seen in the other two images, with the few discernible features consisting mainly of complex galaxies (i.e. AGN jets). There is a discernibly non-Gaussian pattern to the {\minur} image, even with the majority of bright sources subtracted. This is due to the convolution of the ``dirty'' synthesized beam with the faint unsubtracted point sources. The dirty synthesized beams $B_d$ have a maximum positive sidelobe of roughly 0.07 and a minimum sidelobe of $-0.09$. An image of $B_d$ for a snapshot is shown in Fig.~\ref{fig:bms}(a), with the clean beam $B_c$ shown in panel (b).

It is not clear if the presence of the diffuse Galactic emission will interfere with cosmic web detection; however, the removal, by ignoring the shortest baselines, may also have the effect of filtering or removing the cosmic web signal. Since the answer is unknown, at this time, we work with both the {\galr} and {\minur} images, which should yield information on the effect of the Galactic emission on the cross correlation. The {\full} image is not used in the cosmic web analysis, and is just shown for comparison (more explanation on this is given in Sec.~\ref{sec:rdatnoise}). 

The {\full} image has a mean frequency of $168\,$MHz and the {\galr} images have mean frequencies of $180\,$MHz. While the full width of the primary beam was imaged in each case, we constrain our examination to the central $21.76\degr \times 21.76\degr$, or roughly the $50\,$per cent power of the beam, to avoid the regions of increased instrumental noise further out in the field. Table~\ref{tab:rims} lists the different image properties.

\begin{table*}
\centering
\caption{Radio image properties. All the images are $21.76\degr \times 21.76\degr$, with equal area pixels with sides$=0.51\,$arcmin. The confusion noise $\sigma_{\rm c}$ was estimated using the {\it P(D)} confusion analysis technique, see Sec.~\ref{sec:rdatnoise} for details.}
\label{tab:rims}
\begin{tabular}{lcccccccc}
\hline
Name & $\nu$ &Weighting & $\theta_{\rm B}$ & $\sigma_{\rm n}$ & $\sigma_{\rm c}$ & {\it uv}$_{\rm min}$ & point sources subtracted & limit$_{\rm subtraction}$\\
 & MHz& & [arcmin] & [mJy beam$^{-1}$] &[mJy beam$^{-1}$]  & [m] & &[Jy beam$^{-1}$] \\
 \hline
{\full} & 168 & uniform & 2.3 &0.60 &9.5& 7.7& No & -- \\
{\galr} & 180 & robust 0.25 & 2.9 & 0.96 & 4.5& 7.7& Yes & 0.05 \\
{\minur}& 180 &robust 0.25 & 2.8 &0.96 &4.4 & 34& Yes & 0.05 \\
\hline
\end{tabular}
\end{table*}

\subsection{Image noise}
\label{sec:rdatnoise}

\begin{figure}
\includegraphics[scale=0.365]{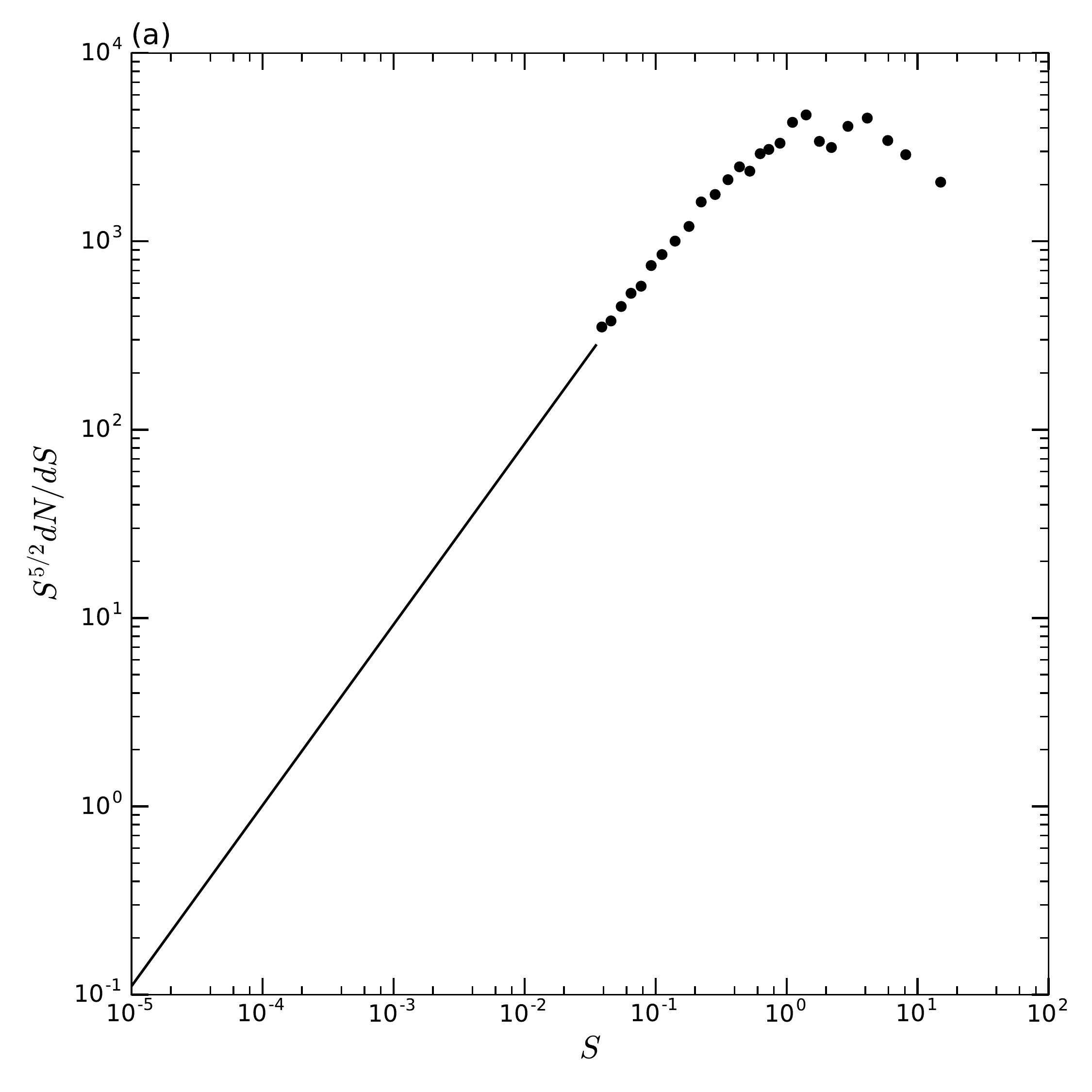}
\includegraphics[scale=0.365]{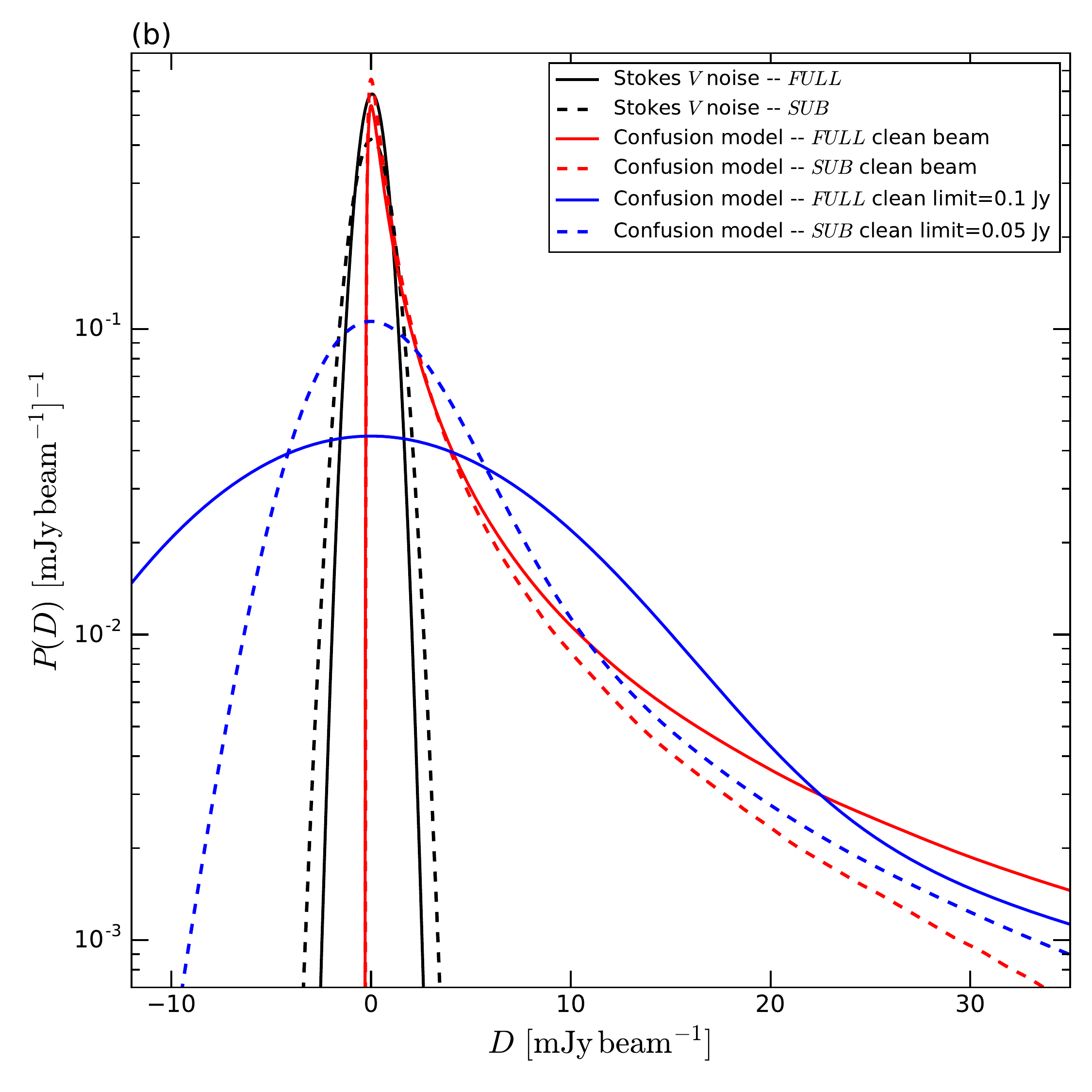}
\caption{Source count and noise distributions. Panel (a) shows the $150\,$MHZ Euclidean normalized source count. The black points are from \citet{Franzen16} and the black solid line is the power-law model from eq.~(\ref{eq:pwl}). Panel (b) shows the probability distributions of instrumental and confusion noise. All solid lines show {\full} image values, while dashed lines show {\galr} values. The black lines are the pixel distributions from the stokes $V$ noise maps, with $\sigma_{\rm n}=0.60\,$mJy beam$^{-1}$ ({\full}) and $\sigma_{\rm n}=0.96\,$mJy beam$^{-1}$ ({\galr}). The red lines are the noiseless confusion {\it P(D)} models (solid line for {\full} with $S_{\rm max}=5\,$Jy and dashed line for {\galr} with $S_{\rm max}=0.05\,$Jy) using the source count shown in panel (a) and clean synthesized beams, which have confusion noise values of $\sigma_{\rm c}=4.4\,$mJy beam$^{-1}$ ({\full}) and $\sigma_{\rm c}=3.5\,$mJy beam$^{-1}$ ({\galr}). The blue lines show the confusion {\it P(D)}s with the dirty beam (panel (a) of Fig.~\ref{fig:bms}) for sources below the clean limit. For the {\full} image, $S_{\rm max}=5\,$Jy with a clean limit of $0.1\,$Jy and $\sigma_{\rm c}=9.5\,$mJy beam$^{-1}$ (solid line). For the {\galr} image, $S_{\rm max}=0.05 \,$Jy with a clean limit of $0.05\,$Jy and  $\sigma_{\rm c}=4.5\,$mJy beam$^{-1}$ (dashed line).}
\label{fig:pdnoise}
\end{figure}

The total image noise $\sigma_{\rm tot}$ is a combination of instrumental noise $\sigma_{\rm n}$ and confusion noise $\sigma_{\rm c}$ from the point sources being convolved with the beam.\footnote{In this paper $\sigma_{\rm c}$ represents the confusion noise from sources convolved with the beam, whether that be the dirty beam or clean beam or a combination of both. In the case of the dirty beam it is the combination of the classical confusion noise (i.e. convolved with the clean beam) and sidelobe confusion. Unless specified otherwise, $\sigma_{\rm c}$ estimates throughout this paper come from using the dirty beam for sources below the clean limit and the clean beam for sources above the clean limit. } As an estimate of the instrumental noise we use the Stokes $V$ images, as the Stokes $V$ should be free from source signal (assuming minimal leakage). The {\minur} Stokes $V$ image is shown in Fig.~\ref{fig:radims}(d). The estimated instrumental noise values for the different images are listed in Table~\ref{tab:rims}.

As discussed by \citet{Franzen16}, the MWA is essentially confusion limited (or close to) in a single snapshot. To estimate the confusion noise, we use the source count of this field obtained by \citet{Franzen16}, along with a power-law model for the fainter sources
\begin{equation}
\frac{dN}{dS}=6998 S^{-1.54} \, {\rm Jy}^{-1} \, {\rm sr}^{-1}.
\label{eq:pwl}
\end{equation}
The Euclidean-normalized differential source count is shown in Fig.~\ref{fig:pdnoise}(a). 

We use the {\it P(D)} confusion analysis technique discussed by \citet{Vernstrom13} to estimate the confusion noise given the source count for different beams. This method uses the image pixel size and beam shape to predict an image probability distribution (i.e. a normalized image histogram) for a given input source count. We calculated noiseless {\it P(D)} distributions assuming only the clean beam for all sources with $S<S_{\rm max}$, for the {\full} radio image with $S_{\rm max}=5\,$Jy and for the {\galr} images with $S_{\rm max}=0.05\,$Jy. We then calculated noiseless {\it P(D)} distributions using the clean beams above the clean limit and the dirty beams below the clean limit. For the {\full} image $S_{\rm max}=5\,$Jy and the clean limit equal to $0.1\,$Jy, and for the {\galr} radio images the clean limit$=S_{\rm max}=0.05\,$Jy. A comparison of the stokes $V$ noise distributions from the {\full} and {\galr} images, and the confusion models is shown in Fig.~\ref{fig:pdnoise}(b).

The instrumental noise for the full $45 \,$hours is estimated as $0.60\,$mJy beam$^{-1}$ and after $5\,$hours is estimated as $0.96\,$mJy beam$^{-1}$. These values are estimated from the central regions of the Stokes $V$ images, and may be somewhat overestimated if there is leakage from the other Stokes parameters \citep{Lenc16}. The actual image rms values may include more than just purely instrumental noise and confusion noise, such as image artefacts. The confusion noise values, even with only clean beams, are $\sigma_{\rm c}=4.4\,$mJy beam$^{-1}$ ({\full}) and $\sigma_{\rm c}=3.5\,$mJy beam$^{-1}$ ({\galr}). The confusion noise values increase to $\sigma_{\rm c}=9.5\,$mJy beam$^{-1}$ ({\full}) and $\sigma_{\rm c}=4.5\,$mJy beam$^{-1}$ ({\galr}) when taking into account the dirty beam for uncleaned sources. It should be noted here that the {\it P(D)} method does not take into account any source clustering, which may affect the predicted confusion noise estimates. 

Clearly, the confusion noise is the dominant source of noise in the images (again, here confusion noise refers to the combination of classical confusion and sidelobe confusion). The large confusion noise demonstrates why it is important to subtract, or at least clean, as many sources as possible if we are interested in a detecting any faint ($S\la$mJy) signals. We feel confident in using the images made from a subset of the time, rather than the {\full} image. Even though the subset images have slightly higher instrumental noise, they have fewer point sources (from the point source subtraction) and thus lower confusion noise.

\section{Tracers of large-scale structure}
\label{sec:odata}

\begin{figure*}
\includegraphics[scale=0.395]{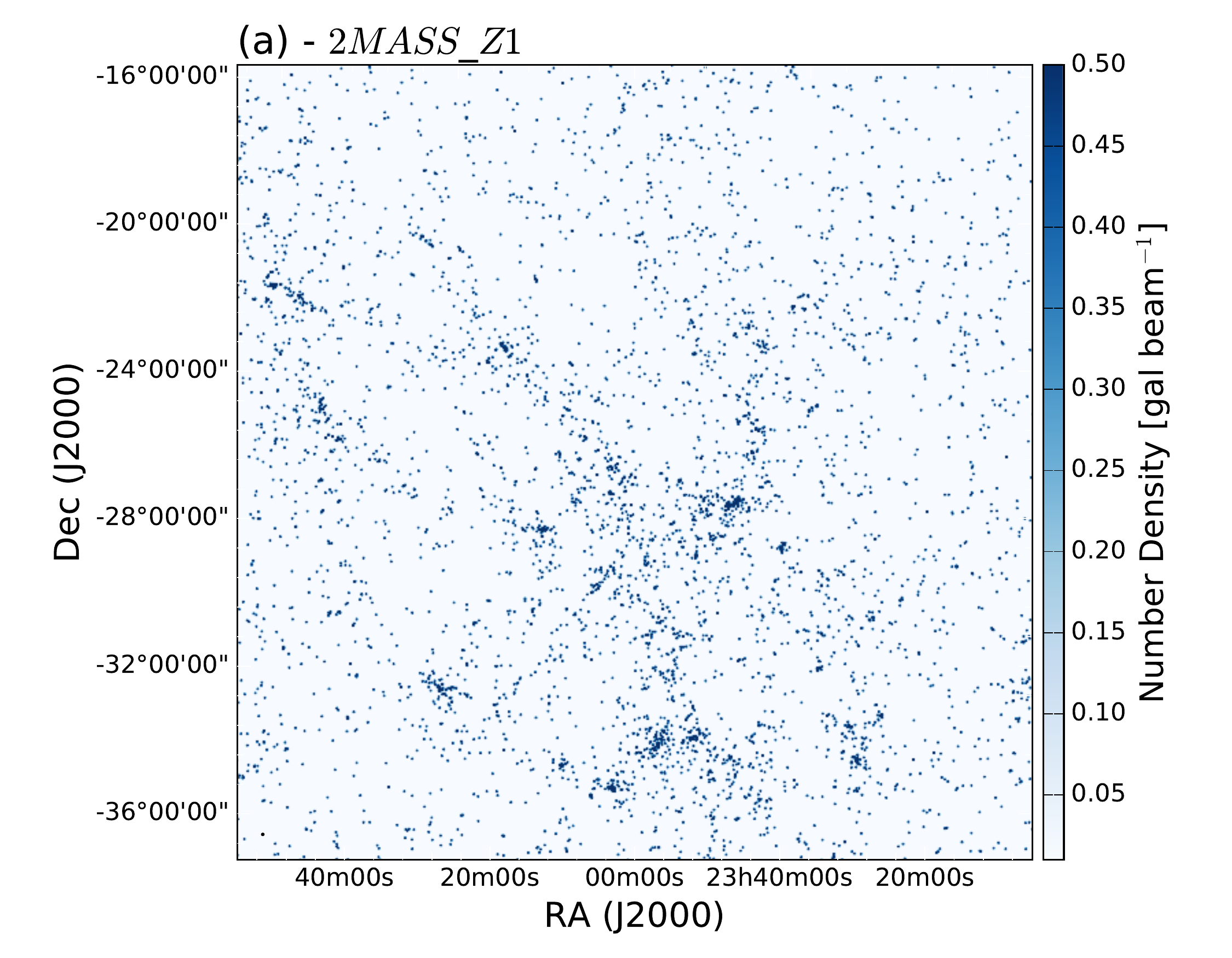}\includegraphics[scale=0.395]{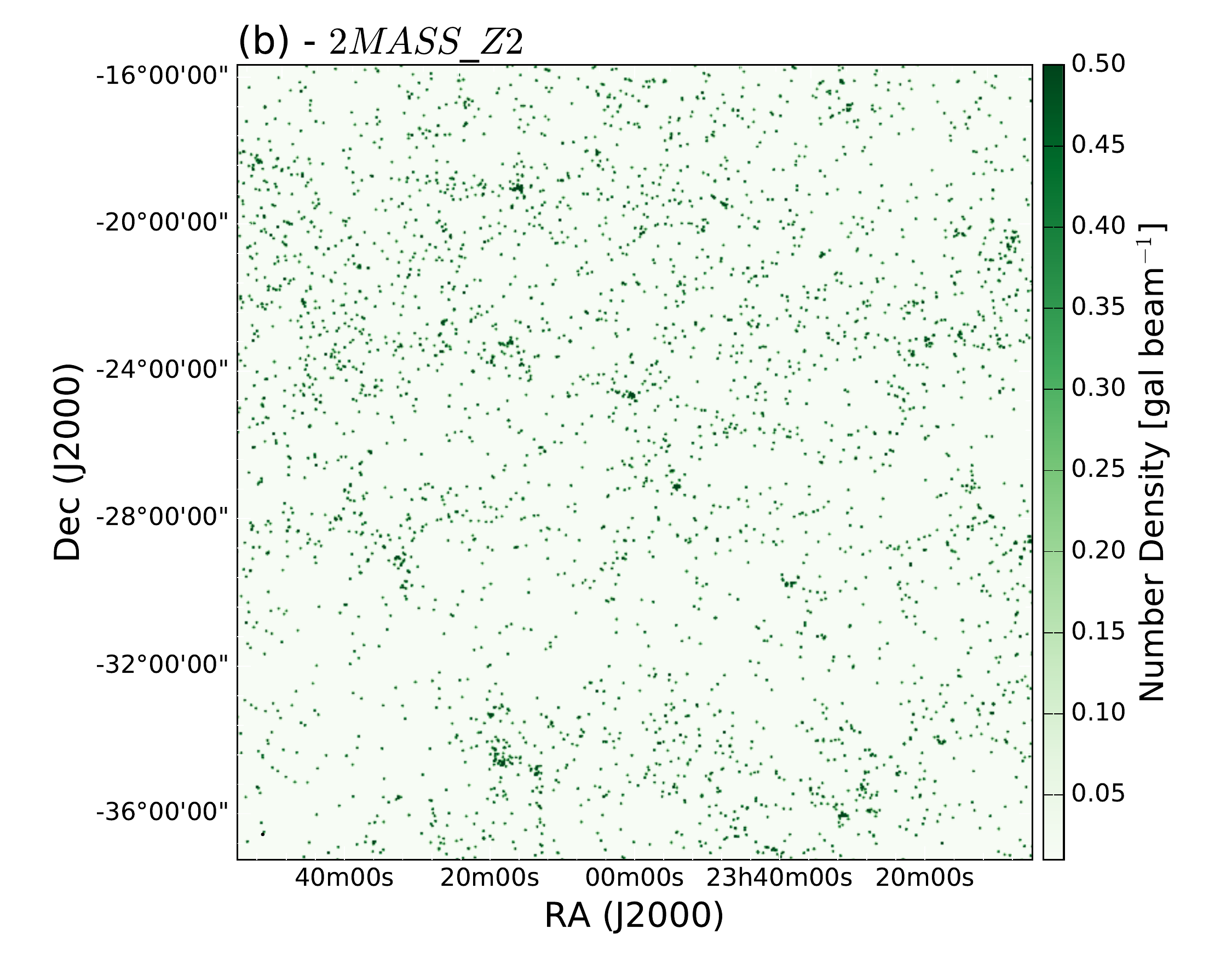}
\includegraphics[scale=0.395]{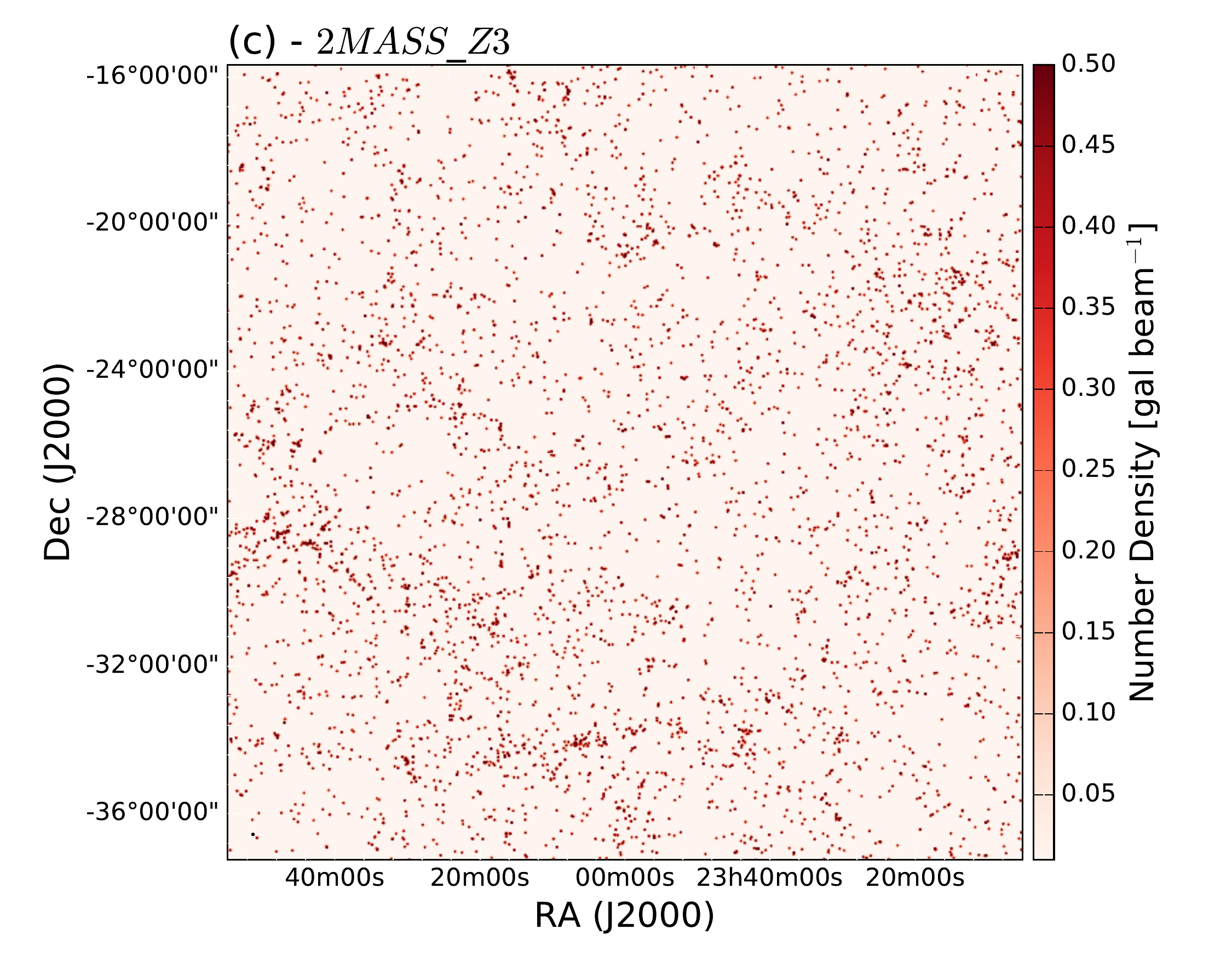}\includegraphics[scale=0.395]{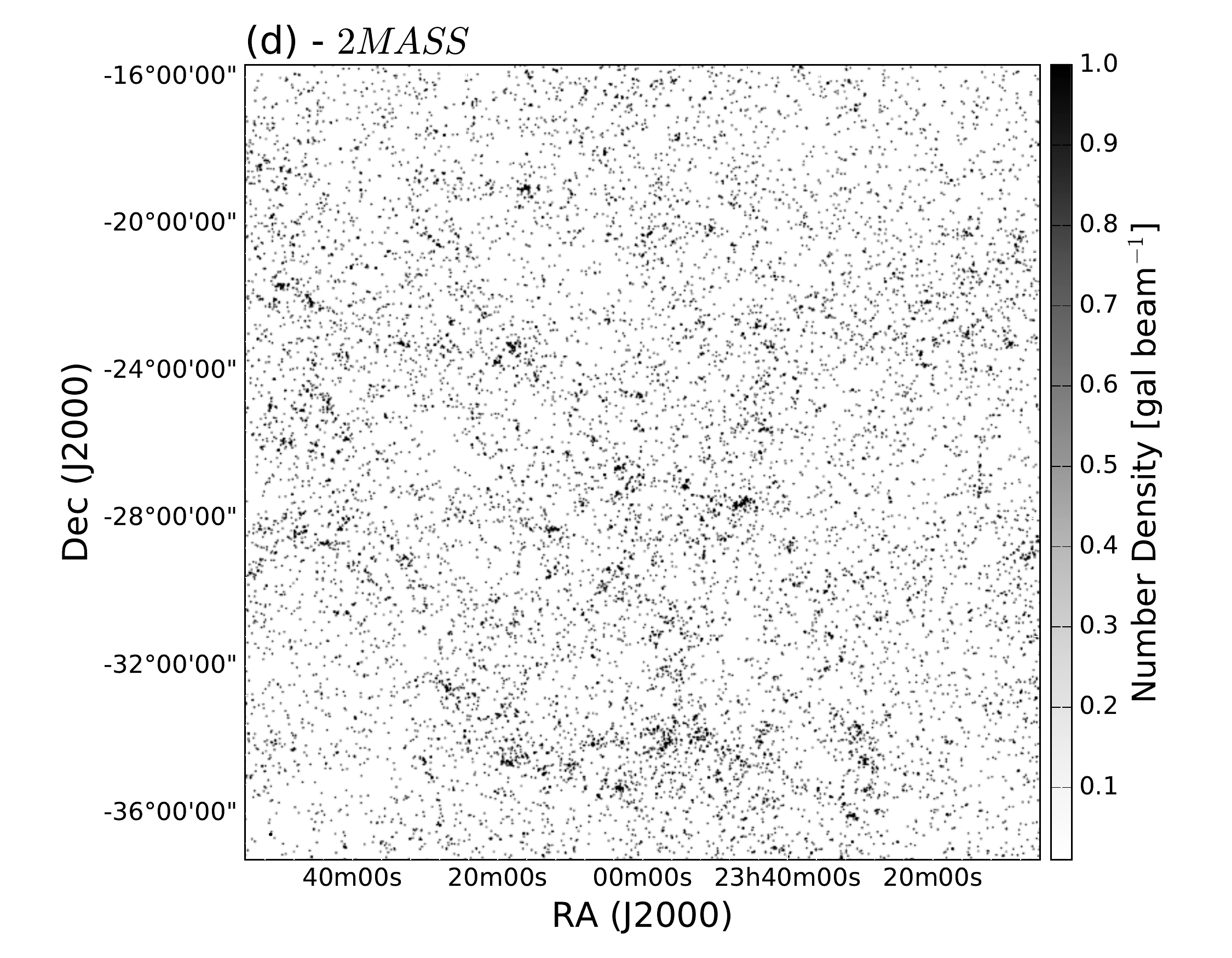}
\caption{2MASS galaxy number density maps of the EoR0 field. The maps used in the analysis cover the area of the main MWA primary beam lobe, approximately $43.5\degr$ on a side, however, only the central $21.76\degr \times 21.76\degr$ are shown here to match the radio images in Fig.\ref{fig:radims}. These maps have been convolved by a $3\,$arcmin beam for better visualization (the maps used for the analysis are not beam convolved). Panels (a), (b), and (c) show redshift slices corresponding to $0\le z_{\rm a} \le 0.07 \le z_{\rm b} \le 0.10 \le z_{\rm c} \le 0.41$. Panel (d) shows the catalogue sources combined across all redshifts. Table~\ref{tab:tmassim} provides details of the images.}
\label{fig:2massim}
\end{figure*}

\begin{figure*}
\includegraphics[scale=0.395]{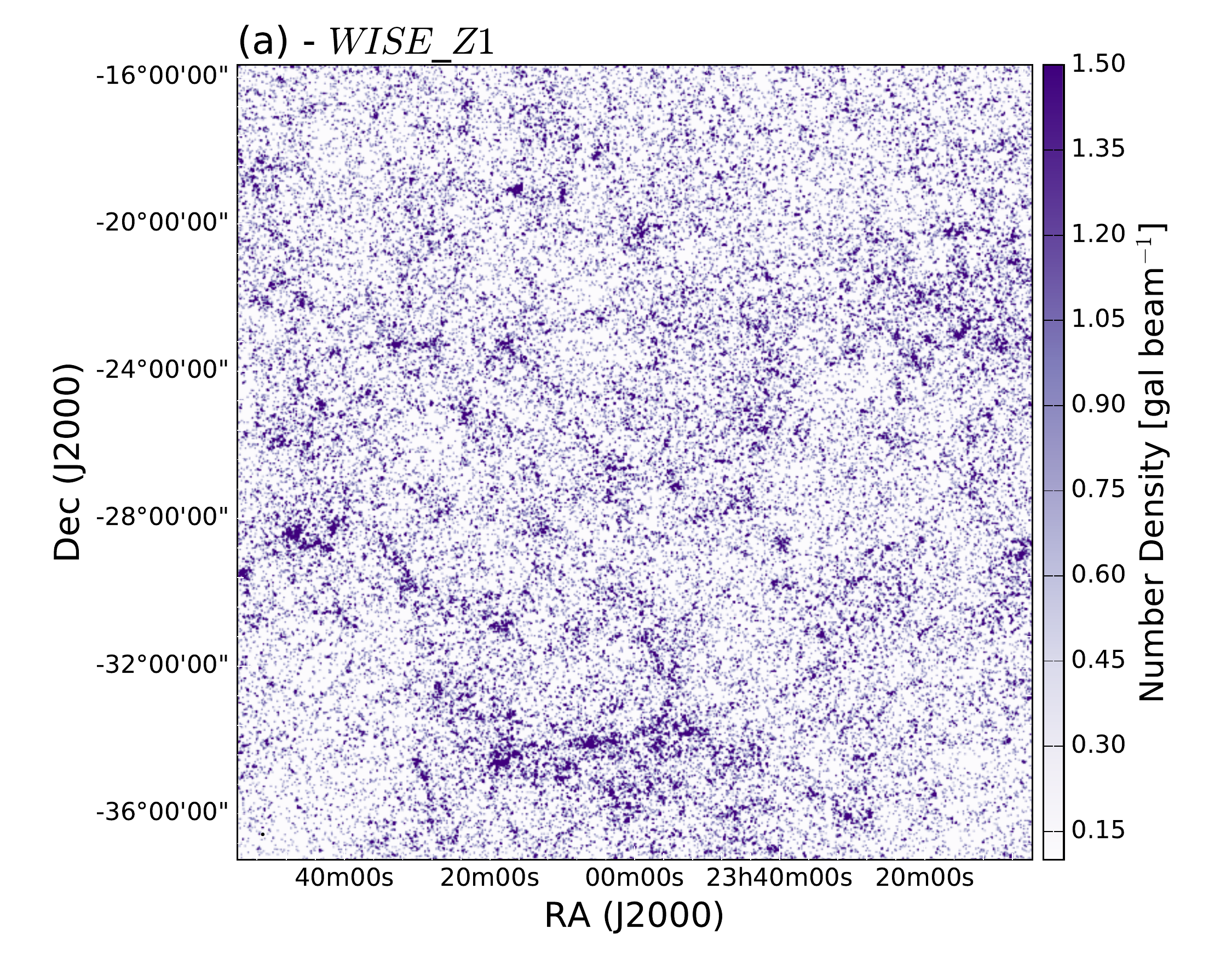}\includegraphics[scale=0.395]{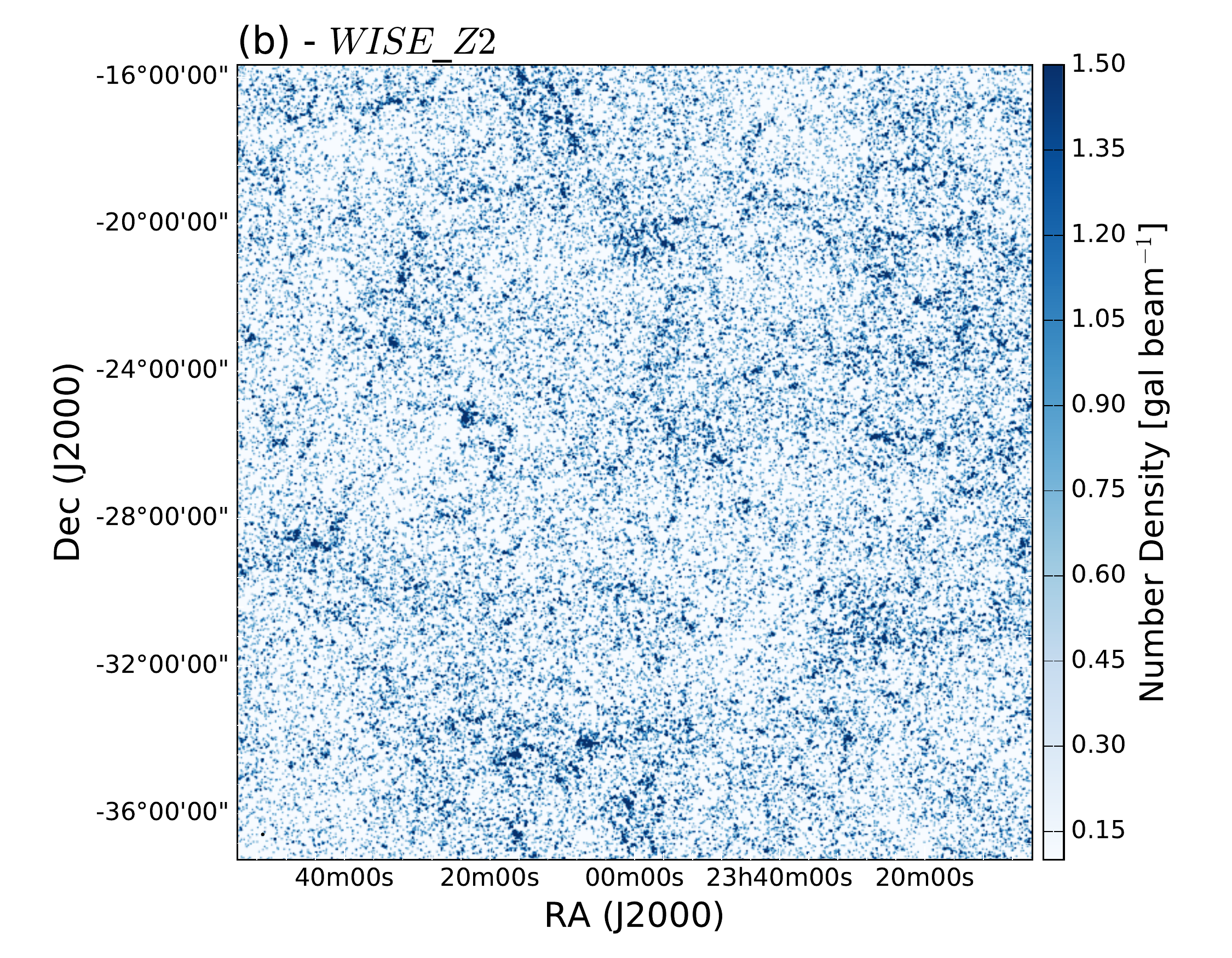}
\includegraphics[scale=0.395]{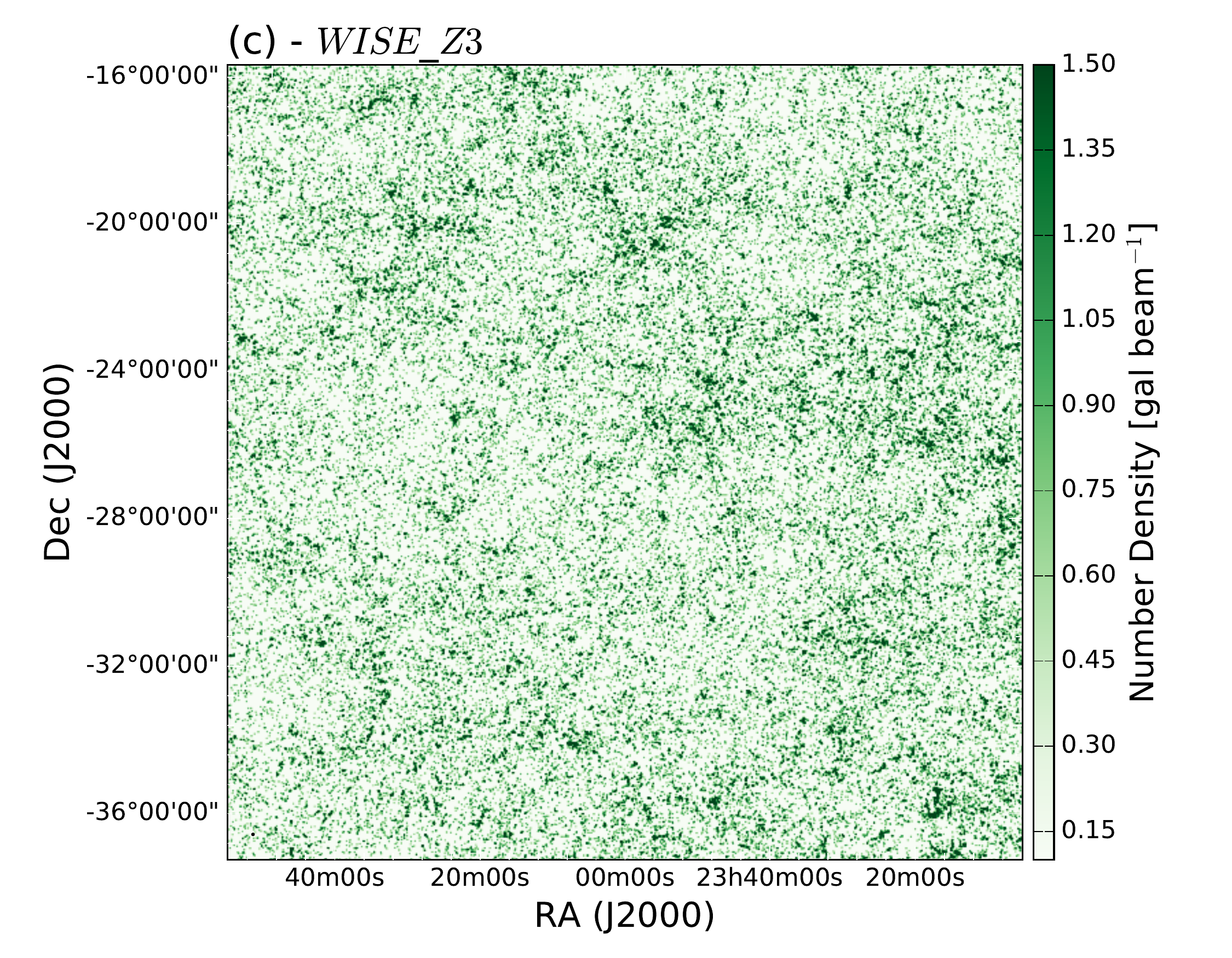}\includegraphics[scale=0.395]{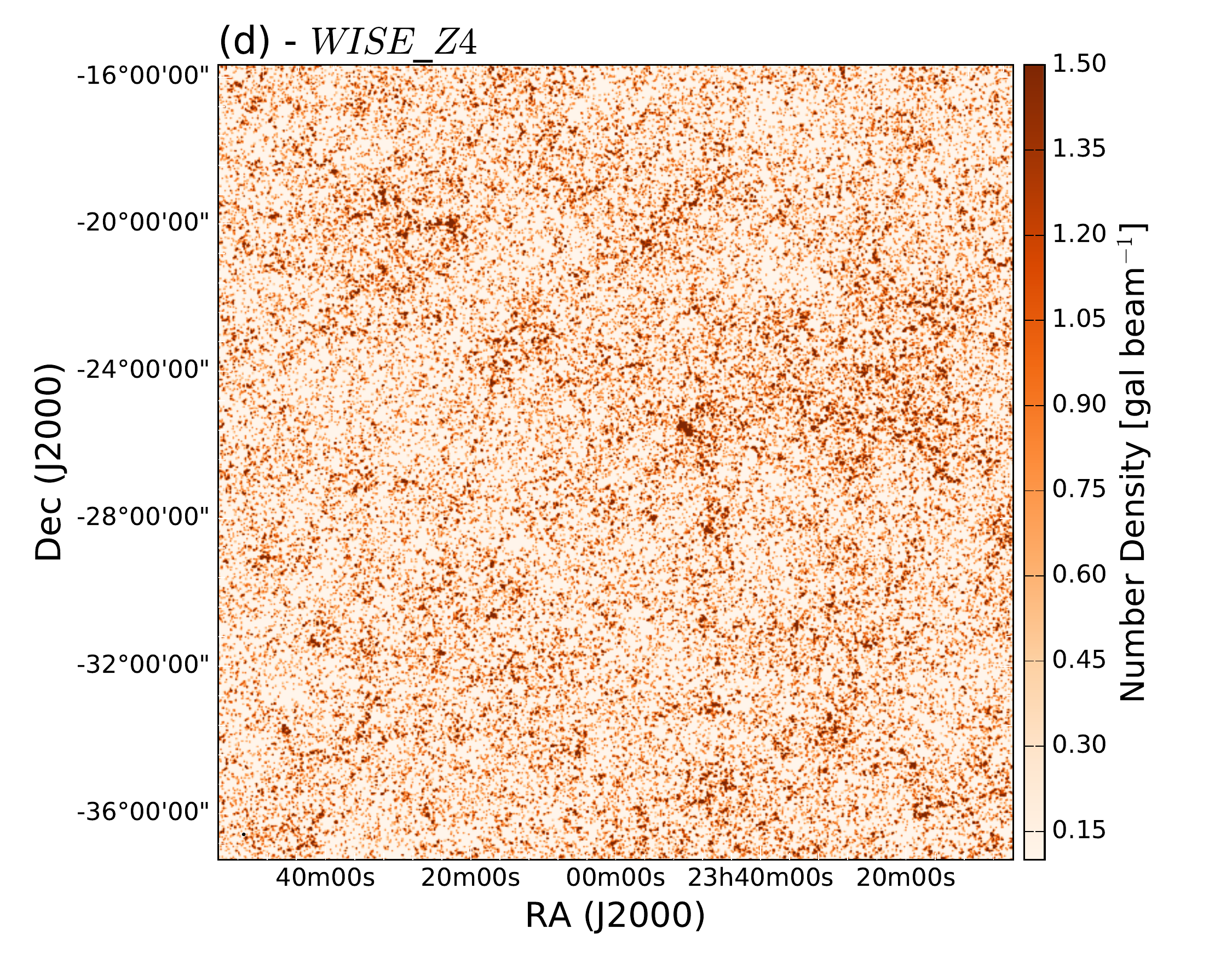}
\includegraphics[scale=0.395]{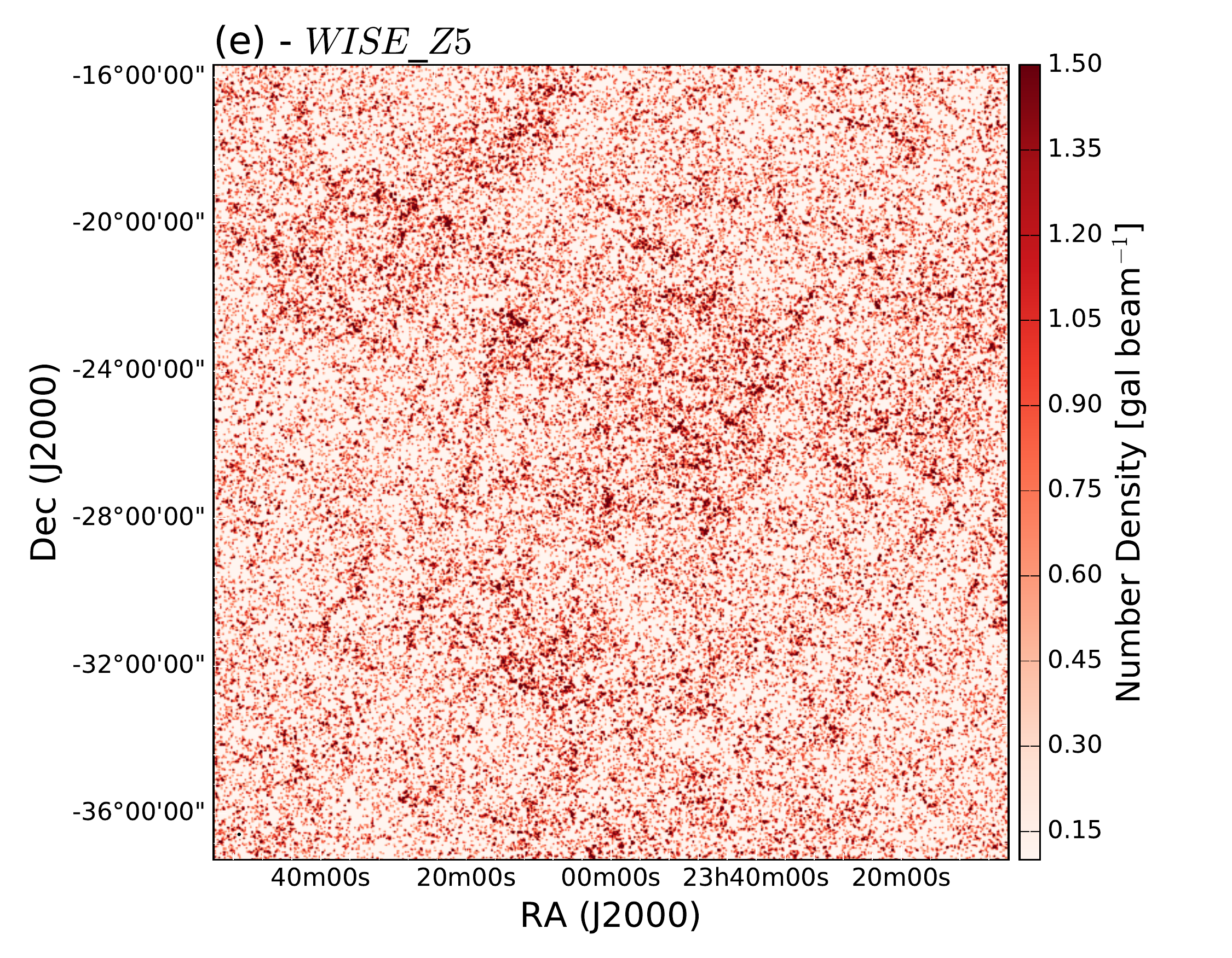}\includegraphics[scale=0.395]{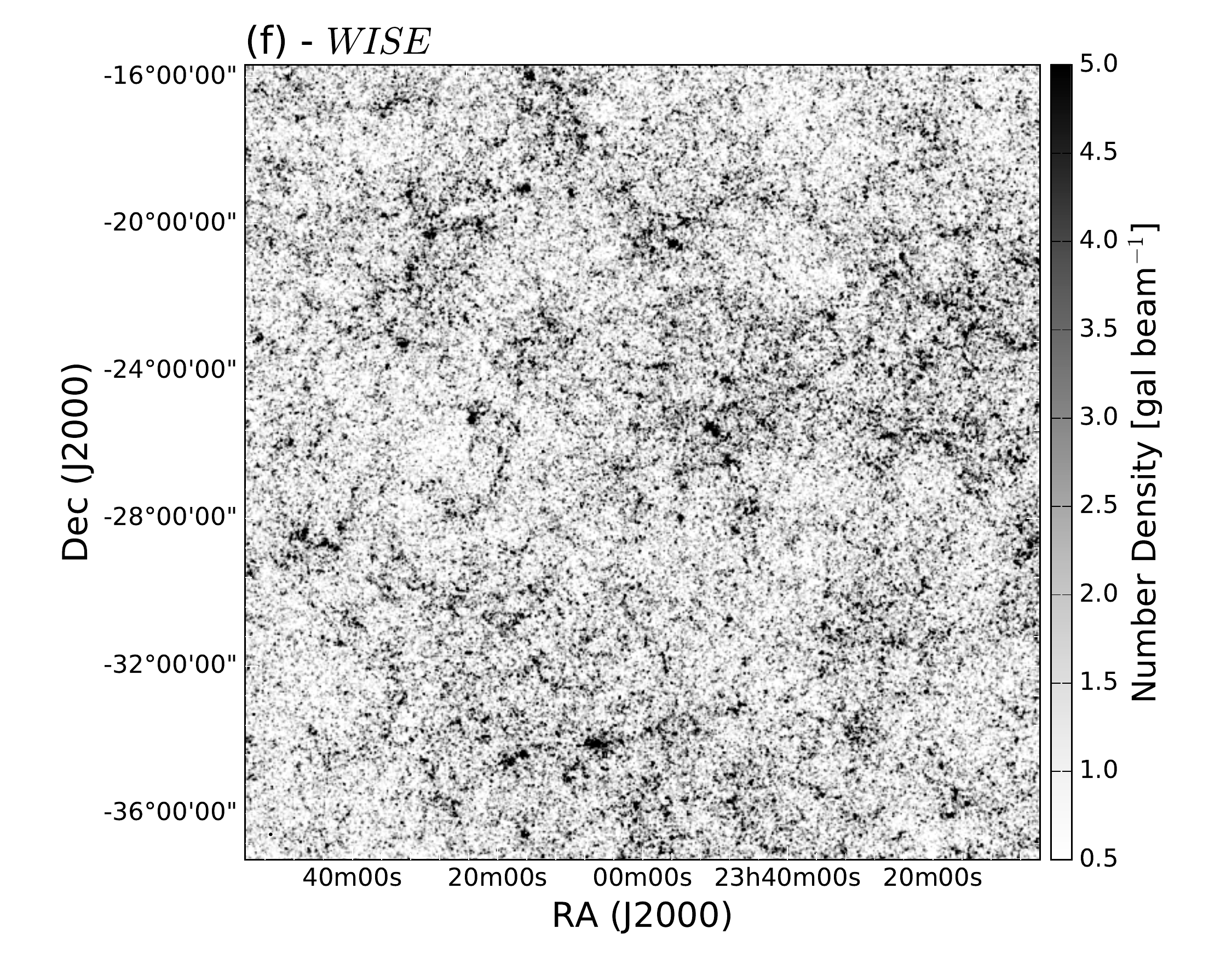}
\caption{As for Fig.~\ref{fig:2massim}, but for WISE galaxy number density maps of the EoR0 field. Panels (a) through (e) show redshift slices corresponding to $0\le z_{\rm a} \le 0.13 \le z_{\rm b} \le 0.18 \le z_{\rm c} \le 0.23\le z_{\rm d} \le 0.27 \le z_{\rm e} \le 0.57$. Panel (f) shows the catalogue sources combined across all redshifts. Table~\ref{tab:twiseim} provides details of the images.}
\label{fig:wiseim}
\end{figure*}

\begin{figure}
\includegraphics[scale=0.365]{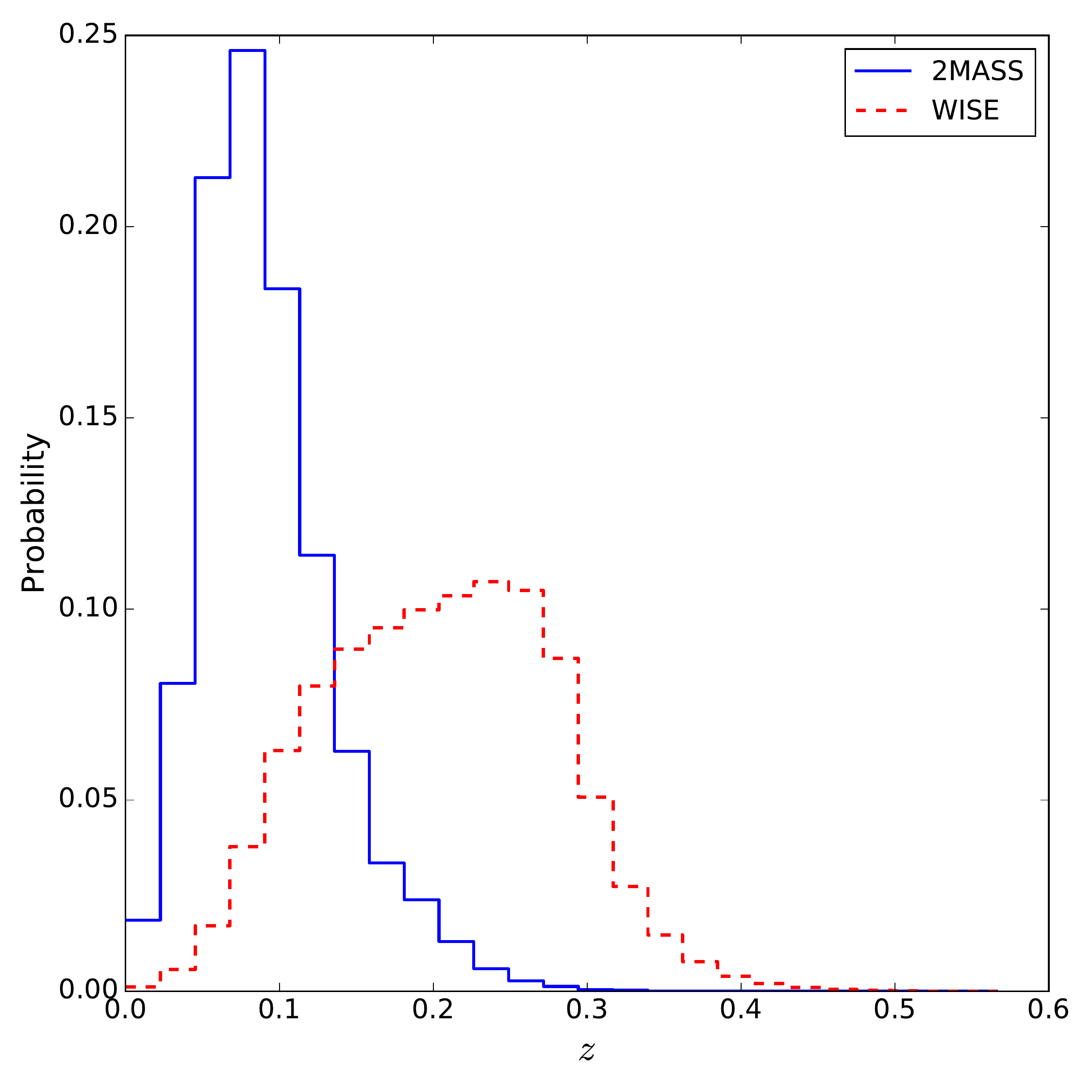}
\caption{Normalized redshift distributions for the 2MASS and WISE galaxy samples in the full EoR0 field. The solid blue line shows the 2MPZ  2MASS catalogue while the red shows the WISExSCOS catalogue. The median redshifts are $\langle z \rangle=0.08$ for 2MASS and  $\langle z \rangle=0.205$ for WISE.}
\label{fig:zhist}
\end{figure}

There are two near infrared (NIR) galaxies catalogues that have uniform coverage over the EoR0 field and provide spectroscopic and photometric redshift estimates that can be used as tracers of large-scale structure. For the first we use the publicly available data from the Two-Micron All-Sky Survey \citep[2MASS,][]{Skrutskie06}. Rather than using the full 2MASS catalogue we use the Two Micron All Sky Survey Photometric Redshift Catalog \citep[2MPZ,][]{Bilicki14a}, which contains spectroscopic and photometric redshift estimates for 1 million 2MASS sources.\footnote{Available from \url{http://ssa.roe.ac.uk/TWOMPZ} .} The median redshift of the distribution is $\langle z \rangle=0.08$. We created four galaxy number density maps from this catalogue. We made one map using all available sources in the field and three additional maps in redshift slices. The details of the four maps, including the redshifts, are given in Table~\ref{tab:tmassim}. Images of the four maps are shown in Fig.~\ref{fig:2massim} (while the images in Fig.~\ref{fig:2massim} have been convolved by a $3\,$arcmin beam for better visualization the maps used in the analysis are left in units of galaxies pixel$^{-1}$). 

\begin{table*}
\centering
\caption{2MASS galaxy number density map properties. All the images are $43.5\degr$ on a side, with equal area pixels with sides$=0.51\,$arcmin. Here $\langle z \rangle$ is the median redshift, $N_{\rm G}$ is the total number of galaxies in the map, $\mu_{\rm G}$ is the average number of galaxies per pixel, and $\sigma_{\rm G}$ is the standard deviation on the mean.}
\label{tab:tmassim}
\begin{tabular}{lcccccc}
\hline
Name & $z_{\rm low}$ &$z_{\rm high}$ & $\langle z \rangle $& $N_{\rm G}$ &$\mu_{\rm G}$&$\sigma_{\rm G}$\\
 & & & & [galaxies] & [galaxies pixel$^{-1}$]&[galaxies pixel$^{-1}$]\\
 \hline
2MASS\_Z1 & $0.00$ & $0.07$ & $0.05$&15658&$5.9\times10^{-4}$&$0.024$\\
2MASS\_Z2 &$0.07$ & $0.10$ & $0.08$&15651&$5.9\times10^{-4}$&$0.024$\\
2MASS\_Z3 &$0.10$ & $0.41$ & $0.13$&16127&$6.1\times10^{-4}$&$0.025$\\
2MASS &$0.00$ &$0.41$ & $0.08$ & 47436&$1.8\times10^{-3}$&$0.042$\\
\hline
\end{tabular}
\end{table*}

The second catalogue we make use of is from the Wide-Field Infrared Survey Explorer \citep[WISE,][]{Wright10} all-sky survey, specifically the WISExSCOS photometric redshift catalogue \citep{Bilicki16}, which contains 18.5 million sources.\footnote{Available from \url{http://ssa.roe.ac.uk/WISExSCOS}.} This catalogue was constructed by cross-matching the ALLWISE catalogue and the SuperCOSMOS all-sky samples and employing the artificial neural network approach \citep[the ANNz algorithm,][]{Collister04}, which was also performed for the 2MPZ catalogue with the 2MASS XSC catalogue (instead of the WISE catalogue). The WISE catalogue only contains photometric redshifts, with a median redshift of $\langle z \rangle=0.2$. Due to the fact that the WISE catalogue contains a larger number of sources over a larger redshift range, we split the catalogue up into five redshift slices, as well as a combined map (with the number of bins chosen in an effort to balance not having too many bins for computation reasons and not having too many sources per redshift bin). The six maps are shown in Fig.~\ref{fig:wiseim}, with map details presented in Table~\ref{tab:twiseim}.

While these datasets likely contain overlap in the sources, each covers a different redshift range. The redshift distributions for each catalogue (in the EoR0 field) is shown in Fig.~\ref{fig:zhist}. 

\begin{table*}
\centering
\caption{WISE galaxy number density map properties. All the images are $43.5\degr$ on a side, with equal area pixels with sides$=0.51\,$arcmin. Here $\langle z \rangle$ is the median redshift, $N_{\rm G}$ is the total number of galaxies in the map, $\mu_{\rm G}$ is the average number of galaxies per pixel, and $\sigma_{\rm G}$ is the standard deviation on the mean.}
\label{tab:twiseim}
\begin{tabular}{lcccccc}
\hline
Name & $z_{\rm low}$ &$z_{\rm high}$ & $\langle z \rangle $& $N_{\rm G}$ &$\mu_{\rm G}$&$\sigma_{\rm G}$\\
 & & & & [galaxies] & [galaxies pixel$^{-1}$]&[galaxies pixel$^{-1}$]\\
 \hline
WISE\_Z1 & $0.00$ & $0.13$ & $0.11$&231666&$8.8\times10^{-3}$&$0.095$\\
WISE\_Z2 &$0.13$ & $0.18$ & $0.16$&2316668&$8.8\times10^{-3}$&$0.095$\\
WISE\_Z3 &$0.18$ & $0.23$ & $0.20$&231665&$8.8\times10^{-3}$&$0.094$\\
WISE\_Z4 &$0.23$ & $0.27$ & $0.25$&231667&$8.8\times10^{-3}$&$0.094$\\
WISE\_Z5 &$0.27$ & $0.57$ & $0.30$&231663&$8.8\times10^{-3}$&$0.094$\\
WISE &$0.00$ &$0.57$ & $0.20$ & 1158329&$4.4\times10^{-2}$&$0.21$\\
\hline
\end{tabular}
\end{table*}

\section{Cross correlation method}
\label{sec:crx}
\subsection{The cross-correlation function}
\label{sec:ccf}

The cross-correlation function (CCF) of two images or maps, $R$ and $G$, is defined as
\begin{equation}
{\rm CCF}(\Delta x, \Delta y)_{RG}= \sum_{i,j} (R_{i,j}-\bar{R})(G_{i,j}(\Delta x, \Delta y) -\bar{G}),
\label{eq:crx1}
\end{equation}
where $R$ is the radio map, $G(\Delta x, \Delta y)$ is the galaxy number density map shifted in relation to $R$, and $\bar{R}$ and $\bar{G}$ are the respective map means. The normalized cross correlation is just ${\rm CCF}_{RG}/(\sigma_R \sigma_G)$, where $\sigma$ is the standard deviation of the respective map. The sum is over all pixels. The 1D cross-correlation function, CCF($\Delta r$)$_{RG}$, is just the radial average of the 2D CCF, with $r=\sqrt{\Delta x^2 +\Delta y^2}$. 

This 1D averaging assumes radial symmetry in the 2D function. When the 2D function is not  radially symmetric, performing the radial average may yield an inaccurate estimate and should not be used. Over a large enough area, the cosmic web should be isotropic and should produce a radially symmetric 2D cross correlation with the galaxy number density. This may not be case when considering smaller areas or low redshift galaxy number densities, however, looking at the galaxy number density maps shown in Figs.~\ref{fig:2massim} and \ref{fig:wiseim} tells us we are in the regime where the cosmic web is (roughly) isotropic. 

The redshift range of the LSS tracers matters because the synchrotron emission in filamentary LSS should have an average width, or characteristic size. If that size is  $\gg$ the beam size (degrees compared to an arcmins) then the cross correlation signal is spread out over a large range of angular shifts, likely producing lower amplitudes and making it more difficult to detect. If the filament size is $\ll$ than the beam size the emission would appear point source like and be more difficult to disentangle from the correlation signal from actual point sources. For these reasons ideally we would like to have a minimum of two beam widths across a filament and probably not more than $\sim 10$ (though the upper limit on this is difficult to know and depends on the underlying physics of the cosmic web and how it correlates with the galaxy number densities, which is not well known). Thus there is an ideal redshift range for a given beam size (and assumed filament size) to enhance the cosmic web signal in the cross correlation. 

\subsection{CCF from random galaxy number density maps}
\label{sec:crxnoise}

\begin{figure}
\includegraphics[scale=0.37]{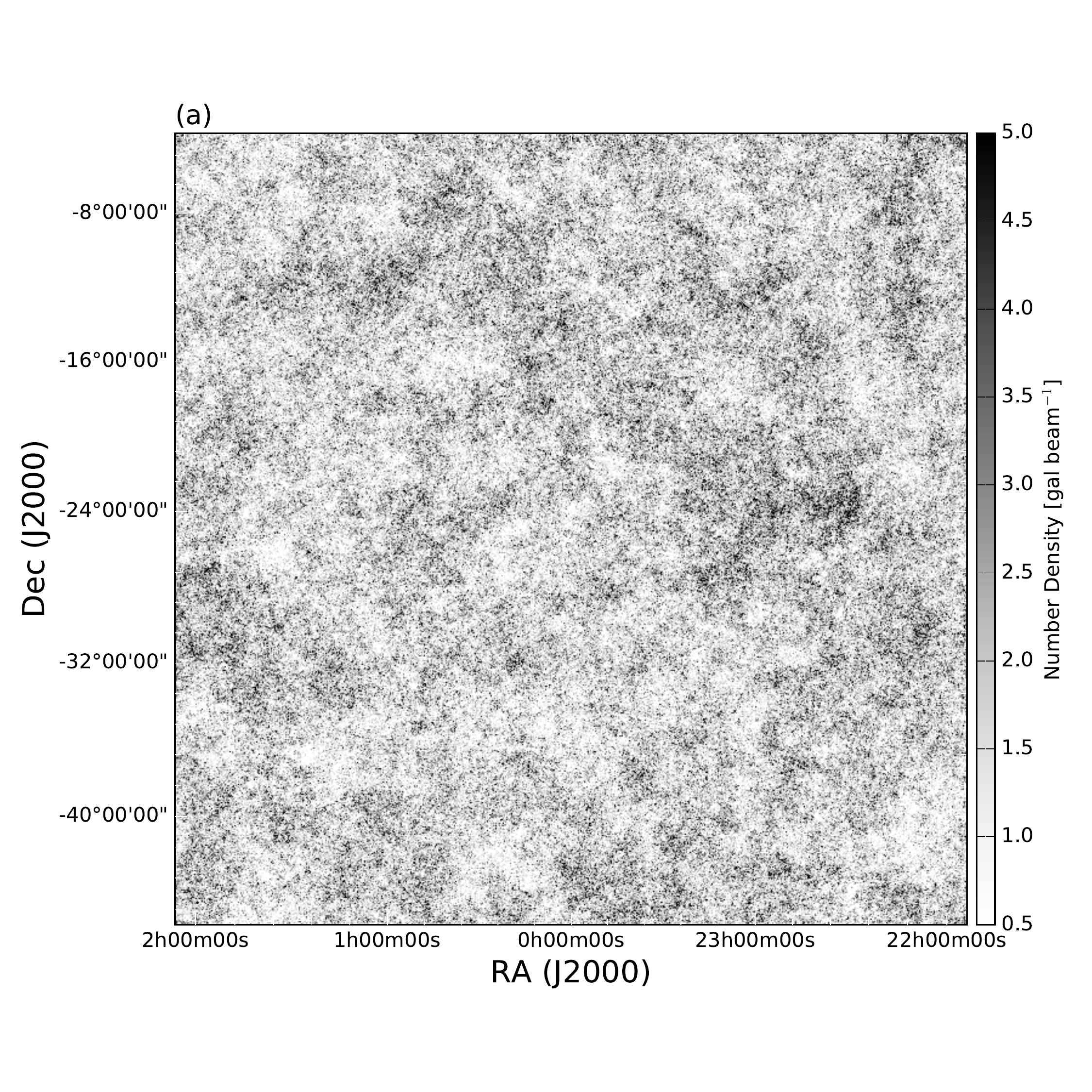}
\centering
\includegraphics[scale=0.34]{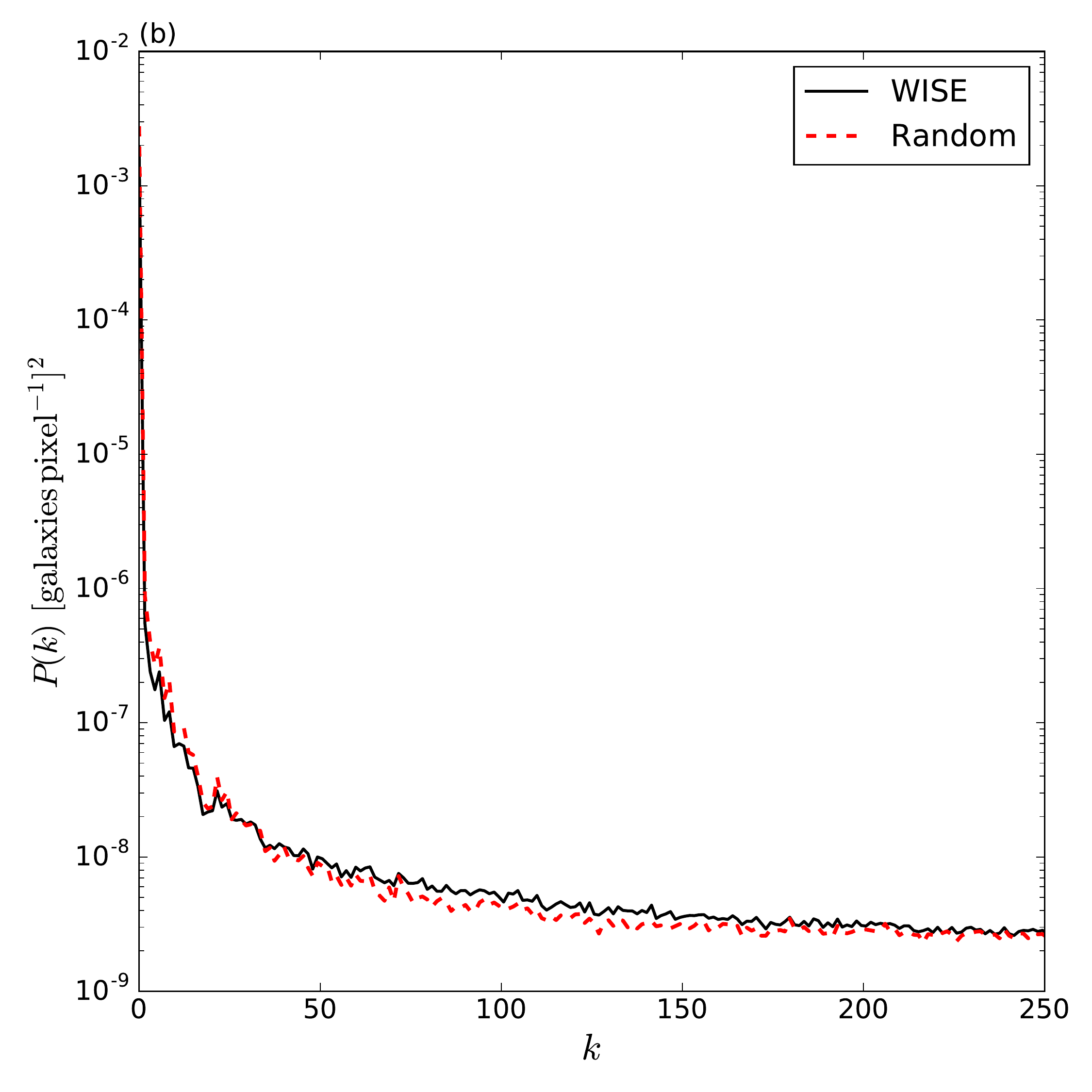}
\caption{Randomly generated WISE galaxy number density map and power spectrum. Panel (a) shows one realization of a randomly generated galaxy number density map designed to match the WISE power spectrum; compare with Fig.~\ref{fig:wiseim}(f). This map has been convolved with the $3\,$arcmin MWA beam to better visualize the points. The panel (b) shows the WISE power spectrum $P(k)$ (solid black line) compared with the power spectrum from the random galaxy number density map (dashed red line).}
\label{fig:rands}
\end{figure}

In order to interpret the significance of the cross correlation of real images and maps we needed to know the expected results from a null correlation. To find this we cross correlate the radio images with galaxy number density maps that should have a zero correlation. For these null maps, we could use different parts of the sky from each catalogue, however, that could introduce inconsistencies caused by Galactic extinction or uneven survey coverage. The other option is to generate random galaxy number density maps.

The process for generating random maps is not as straightforward as populating maps with randomly generated Poisson noise. Randomly generated Poisson noise on its own can match the mean and variance of the real density maps but would be lacking any spatial clustering. The process to generate random galaxy number density maps with the same clustering properties as the true galaxy number density map, and compute the corresponding cross correlation functions, entails generating random Poisson noise with mean and variance matching the galaxy number density map of interest and then using the power spectra of the random map and the actual galaxy number density map to modify the clustering properties of the random map. The process is defined as follows.

\begin{enumerate}[label={\arabic{enumi}.},leftmargin=*]
\item Calculate the power spectrum $P(k)_{G}$ for the galaxy number density map $G$. Here $k=\sqrt{u^2+v^2}$, where $u$ and $v$ are the Fourier conjugates of $x$ and $y$. 
\item Make a map $W$, with same area and pixel size as map $G$, with randomly generated Poisson noise that has the same mean and variance as $G$.
\item Compute the Fourier transform of $W$, ${\cal F} [ W]$.
\item Compute the power spectrum of $W$, $P(k)_W$.
\item For each $k$ bin compute the factor $h(k)$ such that $h(k)=P(k)_{G}/P(k)_{W}$.
\item For the pixels $u,v$ in the $i$th $k$ bin multiply the Fourier transform of $W$ by $h(k)$, ${\cal F} [ W(u,v)_i^*]= {\cal F} [ W(u,v)_i]h(k)_i$.
\item Take the inverse Fourier transform of ${\cal F} [ W(u,v)^*]$, $W^*={\cal F}^{-1}[{\cal F} [ W]^*]$. This inverse transform now has the same mean, variance, and clustering spectrum as $G$, but no longer has the properties of only having zero or a positive integer number for the pixel values (as is the case for the real galaxy number density maps).
\item Convert the values of $W^*$ to probabilities by shifting the values by the minimum of $W^*$ if the minimum value is less than zero and then dividing by the sum.
\item Make a new map $G_{\rm rand}$ by selecting $N_{G}$ pixels pseudo-randomly (with repeats) using the pixel values of $W^*$ as weights in the selection process.
\end{enumerate}

We repeated steps 1--9 1000 times for each radio image and galaxy number density map combination and compute the cross correlation functions. An example of one random galaxy number density map generated from the WISE galaxy number density map is shown in Fig.~\ref{fig:rands}(a) (which can be compared with the real WISE galaxy number density map shown in Fig.~\ref{fig:wiseim}f), with the two 1D power spectra compared in Fig.~\ref{fig:rands}(b). 

For each random galaxy number density map and radio image the CCF was computed, and from those 1000 CCFs we computed the median and confidence intervals for the median at each $\Delta \theta$. To check that the random procedure worked we also chose 10 random positions in the WISE and 2MASS catalogues and cross correlated those galaxy number density maps with the radio images. The values from the random catalogue position maps indicated a similar spread to that found with the randomly generated galaxy number density maps.

\section{Results of MWA and LSS cross-correlation}
\label{sec:results}

\begin{figure*}
\centering
\includegraphics[scale=0.49]{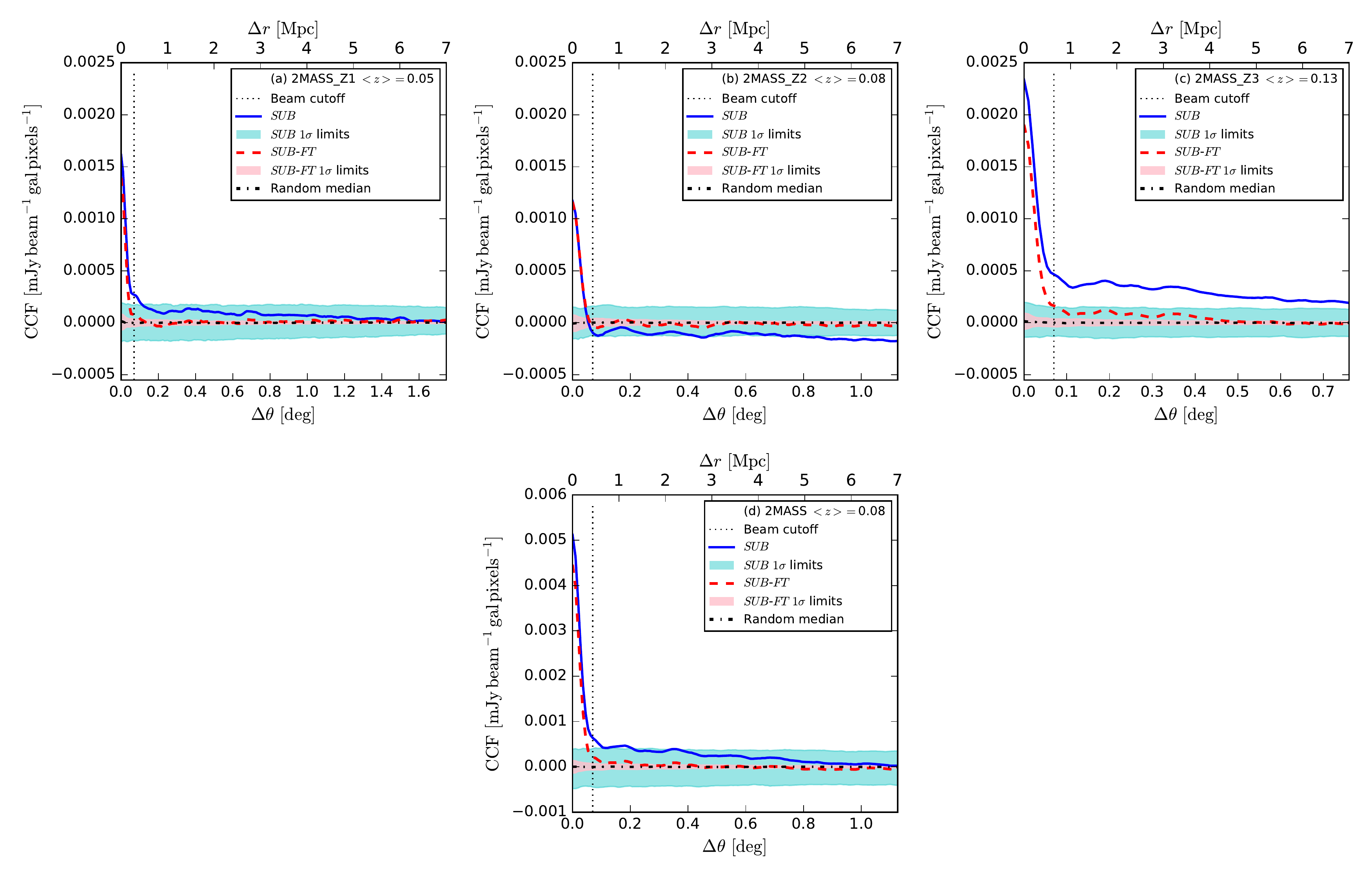}
\caption{Cross correlations of 2MASS galaxy number density maps with radio images. From left to right top to bottom the plots show the CCF of the radio images with 2MASS\_Z1, 2MASS\_Z2, 2MASS\_Z3, and the sum map 2MASS, which are given as the first line of the subplot legends along with the corresponding median redshifts. The blue solid lines show the CCF with the {\galr} radio image and the red dashed lines show the {\minur} results. The black dot-dashed lines show the medians from cross correlating with random galaxy number density maps (see Sec.~\ref{sec:crxnoise}) and the corresponding coloured regions show the $68\,$per cent confidence regions. The vertical black dotted lines show the cutoff point of the main synthesized beam lobe. The bottom horizontal axis shows $\Delta \theta$ in degrees, whereas the top horizontal axis shows $\Delta r$ in Mpc, computed using the median redshift of each galaxy number density map.}
\label{fig:crxs1_2mass}
\end{figure*}

\begin{figure*}
\centering
\includegraphics[scale=0.49]{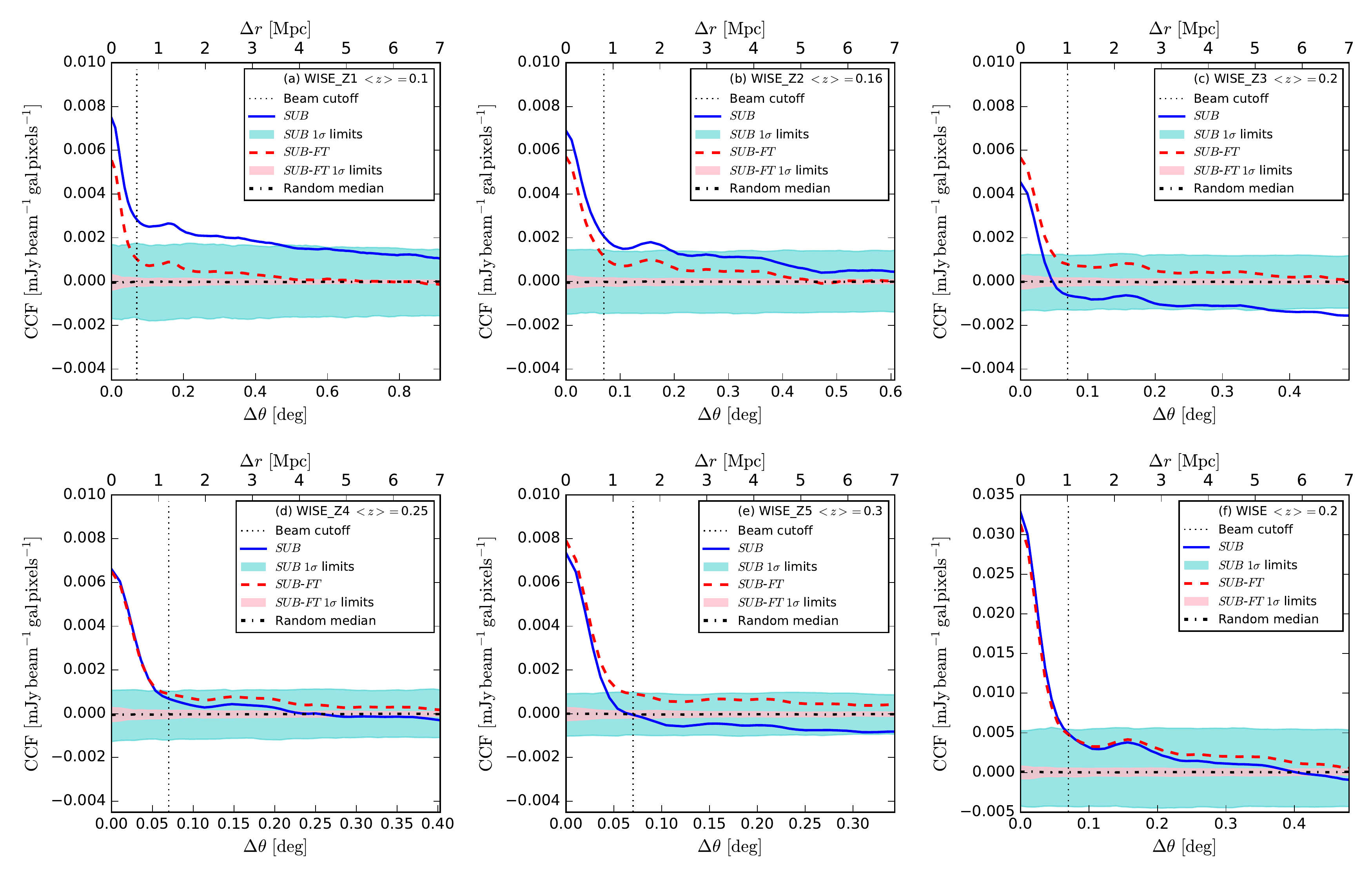}
\caption{As for Fig.~\ref{fig:crxs1_2mass}, but for the WISE galaxy number density maps with radio images. From left to right top to bottom the plots show the CCF of the radio images with WISE\_Z1, WISE\_Z2, WISE\_Z3, WISE\_Z4, WISE\_Z5, and the sum map WISE. }
\label{fig:crxs1_wise}
\end{figure*}

The results of cross correlating the radio images with the 2MASS galaxy number density maps are shown in Fig.~\ref{fig:crxs1_2mass}, while the results from the WISE galaxy number density maps are shown in Fig.~\ref{fig:crxs1_wise}. The median of the random galaxy number density map CCFs and the $68\,$per cent confidence intervals as defined in Sec.~\ref{sec:crxnoise} are also shown. The cross correlation was computed over the full range of possible shifts (max $\Delta \theta  =15.4\degr$), however, we restrict the plot range to a smaller region in order to limit the range to where the majority of the signal is present and show all of the CCFs on the same axis for comparison. We chose this region to be within $\Delta r=7\,$Mpc (as determined by the median redshift), as it shows at least several Mpc without being too large that any shape after the main beam lobe would be difficult to see on any of the plots. 

All of the random CCF medians are approximately zero. In all cases the cross correlations show a pronounced peak at $\Delta r=\Delta \theta = 0$ and then drop off towards zero with varying slopes, with all dropping to zero or below by $\Delta \theta=1\degr$. The amplitudes of the peaks range from $0.001 \,$mJy beam$^{-1}$ galaxies pixels$^{-1}$ ({\minur} and $2MASS\_Z2$) to $0.03 \,$mJy beam$^{-1}$ galaxies pixels$^{-1}$ ({\galr} and $WISE$).

 If we compare the central peaks with the beam profiles in Fig.~\ref{fig:bms}(c), we can see that the beam ACF shapes have central peaks and drop toward zero (or a minimum) at $\Delta \theta \simeq 0.07\degr$ (shown in the figures as a vertical dotted line). A positive correlation of the galaxy number density maps with the radio point source emission should follow this beam shape; having little to no signal past $\Delta \theta \simeq 0.07\degr$, besides signal due to the beam sidelobes. Therefore, it is likely that the peaks seen in the cross correlations at $\Delta \theta \le 0.07\degr$ are due to a positive correlation of the number density with faint unsubtracted point sources. A signal from the cosmic web, which would be diffuse rather than point source like, should show correlation signal on shifts larger than the main beam lobe. The point source contribution and its effect on our ability to constrain the cosmic web signal is discussed in Sec.~\ref{sec:dis_ps}. In almost all cases the CCFs also show a second peak near $\Delta \theta=0.15\degr$, which is from the beam sidelobes. 

The unnormalized 2MASS CCFs all have lower amplitudes than the WISE CCFs. This is due to the large difference in the number of objects between the WISE and 2MASS maps. This effect goes away, at least in terms of the central peak, when considering the normalized CCFs (i.e. dividing the CCFs by the standard deviations of the radio image and galaxy number density map, which are not shown as they only show the relative correlation strength rather than information on the rms fluctuations in Jy beam$^{-1}$ galaxies pixel$^{-1}$).  

However, the 2MASS CCFs also have lower amplitudes (relative to the central peaks) than the WISE maps when looking at shifts greater than the main beam lobe. This could be due to a selection bias, i.e. the galaxies in WISE maps being more clustered than the ones in the 2MASS maps. It could also be due to the fact that the 2MASS objects are at lower redshifts than the WISE objects. At these lower redshifts the characteristic size of the diffuse emission could be much larger than MWA beam, resulting in the diffuse correlation signal being spread across larger $\Delta \theta$s (i.e. having a lower amplitude at a given shift). Also, as shown in Figs.~\ref{fig:2massim} and \ref{fig:wiseim}, it is clear that the WISE galaxies are more clustered than the 2MASS galaxies, which would produce a higher cross correlation signal at larger shifts. 

One of the most noticeable features in Fig.~\ref{fig:crxs1_2mass} and Fig.~\ref{fig:crxs1_wise} is the width of the uncertainty regions of the {\galr} images compared to those of the {\minur} images, with the {\galr} image confidence intervals being 2--4 times larger than the {\minur} intervals, regardless of the galaxy number density map. We believe this is due to large-scale diffuse Galactic emission, which we discuss further in Sec.~\ref{sec:dis_gal}. This does affect our ability to constrain the cosmic web signal, which is discussed further in Sec.~\ref{sec:limits}.

In the following section we explore these results in more detail, looking into systematics and possible physical interpretations.

\section{Discussion}
\label{sec:discussion}
Interpreting the results shown in the previous section is not straightforward. There are many factors to consider which affect the cross correlation signals and the uncertainties. In this section we discuss those factors in greater detail.

\subsection{Effects on the cross correlation due to point sources}
\label{sec:dis_ps}
There are two issues to discuss when considering the effects of point sources. First is the remaining, or unsubtracted, point sources in the image. Second are the point sources that were subtracted. We examine the unsubtracted point sources first. 

\begin{figure}
\includegraphics[scale=0.35]{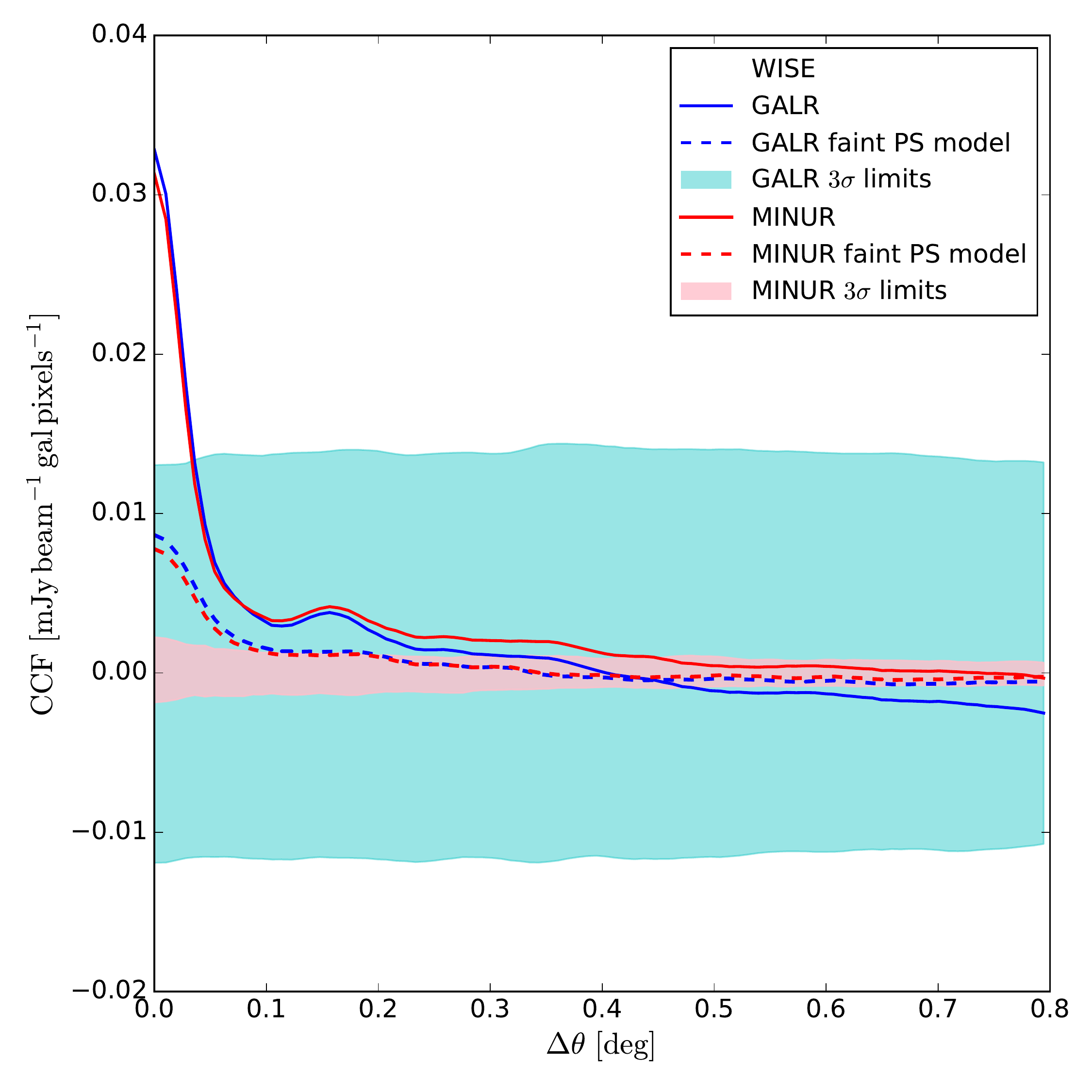}
\caption{As for Fig.~\ref{fig:crxs1_2mass}, but showing the cross correlation functions of radio point source models with galaxy number density maps. This shows the CCFs of the MWA radio images (solid lines) and point source models (dashed lines) with the WISE image (blue lines are the {\galr} images, red lines are the {\minur} images), with the shaded regions showing the $99.7\,$per cent confidence regions. The dashed lines are the point source models that were subtracted during the imaging stage, convolved with the dirty beams, scaled to have variance equal to that found with eq~(\ref{eq:pshot}) using the source count in Sec.~\ref{sec:rdatnoise} and $S_{\rm max}=0.05\,$Jy, and then cross correlated with the WISE map.}
\label{fig:pointsrc}
\end{figure}

\begin{figure}
\includegraphics[scale=0.47]{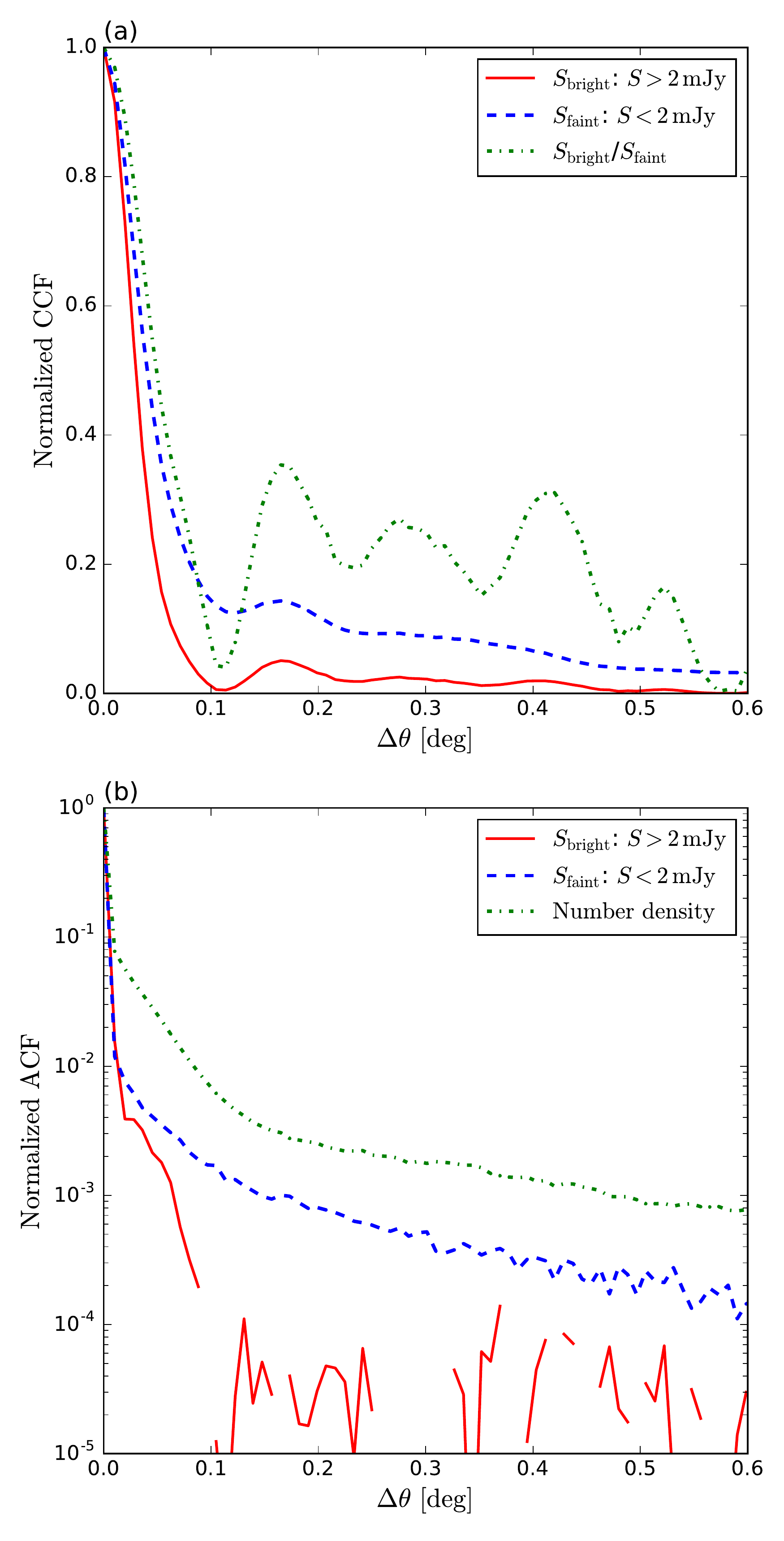}
\caption{Correlation functions of point source models with galaxy number density maps from simulated data. Panel (a) shows the normalized CCFs from the SKADS S$^3$ simulation $150\,$MHz catalogue \citep[][see Sec.~\ref{sec:dis_ps} for details]{Wilman08}. The red solid line shows the $S_{\rm bright}$ CCF for point sources with $S>2\,$mJy, the blue dashed line shows the $S_{\rm faint}$ CCF from point sources with $S<2\,$mJy (convolved with the MWA dirty beam) and the corresponding galaxy number density map, and the green dot-dashed line shows the ratio of $S_{\rm bright}/S_{\rm faint}$. Panel (b) shows the normalized auto correlation functions (with no beam convolution) for $S_{\rm bright}$ (red solid line), $S_{\rm faint}$ (blue dashed line), and the galaxy number density map (green dot-dashed line).  }
\label{fig:pointsrc_sk}
\end{figure}

As was mentioned in Sec.~\ref{sec:results}, the cross correlations still contain a component from unsubtracted point sources. However, it is not known how strong this correlation is or what shape the CCF takes. One possible way to estimate this is to use the point source models (or clean component images) that were generated during the imaging process and were subtracted from the data. These model images can be mosaicked together and scaled to have the same estimated variance of the unsubtracted point sources. The source count can be used to estimate the point source variance $\sigma_{\rm PS}$ by 
\begin{equation}
\sigma_{\rm PS}=\int_{S_{\rm min}}^{S_{\rm max}} \, S^2 \frac{dN}{dS} \, dS .
\label{eq:pshot}
\end{equation}

In our case $dN/dS$ is the differential source count given in Sec.~\ref{sec:rdatnoise} with $S_{\rm max}$ set to $0.05\,$Jy, the cleaning subtraction limit that was used for the radio images. This yields $\sigma_{\rm PS}=1.18\,$mJy. We took the point source clean component images and scaled them such that their un-convolved rms was equal to $\sigma_{\rm PS}$. When the scaled point source images are convolved with the MWA dirty beams the noise becomes $\sigma_{\rm c}=9.1\,$mJy beam$^{-1}$ and $\sigma_{\rm c}=8.2\,$mJy beam$^{-1}$, for the {\galr} and {\minur} beams respectively. The values of $\sigma_{\rm tot}$ for the radio images are $\sigma_{\rm tot}=11.5\,$mJy beam$^{-1}$ for the {\galr} image and $\sigma_{\rm tot}=8.7\,$mJy beam$^{-1}$ for the {\minur} image. Using a value of $\sigma_{\rm n}=0.96\,$mJy beam$^{-1}$ for the instrumental noise, $\sqrt{\sigma_{\rm tot}^2-\sigma_{\rm c}^2-\sigma_{\rm n}^2}$ leaves $7.0\,$mJy beam$^{-1}$ unaccounted for in the {\galr} image and $2.7\,$mJy beam$^{-1}$ in the {\minur} image. This could represent the variance from the Galaxy, other instrumental effects, and/or the cosmic web.

Figure~\ref{fig:pointsrc}(a) shows the CCFs of these point source models with the WISE galaxy number density map, compared with the CCFs of the {\galr} and {\minur} images and random galaxy number density map confidence intervals. The plot shows central peaks that fall off after the main beam lobe with amplitudes of only about $30\,$per cent of the {\galr} and {\minur} CCF peaks. This shows that, assuming the faint point sources correlate in a similar way as the bright sources, then the unsubtracted point source contribution does not account for the entire, or even majority, of the cross correlation signal. 

The problem with using the subtracted, or bright, point sources as a model for the unsubtracted sources is that it assumes that the clustering or correlation of the bright point sources is the same shape as the faint point sources. We stated previously that the contribution from point sources to the CCF should trace the shape of the beam, but this is not entirely accurate. The CCF may contain signal from point sources that is outside the main beam lobe (besides that due to side lobes), depending on how clustered the emission from sources is and how that clustering correlates with the galaxy number density maps. 

The spatial clustering of bright ($S\ga 10 \,$mJy) radio sources is known from the two-point correlation functions of the NVSS and FIRST surveys \citep[e.g.][]{Blake02a,Overzier03, Massardi10}. More recent work by \citet{Lindsay14a} used fainter radio sources ($S\ga 90\, \mu$Jy) to look at the auto correlation function of the spatial clustering of the radio sources, as well as the cross correlation of the radio with IR sources. \citet{Lindsay14a} found an increase in the clustering bias for sources at higher redshifts. They also discovered that the radio bias at higher redshifts is greater than that assumed by simulation models. While we have some idea of how clustered the radio point source positions are, it is still not known how clustered the radio power from point sources is, particularly from faint point sources, let alone how well it correlates with NIR galaxy number densities. 

One way to test the difference between faint and bright point sources is to use simulated data. By using a catalogue of simulated point sources with a wide range of flux densities whose positions include clustering, we can compare the difference in the correlation of bright and faint point sources (with themselves and with number densities). 

To do this we used data from the Square Kilometre Array Design Studies (SKADS) SKA Simulated Skies (S$^{3}$) simulation \citep{Wilman08}.\footnote{\url{http://s-cubed.physics.ox.ac.uk}} The S$^3$ simulation is a large-scale semi-empirical model of the extragalactic radio continuum sky at several frequencies covering $20\,$deg$^{2}$. This simulation has realistic approximations of the known source counts and contains both small and large-scale clustering. Using the $150\,$MHz flux densities (ignoring source sizes for simplicity) for sources at $z\le 5$, we made two images (with the same pixel size as the MWA images). The first image contained sources with $S\le 2\,$mJy and the other containing sources with $S>2\,$mJy. These were both convolved with the MWA dirty beams. The S$^3$ catalogue also contains $K$-band magnitudes, which we used as a proxy for the NIR catalogues. We chose $200000$ sources with $z\le0.5$ and $K$ magnitudes $\le 19$ (similar to the WISE or 2MASS data) to create a galaxy number density map. We cross correlated both radio images with the galaxy number density map. The normalized CCFs are shown in Fig.~\ref{fig:pointsrc_sk}(a). 

Looking at the CCFs from the simulation shows that beyond the main lobe of the beam, the CCF for faint sources is on average $30\,$per cent larger than the bright source CCF. This means the faint source flux density is more clustered (or correlated on small distances) than the bright source flux density. This is shown in Fig.~\ref{fig:pointsrc_sk}(b), which displays the normalized auto correlation functions of the bright and faint point source images (with no beam convolution), as well as the autocorrelation of the galaxy number density map. From this we can see that the number density is the most clustered of the three and that the faint point sources are more clustered than the bright point sources. 

While this simple simulation test may not represent the true sky accurately (\citet{Lindsay14a} found the clustering bias of higher redshift sources to be larger than that used in the S$^3$ simulation), it shows that with reasonably realistic assumptions it is possible for faint point sources to show correlation signal on scales larger than the main beam lobe, and that the scaled bright point source flux density image may not be a good model for the distribution of faint point sources.

The discussion up until here has focused on the effect of faint point sources that were not subtracted, and their effect on the cross correlation and its interpretation. There are also issues involving the subtracted point sources, or rather, how the sources are subtracted. In our case only the brightest sources ($S\ga0.25\,$Jy) had models of previously measured positions, flux densities, and spectral indices used to subtract, or in this case ``peel" them out. After the peeling step we simply imaged the data with the smallest beam possible ($2.3\,$arcmin) and created a clean component ``point source'' model from the unresolved image peaks. The problem with this method is that with a 2$\,$arcmin beam, an unresolved peak may not be, and is likely not, a single point source. Rather it is likely a blend of multiple sources or even diffuse emission. 

There are at least four previously discovered diffuse radio cluster objects (haloes and relics) in the EoR0 field: a relic in cluster A4038, and A13 \citep{Slee01} and a relic and radio halo in cluster A2744 \citep{Govoni01}. However, even though these are diffuse objects they may have been interpreted as point source clean components during the imaging (due to the low resolution) and been (at least partially) subtracted out.

Clearly what is needed is to have deep high-resolution data available to construct a proper point source model for subtraction. It is only recently that high-resolution extragalactic surveys in the low frequency regime have been able to reach the necessary depths (mJy and sub-mJy), for example the TIFR GMRT Sky Survey \citep[TGSS, e.g. ][]{Intema16} or those that may be conduct with the LOFAR telescope at higher resolutions. However, there are not currently enough data to construct a proper point source model for subtraction that goes to faint enough flux densities. 

Surveys from higher frequencies, such as $1.4\,$GHz where there are much more data available, certainly reach to fainter flux densities. However, accurate spectral indices would be needed for each source to construct a low frequency model, which are usually not available for all of the sources. A similar type of experiment as this could be conducted using radio data at $1.4\,$GHz, such as with the Evolutionary Map of the Universe \citep[EMU,][]{Norris11} survey with the ASKAP telescope or the MeerKAT International GigaHertz Tiered Extragalactic Exploration survey \citep[MIGHTEE,][]{Jarvis11}. The downside to that is the cosmic web emission, with a steep spectral index \citep[$-0.8\la \alpha \la -1.25$, e.g.][]{Liang02,Feretti04}, will be much fainter at higher frequencies. 

An ideal setup would be to have comparably sensitive high (arcsec) and low (arcmin) resolution data from the same telescope. In the low frequency regime, this could be achievable by LOFAR as well as the MWA, once the MWA upgrade, which will include longer baselines, is complete. However, in this case one would have to be careful with the subtraction of any extended emission (i.e. galaxy lobes) so that it would not be misinterpreted as intergalactic diffuse emission at the lower resolution. 

Short of having an accurate point source model before imaging, the method used in this paper of subtracting the clean components can still provide limits on the cosmic web signal via the cross correlation, but these limits would be improved by the ability to clean deeper in the images. This would require longer integrations; the MWA currently uses $2\,$minute snapshots. Even with many $2\,$minute snapshots mosaicked together, the beam shape does not necessarily improve unless the snapshots have very different {\it uv} coverage, which was not the case with the current data. Longer integrations would improve the dirty beam sidelobes, which would reduce confusion noise and allow an image to be cleaned to fainter flux densities. It would also reduce the sidelobe contribution from point sources in the cross correlation. This may require a different telescope setup, i.e. such as the Jansky Very Large Array (VLA), which can track a field over a longer continuous period. 

\subsection{Effects on the cross correlation due to Galactic emission}
\label{sec:dis_gal}

\begin{figure*}
\includegraphics[scale=0.375]{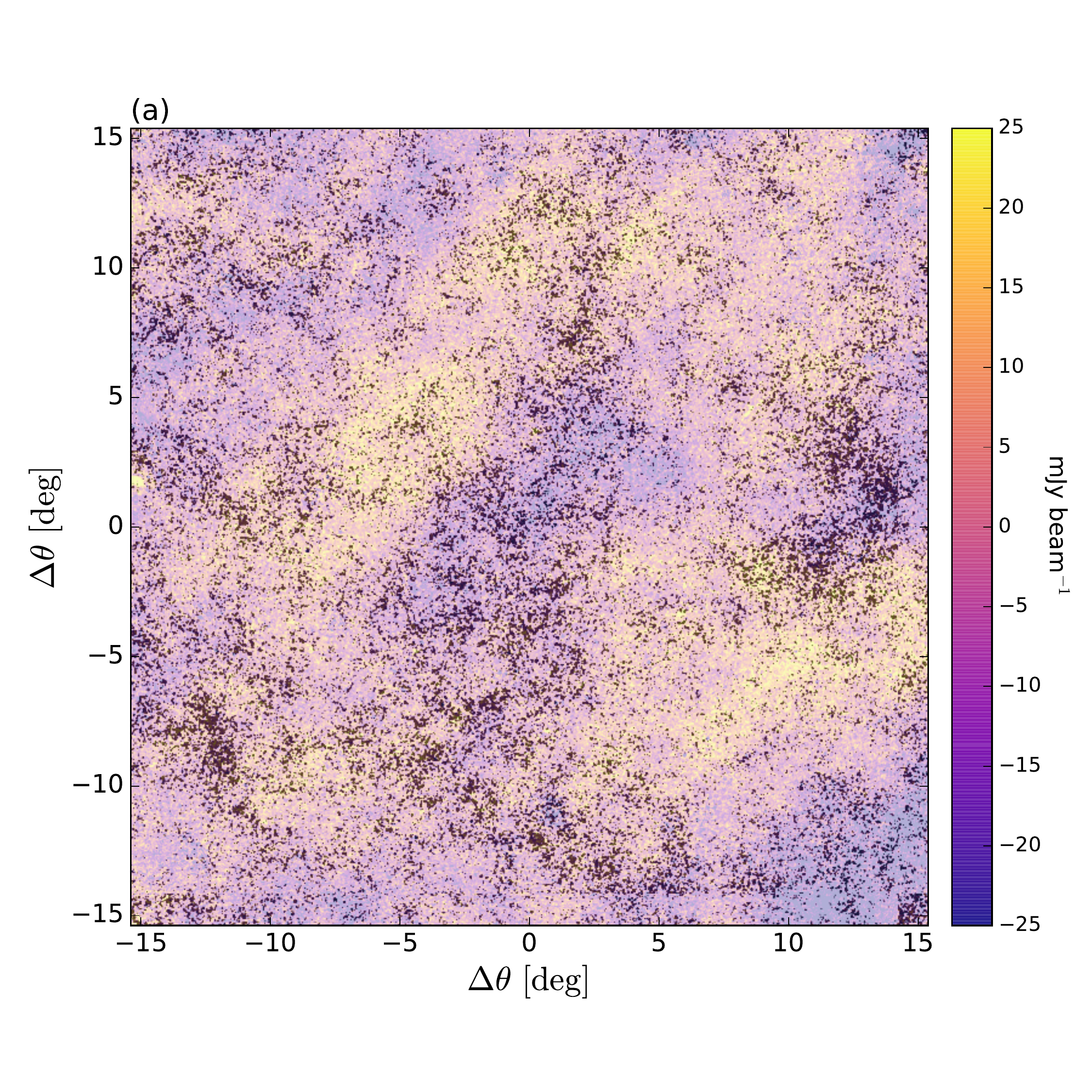}\includegraphics[scale=0.375]{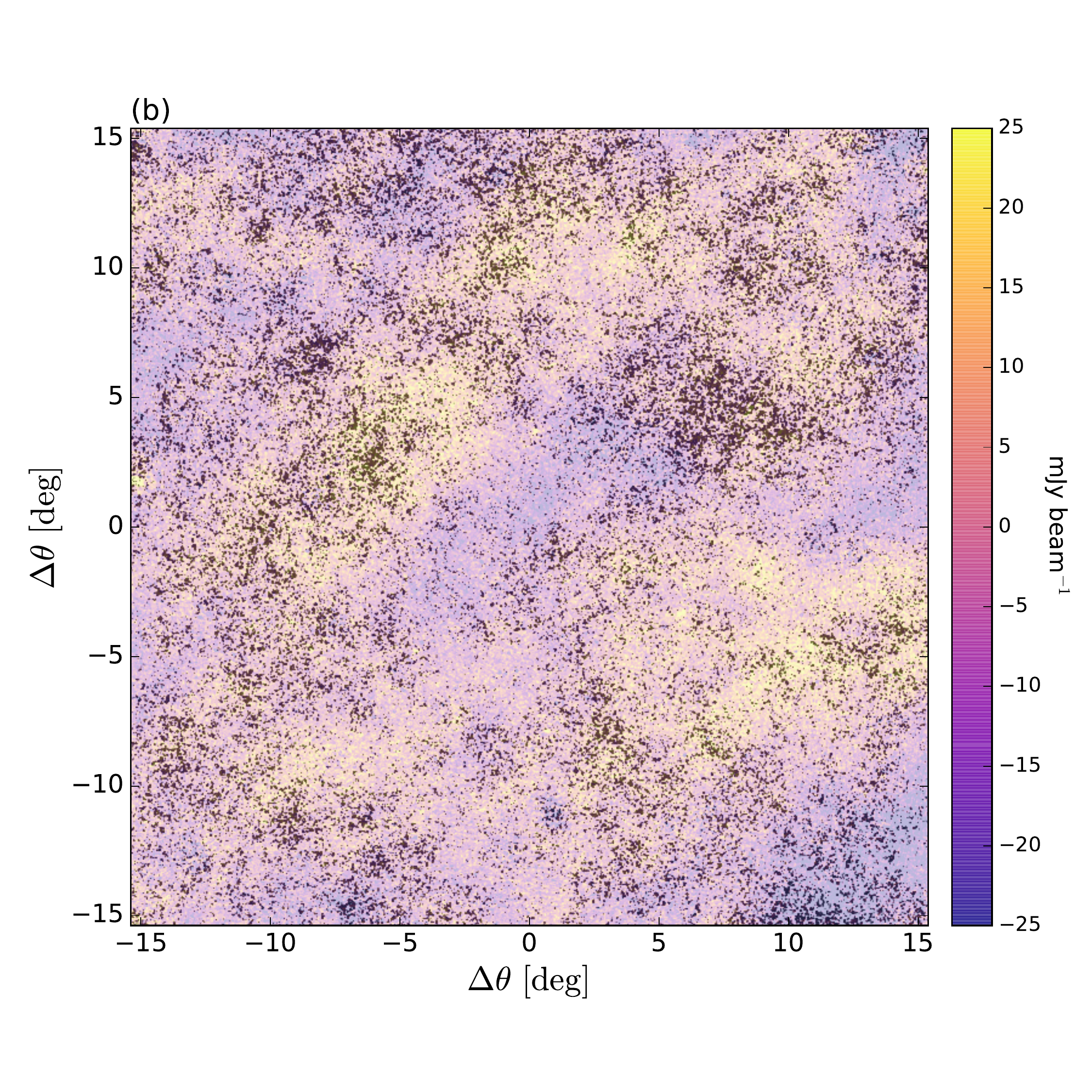}
\includegraphics[scale=0.375]{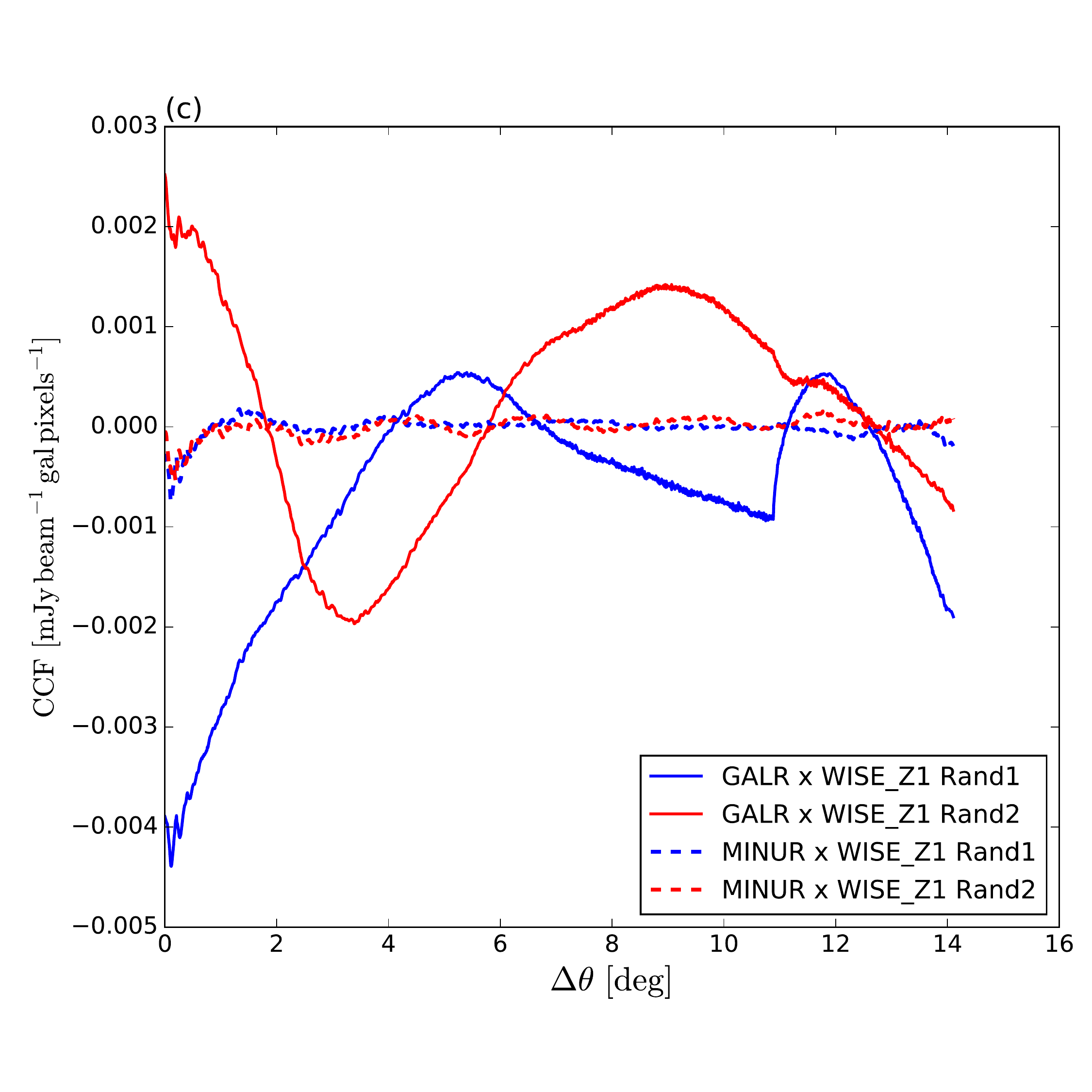}
\caption{Random realizations of WISE\_Z1 example images and cross correlations. Panels (a) and (b) show the {\galr} radio image with the random WISE\_Z1 realization (beam convolved) overlaid (rand1 in panel a and rand2 in panel b). These show how the random clustered Poisson noise can negatively (rand1) and positively (rand2) spatially align (or correlate) with the large regions of diffuse Galactic emission in the {\galr} image. Panel (c) shows the cross correlation functions of these two random galaxy number density maps with the {\galr} radio image (solid blue line for rand1 and dashed blue line for rand2) and {\minur} radio image (solid red line for rand1 and dashed red line for rand2).}
\label{fig:galrand}
\end{figure*}

\begin{figure}
\includegraphics[scale=0.365]{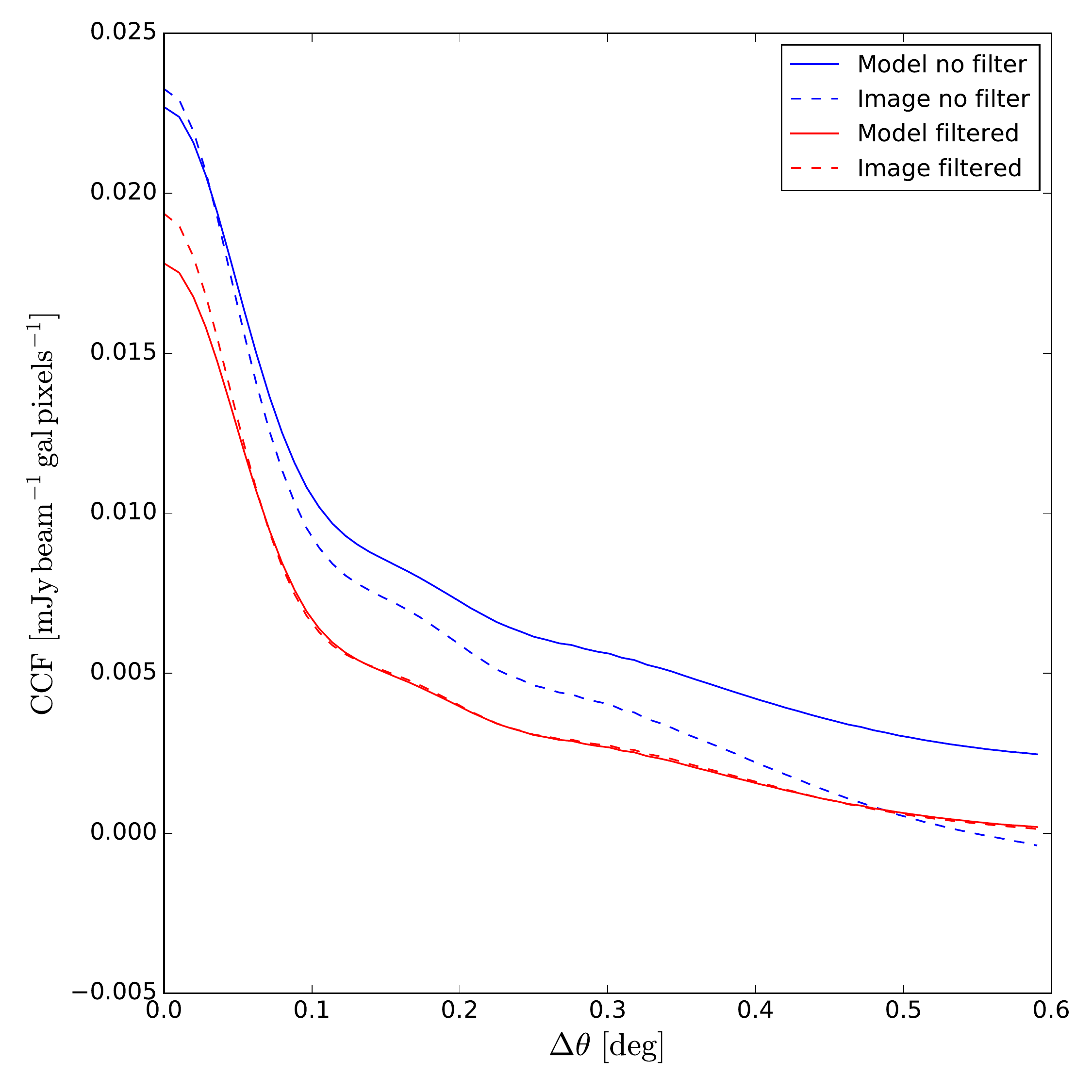}
\caption{Correlation functions of simulated diffuse emission. The solid lines show the CCFs of the models used as input to the imaging simulation (convolved with the corresponding image synthesized beams) cross correlated with the galaxy number density map. The dashed lines show the output images from the simulation cross correlated with the galaxy number density map. The blue lines show the images with no filtering applied, while the red lines show the results when a minimum baseline filter of $34\,$m is applied during the imaging. See Sec.~\ref{sec:dis_gal} for a full description of the simulation.}
\label{fig:dfsim}
\end{figure}

We saw in Figs.~\ref{fig:crxs1_2mass} and \ref{fig:crxs1_wise} that the uncertainties in the correlations are much larger in the {\galr} images than the {\minur} images, where the {\galr} images are the images without the filter on larger degree-scale emission. Larger uncertainties make it more difficult to potentially detect, or at least put tighter constraints on, the cosmic web signal. 

The reason the uncertainty covers such a large range with the {\galr} image is that the random galaxy number density maps are not just Poisson noise, but include clustering. The correlation between the random galaxy number densities and radio point sources is small. However, there are degree-sized areas of bright and faint diffuse, and likely Galactic, emission in the radio image (see Fig.~\ref{fig:radims}b) with which the clustered areas of over- and under-densities in the random galaxy number density maps positively and negatively correlated. We believe these regions to be Galactic as they are too bright to come from something like the cosmic web (the emission is on the order of mJy's and if the cosmic web signal were that bright it would have been detected by now). Also, the regions are too large,  being several degrees across, to come from clusters (unless at a {\it very} low redshift, but they do not correlate spatially with extragalactic source positions). The correlations show little to no change in the CCFs over the range of shifts shown in Figs.~\ref{fig:crxs1_2mass} and \ref{fig:crxs1_wise}, but show large changes if the CCF is examined over the full $15\degr$, which shows the uncertainties in the {\galr} images are dominated by signals at degree sized scales.  

An example using two randomly generated maps from the WISE\_Z1 galaxy number density map are shown in Fig.~\ref{fig:galrand}. Panels (a) and (b) show the random maps overlaid on the {\galr} radio image. It can be seen that higher density areas in panel (a) are aligned with the lower flux density regions in the radio image, whereas the opposite is true in panel (b). Panel (c) shows the full CCFs of these two maps with both the {\galr} and {\minur} radio images. Large changes in amplitude are seen at shifts greater than a few degrees for the {\galr} CCFs, whereas the {\minur} CCFs show much smaller amplitudes at larger shifts, and overall.  

It is possible that the filtering applied to the {\minur} images, and other similar filtering techniques performed in the image plane, may be the solution to this problem. However, it is not known how much, if any, of the cosmic web signal is also filtered out. Other filtering techniques such as spectral filtering could help to alleviate the problem as the Galactic emission around the EoR0 field has a spectral index of $-0.55$ \citep{Guzman11}, whereas we believe the cosmic web spectral dependence to be steeper \citep[with estimates of diffuse cluster emission having $\alpha\la-1.0$, e.g.][]{Feretti12}. Future work may provide further insights into the viability of spectral filtering techniques. 

We carried out a simple test to estimate the impact of spatial filtering (such as applied to the {\minur} image) on the cosmic web cross correlation signal. This test involved creating a model image of diffuse emission based on the galaxy number density maps added to a model of the Galactic emission and using this sky model to simulate a two minute MWA snapshot. The simulated data was then imaged with WSCLEAN with and without the minimum baseline filter.  We computed the the cross correlation functions of the output diffuse images (with and without the filtering) with the galaxy number density map, and compared them with the cross correlation functions of the input diffuse models with the galaxy number density map. In other words, does the output after adding in baseline information and filtering produce the same CCF as the input model?

We started by creating a model of diffuse cosmic web signal by convolving each galaxy number density redshift slice map by a $1.5\,$Mpc FWHM Gaussian function, multiplying by a scale factor of 500 (chosen to give flux density values on the order of mJy beam$^{-1}$) to put it into units of Jy pixel$^{-1}$ and summing all of the convolved images (WISE plus 2MASS) to include emission from across the full available redshift range; this is the diffuse cosmic web emission model. For the Galactic emission we computed the median for each pixel of the {\galr} image over a region of $2\degr$. We added the Galactic emission model to the diffuse cosmic web emission model to create a total sky model. We then applied an MWA primary beam to the total sky model and used \textsc{wsclean}'s {\it predict} feature to simulate {\it uv} values of the total sky model in the {\it uv} data model column for one $2\,$minute dataset. We imaged the data twice; once with no minimum baseline filter and again using a $34\,$m minimum baseline (with robust r=0.25 weighting). We corrected these images for the primary beam and cross correlated them with the summed galaxy number density map (WISE plus 2MASS). 

We compared the CCFs of the imaging output and the galaxy number density map with the CCFs of just the input diffuse model (the one used to generate the simulated MWA dataset) convolved with the two synthesized dirty beams and the galaxy number density map. This yielded CCFs of the diffuse model convolved with each beam (no imaging performed), and CCFs for the diffuse plus Galactic images made with and without the minimum baseline filter. If the filtering and Galaxy have no effect on the cross correlations then the CCFs from the input models (convolved with the corresponding dirty beams) should match those of the CCFs from the output images.

The results of the filter simulation test are shown in Fig.~\ref{fig:dfsim}. In this figure, we are not concerned with the strength of the signal for the filtered compared to the unfiltered case, but rather how the input model CCFs for each case compare to the output image CCFs for each case (comparing the solid lines to the dashed lines rather than the red lines compared to the blue lines). For the filtered case, the input and out CCFs match quite closely, wheres for the unfiltered case the output CCF is lower in amplitude than the input CCF. This shows the filtered image cross correlation actually matches the input model cross correlation better than the image with no filtering. The unfiltered Galactic emission is likely anti-correlated with the galaxy number density map at these shifts producing the decrease in amplitude. 

This test shows that, assuming the cosmic web signal of interest (that associated with the galaxies in the galaxy number density map) is correlated on scales smaller than roughly $3\degr$ (from the $34\,$m baseline filter), then the minimum baseline filtering should not affect the cross correlation result, while filtering the larger scale Galactic emission. The results would likely be different if we were interested in much lower redshift sources ($z\la0.04$), which would have Mpc emission on larger angular scales. This is only a simple test that assumes no point sources or instrumental noise, and a simple model for the cosmic web emission, however, it yields promising results that the smaller uncertainties in the {\minur} image are more accurate for the cosmic web signal. 

We can think of two other possible ways to minimize the effect of the Galaxy in addition to that of filtering. First would be to repeat the experiment using radio data at a higher frequency. However, this option suffers from the same issue mentioned in the previous subsection, that while the Galactic emission would be weaker at a higher frequency so would the cosmic web emission. 

The other possibility is that the random alignments of the galaxy number density maps and the Galactic emission may cancel out over a larger area, such as with a (nearly) all sky survey, e.g. the GaLactic and Extragalactic All-Sky MWA survey \citep[GLEAM,][]{Wayth15}. This is something we consider as a possible next step, but it would require re-imaging of the GLEAM data to subtract point sources and optimize for diffuse emission, which has not yet been performed. 

It is clear that Galactic emission hinders the ability to detect or constrain the cosmic web emission. Given our simple test it appears that filtering, either in the {\it uv} or image plane, helps to mitigate this problem, as long as the particular redshift range is high enough that the cosmic web emission is on scales smaller than the Galactic emission. 

\subsection{Limits on the diffuse cosmic web emission from the cross correlation}
\label{sec:limits}
\subsubsection{Detection thresholds}
\label{sec:detect}

Even though some of the CCFs in Fig.~\ref{fig:crxs1_2mass} and Fig.~\ref{fig:crxs1_wise} show signal above the $99.7\,$per cent confidence intervals, or $3\sigma$ detection thresholds, (such as the WISE {\minur} CCFs at $\Delta r\la 5\,$Mpc; see the redlines and pink shaded regions in Fig.~\ref{fig:crxs1_wise}), we cannot claim a detection of the cosmic web. This is because we know that the signals contain contributions from the correlation of the galaxy number number densities with unsubtracted point sources as well as Galactic emission, in addition to any cosmic web correlation signal. Thus, while there may be real signals detected at $>3\sigma$ confidence, we cannot say how much of the signal is due to the diffuse cosmic web. We can, however, use the information from the CCFs to provide upper limits on the diffuse cosmic web signal.  

There are no functions or models that predict what the cross correlation of the cosmic web with number densities should be based on physical parameters, such as magnetic field strength, synchrotron power, gas density, etc. There have been magnetohydrodynamic (MHD) simulations that predict the diffuse synchrotron emission and the corresponding autocorrelation functions \citep[e.g.][]{Vazza15,Vazza16}, however, these do not include galaxy catalogues to predict a cross correlation function. The results from the MHD simulations are also very dependent on the underlying models, i.e. the strength of the primordial magnetic field and how much emission is contributed from astrophysical sources at different redshifts. At this time there is not a prevailing model for the evolution of cosmic magnetism. Not knowing what we expect, or having a particular model to fit, complicates the derivation of limits on the cosmic web signal. 

With the lack of any preferred or physical model to use, we chose to use the simplest model for diffuse emission: the galaxy number density maps smoothed with a Gaussian function, and assume the emission scales linearly with the galaxy number density. In order to obtain limits on how strong the cosmic web signal would need to be to have a $3\sigma$ detection we performed the following procedure.
\begin{enumerate}[label={\arabic{enumi}.},leftmargin=*]
\item For each galaxy number density map $G$ we generate 2D Gaussian functions $D$ ($D$ for diffuse) with FWHM $\theta_{D}= 1, 2, 3,$ and $4.0\,$Mpc, computed using the corresponding $\langle z \rangle$ of $G$. The function $D$ is normalized to sum to one over the Gaussian function.
\item Convolve $D$ with $G$ to make a diffuse model of the cosmic web $G_{\rm D}$ from the galaxy number density map.
\item Convolve $G_{\rm D}$ with the two MWA beams to make beam convolved models of the diffuse emission $G_{\rm DB}$ ($DB$ for diffuse beam-convolved).
\item Cross correlate $G_{\rm DB}$ with $G$ and compute the 1D CCF$_{DB}$.
\item Find the scale factor ${\cal K}$ such that ${\cal K}\times$CCF$_{DB}$ yields a $3\sigma$ detection for the corresponding radio image and galaxy number density map CCF.
\end{enumerate}
We repeated steps 1--5 for all possible combinations of galaxy number density maps and radio images as described in Sec.~\ref{sec:results}. We chose four Gaussian smoothing sizes in range of 1 to $4\,$Mpc, which seemed like reasonable sizes to try based on sizes of previously measured diffuse radio cluster emission \citep[e.g.][]{Feretti12}.

There are multiple ways the $3\sigma$ detection limit could be defined. For example, t could be the $99.7\,$per cent confidence level at a particular $\Delta \theta, \Delta r$. However, rather than just choosing one position for the model to be greater than the detection limit we chose to use more of the available information. We define the $3\sigma$ detection as where $\sum_{r_{\rm min}}^{r_{\rm max}}{{\rm CCF}_{DB}(r)}$ is greater than the $99.7\,$per cent confidence level of $\sum_{r_{\rm min}}^{r_{\rm max}}{{\rm CCF}_{\rm rand}(r)}$, where ${\rm CCF}_{\rm rand}(r)$ are the CCFs from cross correlating with the random galaxy number density maps. We chose $r_{\rm min}=0.07\degr$ because $0.07\degr$ is the cutoff for the main lobe of the beam and, as previously stated, we expect the cosmic web correlation to be significant at scales larger than the beam. We chose $r_{\rm max}$ to be twice the FWHM ($2\theta_{D}$) for the particular diffuse model. Thus ${\cal K}$ is chosen such that the sum of the CCF signal of the diffuse model from the main beam lobe cutoff to twice the model FWHM is larger than the $99.7\,$per cent level expected from the sum over the same region from the random galaxy number density map correlations. 

\begin{figure*}
\centering
\includegraphics[scale=0.49]{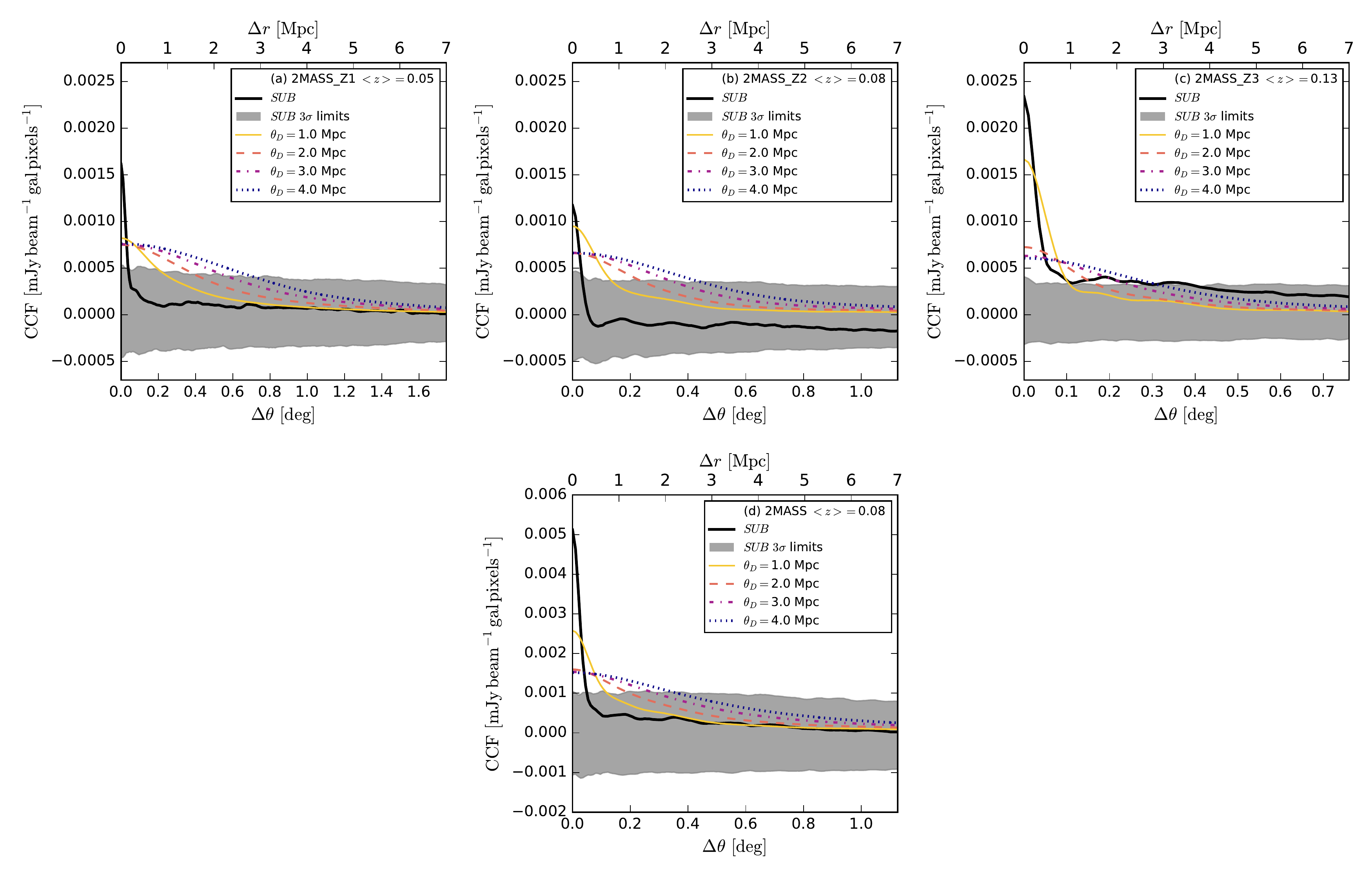}
\caption{As for Fig.~\ref{fig:crxs1_2mass}, but for the cross correlations of {\galr} images and 2MASS maps convolved with diffuse Gaussian functions of 1, 2, 3, and 4$\,$Mpc and dirty beams, scaled by ${\cal K}$ to yield a $3\sigma$ detection at $3\,$Mpc. From left to right top to bottom the plots show the CCF of the radio image with 2MASS\_Z1, 2MASS\_Z2, 2MASS\_Z3, and the sum map 2MASS. The solid black lines are from the {\galr} image CCFs. The grey shaded regions show the $99.7\,$per cent confidence intervals. The blue dotted lines show the diffuse model with $\theta_{D}=1.0\,$Mpc, the purple dot-dashed lines are the $\theta_{D}=2.0\,$Mpc models, the orange dashed lines are for $\theta_{D}=3.0\,$Mpc, and the yellow solid lines are for $\theta_{D}=4.0\,$Mpc.}
\label{fig:dfgal_2mass}
\end{figure*}

\begin{figure*}
\centering
\includegraphics[scale=0.49]{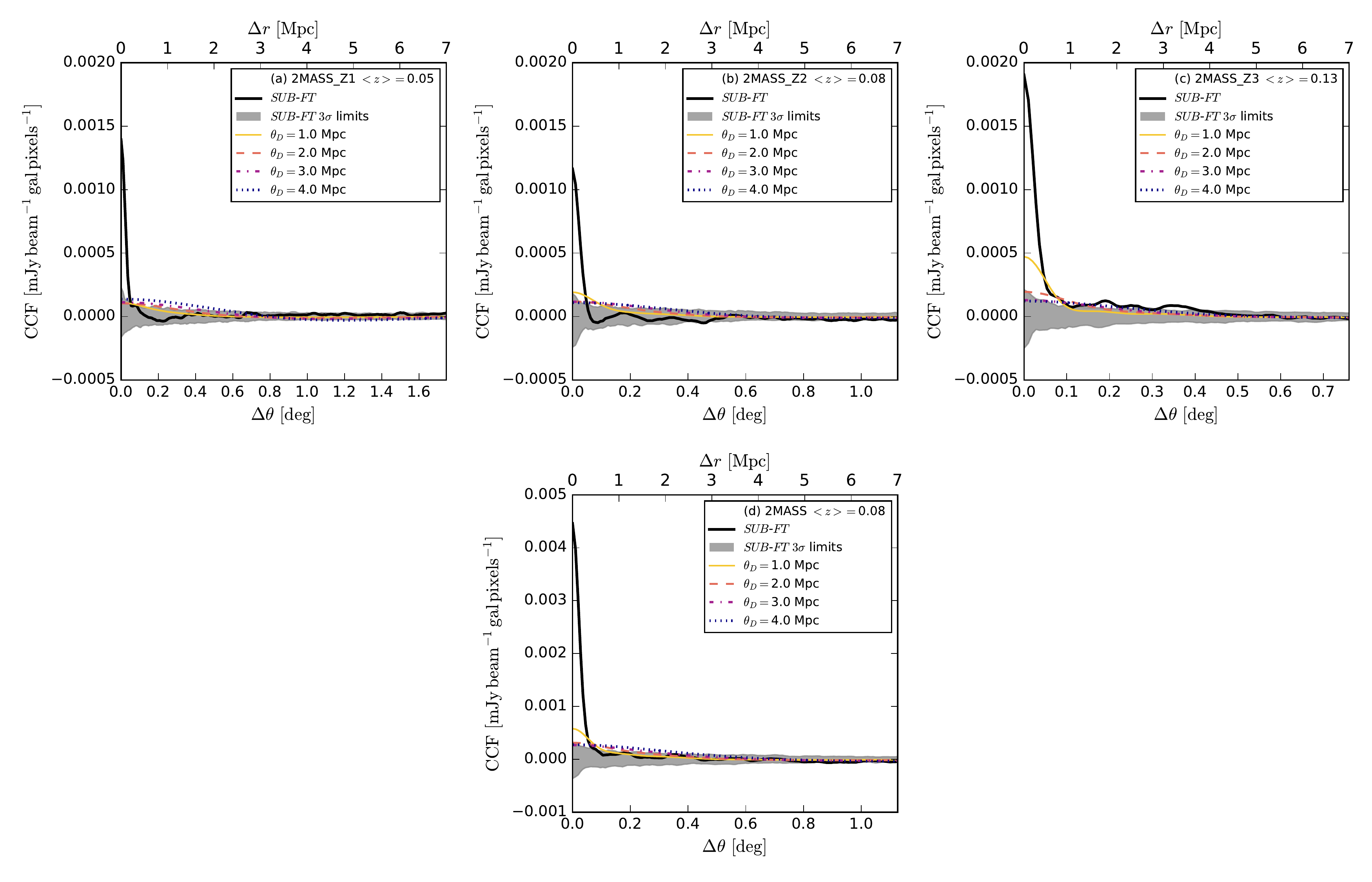}
\caption{As for Fig.~\ref{fig:dfgal_2mass}, but for the cross correlations of {\minur} images and 2MASS maps convolved with diffuse Gaussian functions of 1, 2, 3, and 4$\,$Mpc and dirty beams, scaled by ${\cal K}$ to yield a $3\sigma$ detection at $3\,$Mpc.}
\label{fig:dfmin_2mass}
\end{figure*}

\begin{figure*}
\centering
\includegraphics[scale=0.49]{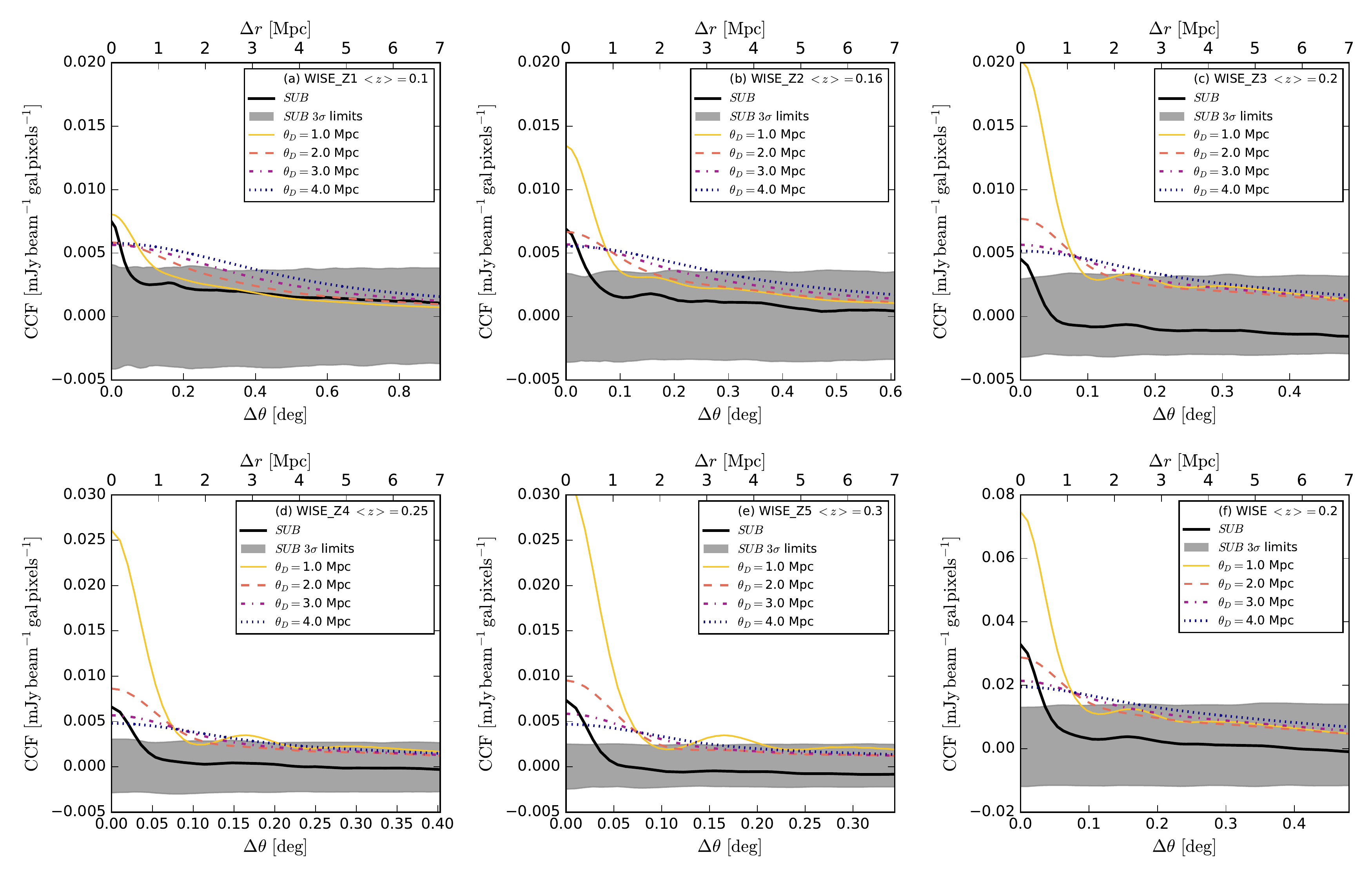}
\caption{As for Fig.~\ref{fig:dfgal_2mass}, but for the cross correlations of {\galr} images and WISE maps convolved with diffuse Gaussian functions of 1, 2, 3, and 4$\,$Mpc and dirty beams, scaled by ${\cal K}$ to yield a $3\sigma$ detection at $3\,$Mpc. From left to right top to bottom the plots show the CCF of the radio image with WISE\_Z1, WISE\_Z2, WISE\_Z3, WISE\_Z4, WISE\_Z5 and the sum map WISE.}
\label{fig:dfgal_wise}
\end{figure*}

\begin{figure*}
\centering
\includegraphics[scale=0.49]{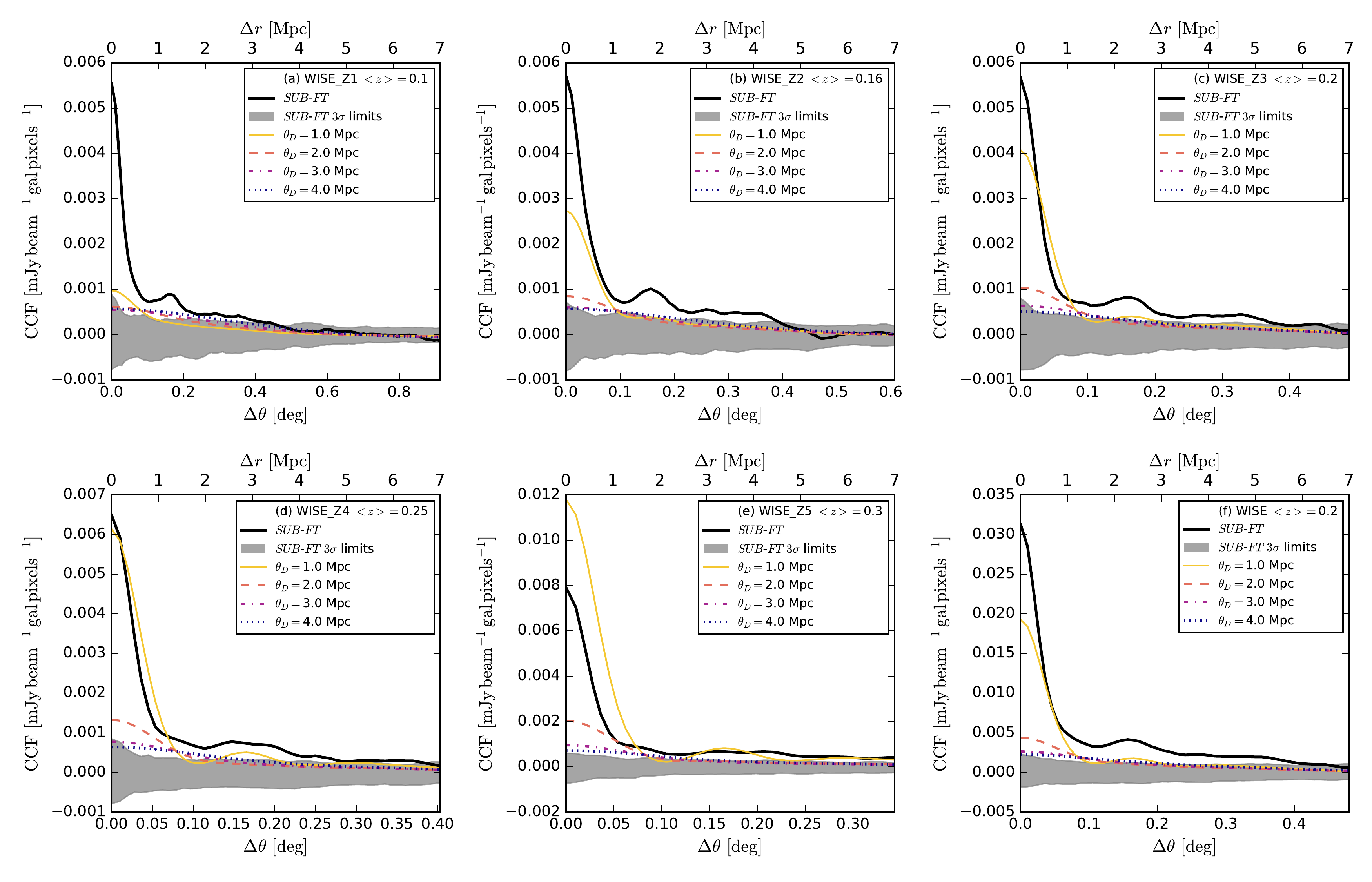}
\caption{As for Fig.~\ref{fig:dfgal_wise}, but for the cross correlations of {\minur} images and WISE maps convolved with diffuse Gaussian functions of 1, 2, 3, and 4$\,$Mpc and dirty beams, scaled by ${\cal K}$ to yield a $3\sigma$ detection at $3\,$Mpc. }
\label{fig:dfmin_wise}
\end{figure*}

The values for ${\cal K}$ and the ${\cal K} \times$ the standard deviations of the $G_{DB}$ maps $\sigma_{G_{DB}}$ are given in Table~\ref{tab:minur_dflims}. The upper limits for ${\cal K}\sigma_{G_{DB}}$ all yield values between roughly 0.09--$2.20\,$mJy beam$^{-1}$. This corresponds to non-beam convolved values, ${\cal K}\sigma_{G_{DF}}$, of $0.01$--$0.30\,$mJy arcmin$^{-2}$. The scaled diffuse model CCFs, with the real image CCFs, for the 2MASS galaxy number density maps are shown in Fig.~\ref{fig:dfgal_2mass} and Fig.~\ref{fig:dfmin_2mass}, for the {\galr} and {\minur} radio images respectively. The WISE galaxy number density map CCFs with the scaled diffuse models are shown in  Fig.~\ref{fig:dfgal_wise} and Fig.~\ref{fig:dfmin_wise}.

\begin{table*}
\centering
\scriptsize
\caption{Cosmic web flux density and magnetic field upper limits from the cross correlation results in Sec.~\ref{sec:results}, models of diffuse emission in Sec.~\ref{sec:detect}, and equation for the magnetic field strength from Sec.~\ref{sec:physical}. Here ${\cal K}$ is the scale factor applied to the diffuse sky models and ${ \cal K} \sigma_{G_{DB}}$ is the scaled rms of the diffuse sky models. The magnetic field strength values $B_{\rm eq0}$ are the upper limits when the ratio of number densities of cosmic ray protons and electrons $K_0=100$, the volume filling factor $\eta=1$, and the spectral index $\alpha=-1.25$. The magnetic field strength values $B_{\rm HA0}$ are the upper limits with a spectral index $\alpha=-1.25$ and electron acceleration efficiency $\xi=5\times10^{-3}$.  }
\label{tab:minur_dflims}
\begin{tabular}{llcccccccccccccccc}
\hline
Radio & Density& \multicolumn{4}{c}{${\cal K}$} & \multicolumn{4}{c}{${\cal K} \sigma_{G_{DB}}$}& \multicolumn{4}{c}{$B_{\rm eq0}$}& \multicolumn{4}{c}{$B_{\rm HA0}$}\\

image&map &  \multicolumn{4}{c}{ [mJy galaxy$^{-1}$]}&  \multicolumn{4}{c}{[ mJy beam$^{-1}$]}&\multicolumn{4}{c}{[$\mu$G]}&\multicolumn{4}{c}{[$\mu$G]}\\

 \hline
\normalsize

& & \multicolumn{4}{c}{FWHM} &\multicolumn{4}{c}{FWHM}&\multicolumn{4}{c}{FWHM}&\multicolumn{4}{c}{FWHM}\\
 & & \multicolumn{4}{c}{[Mpc]} &\multicolumn{4}{c}{[Mpc]}&\multicolumn{4}{c}{[Mpc]}&\multicolumn{4}{c}{[Mpc]}\\
 & & 1 & 2 & 3 & 4 & 1 & 2 & 3 & 4 & 1 & 2 & 3 & 4 & 1 & 2 & 3 & 4 \\

\hline
{\galr} & 2MASS\_Z1 & 14.0 & 21.9 & 30.7 & 38.2 & 0.69 & 0.87 & 1.04 & 1.16 & 0.33 & 0.29 & 0.28 & 0.27 & 0.23 & 0.26 & 0.28 & 0.30 \\
{\galr} & 2MASS\_Z2 & 16.1 & 25.3 & 38.1 & 51.2 & 0.66 & 0.82 & 1.04 & 1.22 & 0.33 & 0.3 & 0.29 & 0.28 & 0.22 & 0.25 & 0.28 & 0.30 \\
{\galr} & 2MASS\_Z3 & 13.2 & 15.9 & 22.3 & 28.9 & 0.70 & 0.61 & 0.73 & 0.84 & 0.35 & 0.29 & 0.27 & 0.26 & 0.23 & 0.21 & 0.23 & 0.25 \\
{\galr} & 2MASS & 9.7 & 15.0 & 21.2 & 27.0 & 0.86 & 1.08 & 1.31 & 1.48 & 0.35 & 0.32 & 0.3 & 0.29 & 0.25 & 0.29 & 0.32 & 0.33 \\
{\galr} & WISE\_Z1 & 3.9 & 5.0 & 6.3 & 7.8 & 1.03 & 1.15 & 1.31 & 1.48 & 0.38 & 0.33 & 0.31 & 0.30 & 0.28 & 0.29 & 0.31 & 0.33 \\
{\galr} & WISE\_Z2 & 4.4 & 4.8 & 5.7 & 6.9 & 1.28 & 1.09 & 1.19 & 1.33 & 0.42 & 0.34 & 0.32 & 0.30 & 0.31 & 0.29 & 0.30 & 0.32 \\
{\galr} & WISE\_Z3 & 5.5 & 4.8 & 5.4 & 6.3 & 1.69 & 1.09 & 1.09 & 1.19 & 0.46 & 0.35 & 0.32 & 0.31 & 0.36 & 0.29 & 0.29 & 0.30 \\
{\galr} & WISE\_Z4 & 5.9 & 4.4 & 4.8 & 5.4 & 1.95 & 1.03 & 0.97 & 1.02 & 0.49 & 0.36 & 0.32 & 0.31 & 0.38 & 0.28 & 0.27 & 0.28 \\
{\galr} & WISE\_Z5 & 6.2 & 4.0 & 4.2 & 4.7 & 2.19 & 0.99 & 0.87 & 0.9 & 0.53 & 0.37 & 0.33 & 0.31 & 0.41 & 0.27 & 0.26 & 0.26 \\
{\galr} & WISE & 1.6 & 1.6 & 1.9 & 2.2 & 1.50 & 1.25 & 1.32 & 1.44 & 0.45 & 0.36 & 0.34 & 0.32 & 0.34 & 0.31 & 0.32 & 0.33 \\

{\minur} & 2MASS\_Z1 & 2.9 & 6.1 & 11.5 & 21.7 & 0.09 & 0.13 & 0.18 & 0.26 & 0.21 & 0.19 & 0.19 & 0.19 & 0.09 & 0.11 & 0.12 & 0.15 \\
{\minur} & 2MASS\_Z2 & 4.1 & 6.7 & 11.3 & 16.3 & 0.13 & 0.14 & 0.19 & 0.21 & 0.23 & 0.20 & 0.20 & 0.19 & 0.10 & 0.11 & 0.13 & 0.13 \\
{\minur} & 2MASS\_Z3 & 4.4 & 5.8 & 6.9 & 9.6 & 0.19 & 0.16 & 0.15 & 0.18 & 0.26 & 0.21 & 0.19 & 0.19 & 0.13 & 0.11 & 0.11 & 0.12 \\
{\minur} & 2MASS & 1.9 & 3.1 & 5.3 & 7.8 & 0.12 & 0.14 & 0.19 & 0.22 & 0.23 & 0.20 & 0.20 & 0.19 & 0.10 & 0.11 & 0.13 & 0.14 \\
{\minur} & WISE\_Z1 & 0.6 & 0.9 & 1.2 & 1.7 & 0.12 & 0.13 & 0.14 & 0.17 & 0.23 & 0.20 & 0.19 & 0.18 & 0.10 & 0.10 & 0.11 & 0.12 \\
{\minur} & WISE\_Z2 & 1.1 & 0.9 & 1.0 & 1.3 & 0.24 & 0.14 & 0.13 & 0.15 & 0.29 & 0.21 & 0.19 & 0.19 & 0.14 & 0.11 & 0.11 & 0.11 \\
{\minur} & WISE\_Z3 & 1.3 & 0.9 & 0.9 & 1.0 & 0.32 & 0.14 & 0.12 & 0.12 & 0.32 & 0.22 & 0.2 & 0.18 & 0.16 & 0.11 & 0.10 & 0.10 \\
{\minur} & WISE\_Z4 & 1.6 & 0.9 & 0.9 & 1.1 & 0.43 & 0.15 & 0.13 & 0.14 & 0.36 & 0.23 & 0.21 & 0.20 & 0.19 & 0.11 & 0.10 & 0.11 \\
{\minur} & WISE\_Z5 & 2.6 & 1.0 & 0.9 & 1.0 & 0.77 & 0.2 & 0.13 & 0.13 & 0.42 & 0.26 & 0.22 & 0.20 & 0.25 & 0.13 & 0.11 & 0.11 \\
{\minur} & WISE & 0.3 & 0.3 & 0.3 & 0.3 & 0.21 & 0.14 & 0.13 & 0.14 & 0.29 & 0.22 & 0.20 & 0.19 & 0.13 & 0.11 & 0.10 & 0.11 \\
\end{tabular}
\end{table*}

\subsubsection{Physical interpretation}
\label{sec:physical}

The upper limits found in the previous subsection tell us about the potential brightness of the cosmic web, as well as informing us about the image requirements necessary for a detection. From those limits we can infer information about cosmic magnetism by working out how those brightnesses translate into magnetic field strengths.

Magnetic fields in clusters and filaments depend on the number density of the cosmic ray electrons, and without observational data (e.g. from X-ray emission by inverse Compton scattering) an assumption about the relation between cosmic ray electrons and magnetic fields has to be made. Two common approaches are the minimum total energy density ($\epsilon_{\rm tot}=\epsilon_{\rm CR}+\epsilon_{\rm B}=$min) and equipartition between the total energy densities of the cosmic rays and the magnetic field ($\epsilon_{\rm CR}=\epsilon_{\rm B}$). These two assumptions generally give similar results and are often used interchangeably. 

The equations for deriving the minimum energy or equipartition magnetic fields from radio observations are worked out in detail by \citet{Pacholczyk70} and \citet{Miley80}. Here, however, we use the revised versions from \citet{Beck05}. For a region in a synchrotron radio source the equipartition magnetic field (in Gauss) is
\small
\begin{equation}
B_{\rm eq}=\left [ \frac{4\pi (1-2\alpha)(K_0+1)E_{\rm p}^{1+2\alpha}(\nu/2c_1)^{-\alpha} I_{\nu} (1+z)^{3-\alpha}}{(-2\alpha-1)\,c_2(\alpha)\, l \, \eta \, c_4(i)}\right ]^{1/(3-\alpha)}.
\label{eq:bme}
\end{equation}
\normalsize
Here $K_0$ is the ratio of number densities of cosmic ray protons and electrons per particle energy interval within the energy range traced by the synchrotron emission (rather than the commonly used ratio of energy in the protons to that in the electrons). The volume filling factor of the emitting region(s) is $\eta$, $z$ is the (median) redshift, $l$ is the path length through the source in the line of sight which we choose to be $\theta_{D}$, $\alpha$ is the spectral index ($S(\nu) \propto \nu^{\alpha}$), and $E_{\rm p}$ is the proton rest energy. The synchrotron intensity of the region at the frequency $\nu$ is $I_{\nu}$, the flux density $S$ converted from Jy beam$^{-1}$ to ${\rm erg} \, {\rm s}^{-1} \,  {\rm cm}^{-2} \, {\rm Hz}^{-1} \,{\rm sr}^{-1}$.
The constants $c_1$, $c_2$, and $c_4$ are described in Appendix A of \citet{Beck05}. 

The values of $l$ are set to the different $\theta_{D}$ of 1, 2, 3, and 4.0$\,$Mpc. The values of $\eta$, $K_0$, and $\alpha$ are not well known. The volume filling factor can range $0 < \eta \le 1$, with the value generally assumed in the literature (for clusters) to be 1 \citep[e.g][]{Govoni04}. The value of $K_0$ depends on the mechanism of generation of relativistic electrons. From \citet{Beck05} some typical values for different injection mechanisms are: Fermi shock acceleration (strong shocks, non-relativistic gas) = $40$ -- $100$; Secondary electrons = $100$ -- $300$; Turbulence  $\simeq$ 100; and Pair plasma = 0. Typically $\alpha$ values for shocks tend to be steeper, with $\alpha\la -1.0$. \citet{Feretti12} found average spectral indices from observations of diffuse cluster haloes and relics of $-1.45$ for haloes, $-1.3$ for elongated relics, and $-2.0$ for rounder relics. 
 
We used a range of values for $K_0$, $\eta$, and $\alpha$ such that $1.0\le K_0 \le 300$, $0.01\le \eta \le 1$, and $-2.25 \le \alpha \le -0.6$ to compute limits on $B_{\rm eq}$ with the limits on ${\cal K}\sigma_{G_{DB}}$ presented in Sec.~\ref{sec:detect} as $S_{\nu}$. The resulting magnetic field limits are in the range $0.03 \le B_{\rm eq} \, [\mu{\rm G}] \le 1.98$. 

If we take the case where $\eta=1.0$, $K_0=100$, and $\alpha=-1.25$ (which we call $B_{\rm eq 0}$) then $0.18 \le B_{\rm eq0} \, [\mu{\rm G}] \le 0.52$, depending on the radio image, galaxy number density map, and $\theta_{D}$ of the diffuse model. The values for $B_{\rm eq0}$ for each radio image and galaxy number density map are given in Table~\ref{tab:minur_dflims}. 

It should be noted that the equipartition argument may not be valid for the case of filaments (or the WHIM in general), as the electrons there should be produced by shocks (so relatively fresh injection from diffusive shock acceleration), where no equipartition argument applies (Vazza, private communication). \citet{Bruggen12} and \citet{Wong09} showed that for cluster emission the difference from equipartition within the virial radius was less than $1\,$per cent and $20\,$per cent at 1.4 the virial radius. Beyond 1.4 the virial radius the non-equipartition effect depends strongly on the non-adiabatic electron heating efficiency. 

Given that the validity of equipartition in this case may be in question we also compute magnetic field strength values from a model derived from the acceleration of relativistic particles at shocks, given in \citet{Vazza15}. The model is a high amplification model denoted in \citet{Vazza15} as the HA model. From eq.~(1) of \citet{Vazza15} the radio power $I$ at frequency $\nu$ is given by
\begin{equation}
I_{\nu}\propto S \, n_d \,  \nu^{-\delta/2} \,  \xi (M,T) \, T_d^{3/2} \frac{B^{1+\delta/2}}{B_{\rm CMB}^2+B^2},
\label{eq:vaz1}
\end{equation}
where $S$ is the shock surface area, $n_d$ is the downstream electron density, $\xi (M, T)$ is the electron acceleration efficiency, $\delta=2 \alpha$, $T_d$ is the downstream electron temperature, and $B_{\rm CMB}$ is the cosmic microwave background magnetic field. Averaging over all redshifts, assuming a mean gas over-density of $n/n_{cr}\simeq 10$, and rearranging eq.~(\ref{eq:vaz1}) can be simplified to eq.~(3) of \citet{Vazza15} giving the mean magnetic field of the WHIM,
\begin{equation}
B_{\rm HA}\simeq 0.05 \, \mu \rm{G} \sqrt{\frac{I_{\rm WHIM}}{5\times10^{-3} \, \rm{Jy} \, \rm{deg}^{-2} \, \left ( \frac{100 \, \rm{MHz}}{\nu} \right )^{\alpha} \left ( \frac{\xi}{10^{-3}} \right ) }}.
\label{eq:bha}
\end{equation}
We use this relation to obtain new magnetic field upper limits by converting the Jy deg$^{-2}$ factor to Jy beam$^{-1}$ and using the ${\cal K} \sigma_{G_{DB}}$ flux density limits in place of $I_{\rm WHIM}$. If we use the same range of values for $\alpha$ as used in the equipartition case and use a range of values for $\xi$ of $5\times 10^{-5} \le \xi \le 0.025$ then we obtain values for $B_{\rm HA}$ in the range $0.03 \le B_{\rm HA} \, [\mu\rm{G}] \le 5.86$. If we again set $\alpha=-1.25$ and $\xi=5\times 10^{-3}$ we obtain $0.09 \le B_{\rm HA 0} \, [\mu\rm{G}] \le 0.41$, which are consistent with the values obtained for $B_{\rm eq0}$. The values for $B_{\rm HA0}$ are also listed in Table~\ref{tab:minur_dflims}.

\begin{figure}
\includegraphics[scale=0.37]{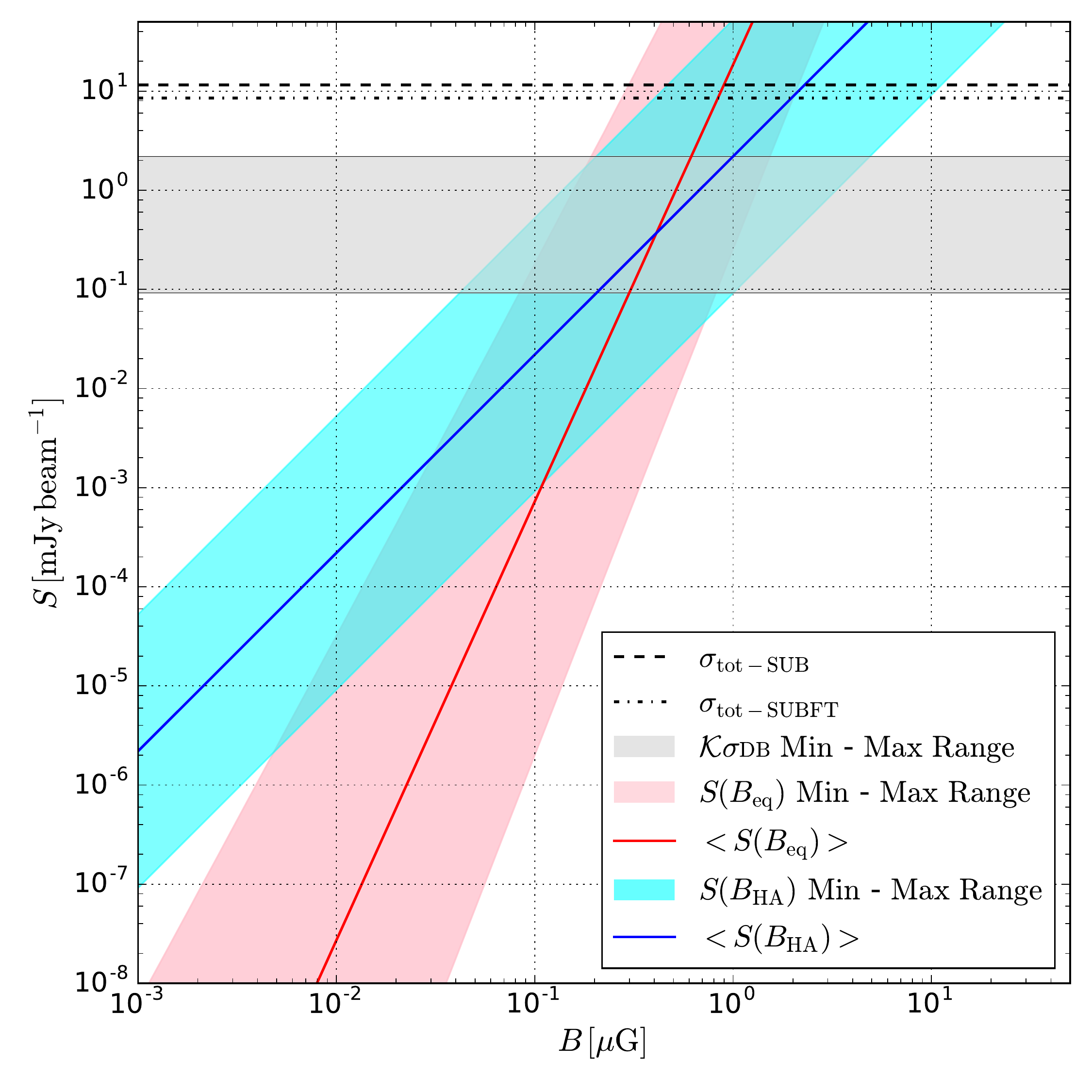}
\caption{Surface brightness values $S$ for given equipartition magnetic field strengths $B_{\rm eq}$ and HA magnetic field strengths $B_{\rm HA}$. The values of $S$ for equipartition were computed using the inversion of eq.~(\ref{eq:bme}) and the values of $S$ for the HA magnetic field strengths were computed using the inversion of eq.~(\ref{eq:bha}). The dashed black line shows the rms standard deviation of the {\galr} radio image, while the dot-dashed is the rms for the {\minur} image. The grey shaded region shows the minimum maximum range of ${\cal K}\sigma_{DB}$ values, given in Table~\ref{tab:minur_dflims}. The red solid line is the mean from all calculated values of $S(B_{\rm eq})$ and the pink shaded region shows the minimum and maximum range, while the blue solid line is the mean from all calculated values of $S(B_{\rm HA})$ and the light blue shaded region shows the minimum and maximum range.}
\label{fig:beqseq}
\end{figure}

While these limits on their own are not enough to allow us to discriminate between any competing magnetism models, they do allow us to investigate future observational requirements for detection. In order to see how sensitive an image would need to be for different $B$ values, we can invert eq.~(\ref{eq:bme}) and eq.~(\ref{eq:bha}) to find the different flux density values for a given magnetic field strength. We computed these flux densities using the same ranges for $K_0$, $\eta$, $\xi$, and $\alpha$ as above and magnetic field strength values of $1\times 10^{-9}\le B_{\rm eq} \, [{\rm G}] \, \le 1\times 10^{-5}$, using all of the $\theta_{D}$ values for $l$, both sets of beam sizes for the two radio images, and the $\langle z \rangle$ from the different galaxy number density maps. Figure~\ref{fig:beqseq} shows the minimum and maximum flux density range as well as the mean flux density for a given $B_{\rm eq}$ or $B_{\rm HA}$ from all of the values. Also shown in Fig.~\ref{fig:beqseq} is the minimum and maximum range of the ${\cal K}\sigma_{DB}$ values, as well as the radio image rms values of $11.5\,$mJy beam$^{-1}$ for the {\galr} radio image and $8.7\,$mJy beam$^{-1}$ for {\minur}.

The cross correlation with LSS tracers has allowed limits on the cosmic web flux density (${\cal K}\sigma_{DB}$) one to two orders of magnitude below the image rms values, with resulting $B_{\rm eq}$ upper limits in the range of $0.03 \le B_{\rm eq} \, [\mu{\rm G}] \le 1.98$ and $B_{\rm HA}$ upper limits in the range of $0.03 \le B_{\rm HA} \, [\mu{\rm G}] \le 5.86$. Values of $B_{\rm eq}\simeq 0.1 \, \mu$G result in values of $1\times10^{-6}\le S \, [{\rm mJy} \, {\rm beam}^{-1}] \le 0.1$, which would require image rms values of $1\times10^{-4}\le S \, [{\rm mJy} \, {\rm beam}^{-1}] \le 0.5$. However, $B_{\rm HA}\simeq 0.1\, \mu$G results in values of $9\times10^{-4}\le S \, [{\rm mJy} \, {\rm beam}^{-1}] \le 0.5$, requiring image rms values of $0.08\le S \, [{\rm mJy} \, {\rm beam}^{-1}] \le 2.0$.  From Fig.~\ref{fig:beqseq} we can see that if the cosmic web has a magnetic field strength in the $1$--$10\,$nG range (or less) then detection via this method (even with the SKA) may not be possible; with the predictions being slightly more optimistic in the HA case. The slope of the $S(B)$ line is most dependent on the value of $\alpha$ and the underlying model for the magnetic field. It is clear that in order to extract more physical information from this method, data (and corresponding CCFs) should be used at multiple frequencies in order to estimate the spectral index.

In the following subsection we see how these limits compare to previous estimates and measurements.

\subsection{Comparison of magnetic field limits with previous estimates}
\label{sec:dis_compare}
To date the only other synchrotron cosmic web constraints from cross correlating come from \citet{Brown10}. That work cross correlated a $34\degr \times 34\degr$  image from the $1.4\,$GHz Bonn survey with a 2MASS galaxy number density map. The reported upper limits on the filament magnetic field are in the range $0.20\eta^{-2/7}$ to $0.74\eta^{-2/7}$ $\mu$G, with $\eta$ being the volume filling factor. If $\eta=1$, upper limits of the magnetic field strength are in the range $0.2 \le B_{\rm eq0} \, [\mu{\rm G}] \le 0.74$, similar to what we report in the previous section. However,  \citet{Brown10} did not use the revised magnetic field strength equation of \citet{Beck05}, which would change the reported limits. The difference between the revised and classical values of $B$ depend on the values of $B$ and $\alpha$, with the revised values being nearly the twice as large as the classical values for $\alpha \simeq -0.5$ and the classical values larger than the revised the values for steeper spectral indices (roughly twice as large for $\alpha =-1.5$, $B=50\, \mu$G, and $K_0=100$).  

There have been observations of diffuse radio emission in clusters in the form of giant and mini radio haloes and radio relics from which magnetic field strengths have been derived. At the centres of dense cool-core clusters magnetic fields have been measured as high as 10--30$\, \mu$G \citep{Kuchar11,Laing08}. Lower density clusters indicate lower central magnetic fields strengths in the range 3--10$\, \mu$G \citep{Feretti99b,Kuchar11,Guidetti10}. \citet{Feretti99a} found magnetic field strengths in cluster halos range from 0.1 to $1\, \mu$G. \citet{Giovannini93} report a minimum energy magnetic field in the Coma cluster (Abell 1656) of $0.4 \, \mu$G, whereas \citet{Brunetti01a}, using a different method, found a smoothly varying field with $2\pm1 \, \mu$G in the cluster centre to $0.3\pm 0.1 \, \mu$G at a distance of $1\,$Mpc. These measurements all come out higher than our estimated magnetic field upper limits, which is expected with these observed clusters having some of the brightest diffuse radio objects. 


One of the only reported detections of a filament was by \citet{Bagchi02}, which reported the detection of a filamentary network in the region of cluster ZwCl 2341.1+0000 stretching over an area of at least $6h^{-1}_{50}\,$Mpc in diameter. Using the synchrotron emission and the minimum energy requirement for the magnetic field they report a magnetic field strength in the range of $0.3$--$1.5\, \mu$G, depending on the assumed filling factor. \citet{Kim89} reported the detection of supercluster-scale radio emission extending between the Coma cluster of galaxies and the Abell 1367 cluster. The estimated strength of the intercluster magnetic field was $0.3\,$--$\,0.6 \, \mu$G. 

Magnetohydrodynamic simulations can also provide estimates for the filamentary magnetic field strength. Simulations by \citet{Ryu08} showed that strong turbulence driven by shocks associated with the WHIM can generate (volume averaged) filament magnetic fields of approximately $0.01\, \mu$G. \citet{Donnert09} only find a filament magnetic field strength of roughly $0.005\, \mu$G. Simulations in \citet{Vazza14a} showed that amplification of primordial magnetic fields via small-scale turbulent dynamos could produce magnetic fields in cluster cores of $0.4\, \mu$G, but only $0.01\, \mu$G in filaments. These estimates are all highly dependent on the primordial magnetic field strength and model for the magnetic field evolution, but are consistent with our upper limits. 

\subsection{Limitations of diffuse emission models}
\label{sec:dis_mod}
The diffuse models used in this paper are simple Gaussian functions convolved with the galaxy number density maps and scaled. They are not physically based, except for the sizes being several $\,$Mpc. The actual case is likely much more complicated. These models assume the radio emission scales linearly with the number density, which may or may not be true, or may be true to some degree. Another option, used in Brown et al. (in preparation, with shallower, wider area, higher frequency data), is to cross correlate the radio emission with results of MHD simulations that have been constrained by observations of the local universe \citep[e.g.][]{Dolag04,Donnert09}. This has the advantage of not dealing with issues related to the number density, or point source, correlation as well as being able to constrain the actual physical parameters of the input magnetism model. However, the disadvantage is the results are dependent on, and potentially only valid for, the specific model of cosmic magnetism used in creating the MHD simulation. 

The MHD simulations presented by \citet{Vazza15} predict radio flux densities in redshift bins and summed along the line of sight. Looking at the autocorrelation function of the diffuse radio emission in these simulations shows multiple peaks at varying distances, depending on redshifts.\footnote{The \citet{Vazza15} simulation results are available to download as images from \url{http://cosmosimfrazza.myfreesites.net/radio-web}.} Implementing a diffuse model with multiple peaks in the autocorrelation in this paper does not seem reasonable with so many free parameters (i.e. the number, size, positions, and amplitudes of the peaks) and so few constraints. 

Additionally, the problem with using previous simulation results \citep[such as][]{Ryu08,Dolag08a,Vazza15} is that they do not tell us how the predicted diffuse emission correlates with galaxy number densities. Ideally, to supplement future similar tests using the cross correlation method, new MHD simulations would be performed that predict the radio flux density of the cosmic web as well as providing a corresponding source catalogue. This would allow us to predict what we expect from the cross correlation given a particular physical model of cosmic (magnetic) evolution and compare that to results with real data. 

Another possible route to use in looking at how to model this signal is to use reconstructions of the dark matter density fields as shown in \citet{Jasche13,Jasche15,Lavaux16}. These works provide models of the underlying dark matter density fields and power spectra as traced by galaxies, and should be traced by the synchrotron cosmic web. It would be a matter of combining these models with different models for cosmic magnetism. For further discussion on the qualitative benefits of different approaches see \citet{Brown11b}.

\subsection{Correlation with X-ray emission}
\label{sec:dis_xr}
These results are based on the assumption that the optical or NIR galaxy number density, or mass density, is a good tracer of the synchrotron diffuse emission, which may not be the case. \citet{Kronberg07} look at the diffuse radio emission in the Coma cluster and found it did not spatially  correlate well with the optical galaxy positions, and suggest that X-ray emission may instead prove a better tracer of the synchrotron cosmic web. 

Cosmic magnetic fields can be derived by comparing inverse Compton (IC) X-ray emission and radio synchrotron \citep{Harris79,Rephaeli87}. The electrons that generate the synchrotron emission can inverse-Compton scatter off of CMB photons to generate X-ray emission. 
The synchrotron and IC power are related by,
\begin{equation}
\frac{L_{\rm IC}}{L_{\rm sync}}=\frac{U_{0}(1+z)^4}{U_{\rm B}}
\label{eq:sync_ic}
\end{equation}
\citep[see e.g.][]{Nath10}. Here $L_{\rm sync}$ and $L_{\rm IC}$ are the synchrotron and IC luminosities, $U_{\rm B}$ is the magnetic field energy density, and $U_{0}=4.2\times 10^{-14}\,$J m$^{-3}$ is the CMB energy density at $z=0$. The Planck function peaks near a frequency of $\nu \simeq 1.6 \times 10^{11}\,$Hz. Inverse Compton X-rays observed at $20\,$keV ($\nu_{\rm IC}=4.8\times 10^{18}\,$Hz) are emitted predominately by electrons with $\gamma \sim 5000$ (with $\gamma$ being the electron Lorentz factor), regardless of redshift. The corresponding synchrotron emission would peak at a rest frame frequency of $\nu_{\rm sync}\sim4.2(B/1 \mu {\rm G})\gamma^2\,$Hz, or $100\,$MHz \citep{Bagchi98}. Given that the radio synchrotron emission and IC X-rays originate from the same relativistic electron population, they should have the same spectral index (assuming a power-law population). If we assume $\alpha=-1$ the magnetic field is 
\begin{equation}
B=1.7(1+z)^2\left ( \frac{S_r \nu_r}{S_x \nu_x} \right )^{0.5} \, \, \mu{\rm G},
\label{eq:icmag}
\end{equation}
where $S_x$ and $S_r$ are the X-ray and radio flux densities at the observed frequencies $\nu_x$ and $\nu_r$. 

The problem with this approach is the sensitivity, resolution, and survey coverage of existing X-ray data and catalogues. There are {\it Chandra} \citep{Evans10}, {\it XMM} \citep{Watson09}, and {\it ROSAT} \citep{Voges99} data and catalogues available in the EoR0 field. {\it Chandra} and {\it XMM} have high resolution and good sensitivity, however, with the limited field of views, the X-ray images cover much smaller areas than the radio data. Thus, while there are existing data in the EoR0 field from these two telescopes, the coverage is incomplete and uneven. {\it ROSAT}, which does have all sky coverage, has much lower resolution and sensitivity. However, the followup to ROSAT is the all sky survey eRASS of the eROSITA telescope \citep{Predehl10}, which should have approximately 30 times better sensitivity. There are also data from the {\it Fermi} telescope over the whole sky, but the sensitivity and resolution are limiting factors with {\it Fermi} as well. This may be motivation for targeted sensitive X-ray and radio observations over a large (few degrees) field.  

\subsection{Other statistical tests}
\label{sec:dis_test}
In this work we have focused on the image plane cross correlation. There are other similar tests that could be performed which may have better results or provide additional constraints \citep{Brown11b}. In \citet{Planck14xix} three separate tests were used to measure the integrated Sachs-Wolfe (ISW) effect. These were the Fourier plane cross power, the image plane cross correlation, and the wavelet space spherical wavelet covariance. While these tests are similar, each has its own advantages and disadvantages and can reveal different or additional information. 

The cross power, or cross angular power spectrum (CAPS), has the advantage of being in the {\it uv} plane, corresponding to the interferometer visibilities. Computing the CAPS removes the need to go to the image plane with the radio data, alleviating effects from image artefacts or gridding and weighting.  Rather than giving information on the correlation as a function of distance, the CAPS gives information on the correlation as a function of angular scale (or spherical harmonic mode $\ell$). In terms of the current data, this may be an easy next step as the EoR0 field has already been analyzed in the {\it uv} plane as part of the MWA EoR experiment, with careful work already done including error bars on every mode, which includes the complex angular modes prior to squaring with error bars on each, the removal of noise bias, and verified calibration \citep[see e.g.][for details]{Trott16,Thyagarajan15a,Thyagarajan15b,Dillon15}. 

Wavelets are ideal kernels to enhance features with a characteristic size, since the wavelet analysis at an appropriate scale $R$ amplifies those features over the background. Wavelets could recover most of the signal-to-noise of the cosmic web signal by just analyzing the appropriate narrow range of scales. The basic idea of this approach is to estimate the covariance of the wavelet coefficients ($W_{\rm cov}$) as a function of the wavelet scale \citep[see e.g.,][for details]{Vielva06}. It has never been tried for the cosmic web, but may present an interesting new approach.

Another test that could be used, either separately or in conjunction with cross correlations, is a stacking analysis. Stacking, or averaging, images at the locations of filaments and/or cluster peripheries would boost the signal-to-noise of the diffuse emission to faint to be directly detected \citep{Brown11b}. However, it is not obvious what images to stack, or how to choose where to stack. The optical and NIR distribution of galaxies can be used to trace large-scale structure and to identify clustered or filamentary regions. However, if the radio emission is not spatially matched up between different images the signal will not add perfectly coherently. 

Each of these methods could provide constraints on the synchrotron cosmic web, and corresponding magnetic fields. In conjunction with each other, they would provide information and constraints on the size, characteristic scale size of the cosmic web, how spatially correlated the cosmic web emission is, and brightness of the cosmic web synchrotron emission as a function of size, distance, and redshift. 

\section{Conclusions}
\label{sec:conclusions}
We have presented flux density and magnetic field strength upper limits on the synchrotron cosmic web from cross correlating radio images of the Murchison Widefield Array EoR0 field with galaxy number density maps from the 2MASS and WISE redshift catalogues. The radio images cover an area of $21.76\degr \times 21.76\degr$ at a frequency of $180\,$MHz. We used two radio images of the MWA EoR0 field: one with a minimum baseline filter of $34\,$m and one with no filter, both having point sources with $S\ga50\,$mJy subtracted. The beam sizes are (on average) $\theta_{\rm B}=2.75\,$arcmin with no filter and $\theta_{\rm B}=2.6\,$arcmin with the filter. 

Using 1000 randomly generated galaxy number density maps, we were able to compute confidence intervals for the cross correlations of the radio images with random galaxy number density maps. We used the $99.7\,$per cent confidence intervals to set the detection thresholds. Using these detection thresholds and models of the diffuse cosmic web derived from smoothing the galaxy number density maps with Gaussian functions we are able to set upper limits on the cosmic web emission. These upper limits are $0.09$--$2.20\,$mJy beam$^{-1}$, depending on which radio images, galaxy number density maps, and Gaussian models are used, which are one to two orders of magnitude below the radio image rms values. These limits correspond to non-beam convolved values of $0.01$--$0.30\,$mJy arcmin$^{-2}$. 

Using the equipartition energy condition for converting synchrotron flux densities to magnetic field strengths we obtain upper limits on the synchrotron cosmic web magnetic field strength values of $0.18\le B_{\rm eq0} \le 0.52\, \mu$G, for values of $\alpha=-1.25$, $\eta=1.0$ (the volume filling factor), and $K_0=100$ (the ratio of number densities if cosmic ray protons and electrons). If $K_0$, $\eta$, and $\alpha$ are in the ranges $1.0\le K_0 \le 300$, $0.01\le \eta \le 1$, and $-2.25 \le \alpha \le -0.6$ we get upper limits on the field strength of $0.03 \le B_{\rm eq} \, [\mu{\rm G}] \le 1.98$. Using the HA magnetism model from \citet{Vazza15} we obtain upper limits of $0.03 \le B_{\rm HA} \, [\mu{\rm G}] \le 5.86$ for $-2.25 \le \alpha \le -0.6$ and $ 5\times 10^{-5} \le \xi \le 0.025$ and $0.09 \le B_{\rm HA0} \, [\mu{\rm G}] \le 0.41$ for $\alpha=-1.25$ and $\xi=5\times10^{-3}$. These upper limits are in the range of previous estimates and measurements of diffuse cluster emission magnetic field strengths. While these limits alone do not alone us to discriminate between cosmic magnetism models, the information does tell us that if the magnetic field strength is $\simeq 0.1\, \mu$G, radio images with rms values of $1\times10^{-4}\le S \, [{\rm mJy} \, {\rm beam}^{-1}] \le 0.5$ could be required for a $3\sigma$ detection via this method; with the true value highly dependent upon the spectral index of the synchrotron emission and the model for the magnetic field strength.
 
We identified and discussed three main obstacles or issues that affect the ability to make a detection or tighten constraints. These are
\begin{enumerate}[label={\arabic{enumi}.},leftmargin=*]
\item Point sources: Point sources, and their corresponding beam sidelobes, cause confusion in the radio image, as well as correlating with the galaxy number densities. The exact shape and amplitude of the point source contribution to the cross correlation is unknown, particularly for faint flux density sources, and therefore, the cross correlation signal due to point sources cannot be easily disentangled from the cross correlation signal due to the cosmic web. Point source subtraction is necessary, as accurately and deeply as possible.
\item Galactic emission: Galactic emission can cause correlations with galaxy number density maps over large angular scales, sometimes dominating over correlation amplitudes on small-scales due to non-Galactic emission. This results in large uncertainties in the cross correlation of the radio images with random galaxy number density maps, with those uncertainties not mainly reflecting the uncertainty due to the Galactic emission. With the detection threshold set by these uncertainties, minimizing them on the scales of interest for the cosmic web is necessary and requires radio images with minimized Galactic emission, or with some form of spatial filtering applied.
\item Synchrotron cosmic web model: Without an actual physical model predicting the cross correlation shape based on physical parameters (such as magnetic field strength, gas pressure, gas density, etc), it is difficult to say if the cross correlation function contains signal from the cosmic web, how much signal is there, and how that signal is physically interpreted. Models from MHD simulations for different magnetic field and galaxy evolution setups are necessary. However, in order to properly physically interpret the cross correlation results, these models must also provide information on how the synchrotron emission correlates spatially with either the gas or galaxy density. 
\end{enumerate}
Improvements in these three areas should greatly enhance our ability to detect or constrain the synchrotron cosmic web with the cross correlation method.

We also discussed several other possible methods for the future statistical detection of the cosmic web which include cross correlating in Fourier or wavelet space, correlating with X-ray data rather than galaxy number densities, and stacking. With the information presented here on the current limits, ways to improve the method, and other possible avenues, coupled with new data from future deep surveys such as EMU or MIGHTEE, or new MWA or LOFAR data, we believe that tighter constraints or a statistical detection of the synchrotron cosmic web should be possible in the near future.

\section{Acknowledgments}
We would like to thank Franco Vazza for the use of his simulations as well as for his helpful feedback. We would also like to thank Andre Offringa for his assistance in the imaging process. The Dunlap Institute is funded through an endowment established by the David Dunlap family and the University of Toronto. T.V. and B.M.G. acknowledge the support of the Natural Sciences and Engineering Research Council of Canada (NSERC) through grant RGPIN-2015-05948, and of the Canada Research Chairs program. This scientific work makes use of the Murchison Radio-astronomy Observatory, operated by CSIRO. We acknowledge the Wajarri Yamatji people as the traditional owners of the Observatory site. Support for the operation of the MWA is provided by the Australian Government (NCRIS), under a contract to Curtin University administered by Astronomy Australia Limited. We acknowledge the Pawsey Supercomputing Centre which is supported by the Western Australian and Australian Governments

\bibliographystyle{mnras}

\bsp

\label{lastpage}
\end{document}